\documentclass{aa}  

\usepackage{graphicx}
\usepackage{txfonts}
\usepackage{color}
\usepackage{multirow}
\begin{document} 

    \title{The TOPGöt high-mass star-forming sample }
    \subtitle{I. Methyl cyanide emission as tracer of early phases of star formation}
   %\title{TOPG\"ot: methyl cyanide emission as tracer\\ of early phases of star formation}

   %\subtitle{I. Overviewing the $\kappa$-mechanism}

   \author{C. Mininni\inst{1,2,3}, F. Fontani\inst{2}, A. S\'anchez-Monge\inst{4}, V.M. Rivilla\inst{5,2}, M.T. Beltr\'an\inst{2}, S. Zahorecz\inst{6,7},  K. Immer\inst{8,9},\\ A. Giannetti\inst{10}, P. Caselli\inst{11},  L. Colzi\inst{5,2}, L. Testi\inst{12,13}, D. Elia\inst{3}}

    \institute{Dipartimento di Fisica e Astronomia, Universit\'a degli Studi di Firenze, Via Sansone 1, 50019 Sesto Fiorentino, Italy 
               \and INAF Osservatorio Astrofisico di Arcetri, Largo E. Fermi 5, 50125 Firenze, Italy 
               \and Istituto di Astrofisica e Planetologia Spaziali, INAF, Via Fosso del Cavaliere 100, I-00133 Roma, Italy
          \and I. Physikalisches Institut, Universit\"at zu K\"oln, Z\"ulpicher Str. 77, 50937 K\"oln, Germany
            \and Centro de Astrobiolog\'ia (CSIC, INTA), Ctra. de Ajalvir, km. 4, Torrej\'on de Ardoz, E-28850 Madrid, Spain
          \and Department of Physical Science, Graduate School of Science, Osaka Prefecture University, 1-1 Gakuen-cho, Naka-ku, Sakai, Osaka 599-8531, Japan
          \and National Astronomical Observatory of Japan, National Institutes of Natural Science, 2-21-1 Osawa, Mitaka, Tokyo 181-8588, Japan
          \and Joint Institute for VLBI ERIC, Oude Hoogeveensedijk 4, 7991 PD Dwingeloo, The Netherlands
          \and Leiden Observatory, Leiden University, Postbus 9513, 2300 RA Leiden, The Netherlands
          \and INAF, Istituto di Radioastronomia, Italian ARC, Via P. Gobetti 101, Bologna, Italy
          \and Max-Planck-Institut für Extraterrestrische Physik, Gießenbachstraße 1, D-85748 Garching bei München, Germany
           \and European Southern Observatory (ESO), Karl-Schwarzschild-Str. 2, D-85748 Garching, Germany
           \and Excellence Cluster Origins, Boltzmannstrasse 2, D-85748 Garching bei M\"unchen, Germany
            }

   \date{Received ; accepted }

% \abstract{}{}{}{}{} 
% 5 {} token are mandatory
 
  \abstract
   {}
   {The TOPG\"ot project studies a sample of 86 high-mass star-forming regions in different evolutionary stages from starless cores to ultra compact HII regions. The aim of the survey is to analyze different molecular species in a statistically significant sample to study the chemical evolution in high-mass star-forming regions, and identify chemical tracers of the different phases.
   
   }
   {The sources have been observed with the IRAM 30m telescope in different spectral windows at 1, 2, and 3 mm. In this first paper, we present the sample and analyze the spectral energy distributions (SEDs) of the TOPG\"ot sources to derive physical parameters such as the dust temperature, $T_{\rm dust}$, the total column density, $N_{\rm{H_2}}$, the mass, $M$, the luminosity, $L$, and the luminosity-to-mass ratio, $L/M$, an indicator of the evolutionary stage of the sources. We use the MADCUBA software to analyze the emission of methyl cyanide (CH$_3$CN), a well-known tracer of high-mass star formation.
 .
   }
   {We built the spectral energy distributions for $\sim80\%$ of the sample and derived $T_{\rm{dust}}$ and $N_{\rm{H_2}}$ values which range between $9-36\,\rm{K}$ and $\sim 3\times10^{21}-7\times10^{23}\,\rm{cm^{-2}}$, respectively. The luminosity of the sources spans over four orders of magnitude from 30 to $3\times10^{5}\,L_{\odot}$, masses vary between $\sim30$ and $8\times10^{3}\,M_{\odot}$, and the luminosity-to-mass ratio $L/M$ covers three orders of magnitude from $6\times10^{-2}$ to $3\times10^{2}\,L_{\odot}/M_{\odot}$.
  
    The emission of the $\rm{CH_3CN(5_{K}-4_{K})}$ K-transitions has been detected towards 73 sources (85\% of the sample), with 12 non-detections and one source not observed in the frequency range of $\rm{CH_3CN(5_{K}-4_{K})}$. The emission of CH$_3$CN has been detected towards all evolutionary stages, with the mean abundances showing a clear increase of an order of magnitude from high-mass starless-cores to later  evolutionary stages. We found a conservative abundance upper limit for high-mass starless cores of $X_{\rm CH_3CN}<4.0\times10^{-11}$, and a range in abundance of $4.0\times10^{-11}<X_{\rm CH_3CN}<7.0\times10^{-11}$ for those sources that are likely high-mass starless cores or very early high-mass protostellar objects. 
    In fact, in this range of abundance we have identified five sources previously not classified as being in a very early evolutionary stage. The abundance of $\rm{CH_3CN}$ can thus be used to identify high-mass star-forming regions in early phases of star-formation.
    }
   {}

\titlerunning{TOPG\"ot }
  \authorrunning{Mininni et al. }
   \maketitle
%________________________________________________________________

\section{Introduction}
High-mass star-forming regions, the cradles of high-mass stars ($M>8\,\mathrm{M_{\odot}}$), are among the most chemically-rich sources in the Galaxy (e.g. \citealt{belloche2019}). Despite their relatively low number, these regions have a deep impact on the physical and chemical evolution of their surroundings and on the evolution of the Galaxy itself \citep{motte2018}.\newline
\indent A unique scenario for the formation of high-mass stars has not been found yet. Several models have been proposed to overcome the theoretical problem  that arises from the fact that protostars with masses  larger than $8\,M_{\odot}$ reach the main sequence before the end of the accretion process, and that the radiation emitted by the newborn star would halt accretion \citep{mckee2003,bonnell2006,keto2007,krumholz2009, motte2018}. Even if a conclusive theory describing the evolutionary sequence of high-mass star-forming regions has not been defined yet, an empirical rough classification has been proposed \citep{beuther2007, motte2018}. The earliest stage is represented by high-mass starless cores (HMSCs), with high H$_2$ volume density ($n\sim10^{5}\,\mathrm{cm^{-3}}$) and low kinetic temperatures $T\sim 15\,\mathrm{K}$, for which no evidence of ongoing star-formation (i.e. powerful outflows, strong IR sources, and/or masers) is present. This stage is followed by that of high-mass protostellar objects (HMPOs), where an accreting protostar is already present and visible at wavelength of $20-70\,\mathrm{\mu m}$, even if still embedded in the natal cloud. HMPOs are warmer than HMSCs, reaching temperatures of a few hundred K close to the central protostar(s). In these conditions, the molecules frozen onto dust grains are released in the gas-phase enriching the gas chemistry and leading to the formation of the so-called hot molecular cores (HMCs). The last  evolutionary stage is represented by ultra compact HII regions (UC HIIs), in which the UV photons from the central protostar have ionized the surrounding gas, creating a photoionized compact region (diameter $\lesssim 0.1$ pc), visible at centimeter wavelengths due to free-free emission.
\begin{table*}[h]
    \centering
     \caption{Spectral windows observed towards Sub-sample I and Sub-sample II (see Sect. 2). Column 3 gives some of the detected molecular species, already analyzed in Sub-sample I and in a few cases also in Sub-sample II, with references in column 4. }
    \begin{tabular}{p{2.5cm}p{2cm}p{8cm}p{2cm}}
    \hline\noalign{\smallskip}
     Freq. range    &Backend  & Molecular species & References\\
      (GHz) & & &\\
       \hline\noalign{\smallskip}
    \multicolumn{4}{c}{\textit{Sub-sample I}}\\
    85.31 - 87.13 & FTS50  &  H$^{13}$CN, HC$^{15}$N, HOCO$^{+}$, CH$_3$OH, SO  & (1)(2)(3)(4)   \\
    88.59 - 90.41 & FTS50 &    H$^{15}$NC, H$^{13}$CO$^{+}$, HC$_3$N    & (1)(2)(5) \\
    89.11 - 96.89 & FTS200  &    HN$^{13}$C,  $^{15}$NNH$^{+}$, N$^{15}$NH$^{+}$, CH$_2$DOH, DC$_3$N & (1)(3)(6)(7)  \\
    93.13 - 93.21 &  VESPA & N$_{2}$H$^{+}$, CH$_3$OH  & (3)(6)(8)    \\
    140.01 - 141.83 & FTS50 &    PN & (4)   \\
    143.29 - 145.11 & FTS50&  CH$_3$OCH$_3$& (9)   \\
    154.18 - 154.25 & VESPA&   N$_2$D$^{+}$, CH$_3$CHO, CH$_3$CDO, CH$_3$COCH$_3$ & (8)(9)  \\
    154.36 - 154.40 & VESPA &  CH$_3$COCH$_3$  &(8)    \\
    216.00 - 223.78 & FTS200&  $^{13}$CN,  C$^{15}$N,CH$_3$OH, CH$_2$DOH   & (3)(6)   \\
    279.48 - 279.55 & VESPA &   N$_{2}$H$^{+}$&(8)\\
    280.91 - 282.73 & FTS50 &  PN  & (4)   \\
    284.19 - 286.01 & FTS50 &  CH$_3$OCHO, CH$_3$COCH$_3$, NH$_2$CHO, C$_2$H$_5$CN & (9)  \\
     \hline\noalign{\smallskip}
     \multicolumn{4}{c}{\textit{Sub-sample II}}\\
       85.7 - 93.5 &FTS200 &$^{13}$CS, H$^{13}$CN, HC$^{15}$N,  H$^{15}$NC, HOCO$^{+}$, $^{15}$NNH$^{+}$, HC$_3$N, CH$_3$OH, CH$_2$DOH  & (10)\\
       141.2 - 149.0 &FTS200 &  CH$_3$OCH$_3$ &\\
    \hline
    \end{tabular}
   \tablefoot{(1) \citet{colzi2018a}, (2) \citet{fontani2018}, (3) \citet{fontani2015a},  (4) \citet{mininni2018}, (5) Socci et al. in prep., (6) \citet{fontani2015b}, (7) \citet{rivilla2020}, (8) \citet{fontani2011},  (9) \citet{coletta2020}, (10) \citet{colzi2018b}.  }
    \label{tab:spw}
\end{table*}

%- run descritto in Colzi et al. (2018):
%
%3mm: 85.31 - 87.13 FTS50
%    : 88.59 - 90.41 FTS50
%
%2mm: 151.75 - 153.57 FTS50
%      148.47 - 150.29 FTS50
%
%########################################
%- run descritto in Fontani et al. (2015):
%
%3 mm: 89.11 - 96.89 FTS200
%
%1 mm: 216.00 - 223.78 FTS200
%
%########################################
%- run descritto in Fontani et al. (2011), ma credo mai
%usato...
%
%3mm: 93.134 - 93.209  VESPA
%
%2mm: 154.181 - 154.253 VESPA
%      154.364 - 154.400 VESPA
%
%1mm: 279.476 - 279.548 VESPA
%
%########################################
%- run descritto in Mininni et al. (2018)
%
%2mm: 143.29 - 145.11  FTS50
%      140.01 - 141.83  FTS50
%
%0.8mm: 284.19 - 286.01  FTS50
%        280.91 - 282.73  FTS50%
\indent It is of particular interest in the field of astrochemistry to characterize the evolution of the chemistry along these evolutionary stages. This will increase the knowledge of the chemical features characterizing each stage, which in turn can shed light on the physical processes that take place, and help constraining the physical evolution. 

To reach this goal it is important to analyze statistically significant samples, in order to derive general properties for each evolutionary stage, not influenced by possible peculiarities of a single source. For this reason, we ideated the TOPG\"ot project\footnote{The name of the project originates from the city in which it has been firstly ideated, G\"othenburg, as part of the GoCAS program “Origins of Habitable Planets” that was held in the Gothenburg Centre for Advanced Studies in Science and Technology from 02/05/16 to 10/06/16: 
https://www.chalmers.se/en/centres/GoCAS/Events/Origins-of-Habitable-Planets/Pages/default.aspx }: the project unites, in the vision of a combined effort, the observations made with the IRAM 30m telescope at 1, 2, and 3mm towards two already statistically significant samples of high-mass star-forming regions (presented in Sect. \ref{sect:sampleselection}), for a total of 86 sources covering all three evolutionary stages described above. Table \ref{tab:spw} lists the spectral windows observed towards the two sub-samples, with examples of molecular species covered by each frequency range.\\ \indent In this paper, we first introduce the sample and focus on the characterization of the physical properties of the sources, analyzing their spectral energy distributions (SEDs) to derive the dust temperature, $T_{\rm dust}$, the H$_2$ column density, $N_{\rm{H_2}}$, the mass, $M$, and the luminosity, $L$, of the sources. The H$_2$ column density allows us to derive the molecular abundances, while the other observables allow us to search for (anti-)correlations between chemical abundances and physical properties of the sources. 
Following the SED analysis, we characterize the emission of methyl cyanide (CH$_3$CN) in the (5$_{\rm{K}}-$4$_{\rm{K}}$) rotational transition, a well-known high-density tracer in high-mass star-forming regions (e.g. \citealt{churchwell1992}). This molecule, firstly detected in its (6$_{\rm K}-$5$_{\rm K}$) transition towards the Galactic Center by \citet{solomon1971}, is a symmetric top and its transitions are characterized by the two quantum numbers $J$, the total angular momentum, and $K$, the projection of the total angular momentum on the symmetry axis. Since quantum mechanic selection rules 
allow only $\Delta J=\pm1$ and $\Delta K=0$ transitions, the population of the different K components can be used to infer a reliable estimate of the kinetic temperature. The analysis of methyl cyanide is also favoured by the fact that the lines with different $K$ in a given rotational transition are separated by a few MHz (see Table 2), i.e. much less than the bandwidth of any spectrometer available at modern radio telescopes. Therefore, these different $K$-components can be observed simultaneously in the same spectral window, and their intensities are independent of relative calibration uncertainties. 
\newline\indent After its first detection several studies detected methyl cyanide in high-mass star-forming regions \citep{bergman1989,churchwell1992,olmi1993,olmi1996singledish,zhang1998,kalenskii2000,pankonin2001, remijan2004,araya2005,purcell2006, rosero2013,minh2016,hung2019} showing emission associated with HMPOs and UC HII regions. \citet{olmi1996interf}, with interferometric observations, and \citet{giannetti2017}, observing a large sample of cores (including HMSCs), found that, while high-energy $K$ components of methyl cyanide trace only the most compact and warmer cores, low-energy $K$ components  trace also more extended and colder gas.

Other high-angular resolution studies have highlighted that methyl cyanide is a good tracer of rotating toroids and infall in high-mass protostars \citep{cesaroni1999,beltran2004,beltran2005,beltran2011,beltran2018,furuya2008}. 
Finally, this species has also been found in gas affected by the passage of a shock wave, as in L1157-B1 \citep{arce2008,codella2009}, in circumstellar envelopes of evolved stars (i.e. \citealt{agundez2015}), in disks \citep{oberg2015,johnston2015,bergner2018,loomis2018}, and cold dense cores (e.g. \citealt{potapov2016,spezzano2017}).\newline
\indent The paper is structured as follows: in Sect. 2 we present the sample and its selection criteria. In Sect. 3 we describe the observations of the CH$_3$CN(5$_{{K}}-$4$_{{K}}$) lines with the IRAM 30m telescope. In Sect. 4 we present the analysis of the SEDs towards the 86 objects in the sample, how we derived the physical quantities of $T_{\rm dust}$, $N_{\rm{H_2}}$, $L$ and $M$, and the analysis of the methyl cyanide ($5_{{K}}-4_{{K}}$) rotational transitions. In Sect. 5 we discuss the results.  
Finally in Sect. 6 we summarize the conclusions. 

\section{The sample}
\label{sect:sampleselection}

The sample of the TOPG\"ot project arises from the combination of two separate sub-samples of high-mass star-forming regions containing 86 targets in total. \newline \indent

\subsection{Sub-sample I}
The first sub-sample (hereafter Sub-sample I), firstly presented in \citet{fontani2011}, consists of 27  sources ($\sim$31\% of the total sample) for which we already have an evolutionary classification: 11 are
HMSCs, 9 are HMPOs, and 7 are UC HII regions. 
\subsubsection{Selection criteria for HMSCs}
The HMSCs have been selected as massive cores embedded in infrared dark-clouds or other massive star-forming regions in which no evidence of ongoing star formation was present. \citet{fontani2011} checked that no embedded infrared sources, powerful outflows or maser emission were detected towards these sources. Later, \citet{tan2016} found that a source classified as HMSC, G028-C1, is associated with a highly-ordered outflow. However, given that this outflow is still in an initial phase with no powerful emission
and that the source shows bright emission of N$_2$D$^+$(3--2), which is a tracer of dense and cold gas of pre-stellar cores (as discussed in  \citealt{tan2016}), we decided to still classify G028-C1 as HMSC. In fact, the three different classes are not sharply separated, and all the other observational parameters checked for this source are more similar to  those of other objects classified as HMSCs than those of HMPOs. Among the HMSCs, three sources (AFGL5142-EC, 05358-mm3, and I22134-G) have been defined as “warm” (HMSC$^{w}$), since they have temperatures $T_{\rm k}> 20\,\mathrm{K}$ (see Table A.3 of \citealt{fontani2011}), derived from ammonia rotation temperatures following \citet{tafalla2004}. This is confirmed by high-angular resolution studies that indicate that they could be externally heated by nearby protostellar objects \citep{zhang2002, busquet2010,sanchezmonge2011,colzi2019}. 
\subsubsection{Selection criteria for HMPOs and UC HII regions}
HMPOs have been selected as high-mass sources associated with infrared sources, and/or powerful outflows and/or faint (S$_{\nu}$ at 3.6 cm $<1$ mJy) radio continuum emission likely tracing a thermal radio jet. 
 UC HIIs are associated with strong radio continuum emission
(S$_{\nu}$ at 3.6 cm $\geq 1$ mJy) tracing gas ionized by the UV photons emitted by a young massive star. More evolved sources, in which HII regions have already dissipated the associated molecular cores, were not included.\\\newline \indent A small part of the observations towards these sources has already been analyzed and published in previous works covering different topics, from deuteration  in selected molecules (N$_2$H$^{+}$, CH$_3$OH, NH$_3$, HC$_3$N; \citealt{fontani2011,fontani2015a,rivilla2020}), to nitrogen fractionation of HCN, HNC and N$_2$H$^+$ \citep{fontani2015b, colzi2018a}, to prebiotic and complex organic molecules (COMs, \citealt{mininni2018,coletta2020}).\newline\indent 
\subsection{Sub-sample II}
The second sub-sample (hereafter Sub-sample II) consists of 59 high-mass star-forming regions ($\sim$69\% of the total sample) located in the northern hemisphere. The sources were identified from a number of available mm-continuum emission surveys of star-forming regions \citep{molinari1996, beuther2002, mueller2002, sridharan2002, sanchezmonge2008}, with the goal of studying the chemical evolution of star-forming regions and statistically search for the presence of chemically-rich hot molecular cores. The sources were selected primarily based on two aspects: (i) sources with kinematic distances\footnote{Further detailed studies of the distances of the sources have placed some of them at farther distances.} $\le5.5$~kpc,to reduce beam-dilution effects and therefore increase the probability to detect molecular emission, and (ii) presence of a bright and dense gas and dust condensation, mainly evaluated on the basis of bright peak intensities ($\ge0.5$~Jy~beam$^{-1}$) at millimeter wavelengths, suggestive of high-mass young stellar objects embedded in dust.\newline\indent For the sources in Sub-sample II a previous evolutionary classification has not been performed yet, however we expect that most of the sources are in the evolutionary stages of HMPOs or UC HII regions, from criterion (ii). A first attempt of classification based on the abundances of methyl cyanide, together with other evolutionary indicators, will be presented in Section 5. 
\newline\newline
The two sub-samples have been separately observed in several spectral windows with the IRAM30m telescope, leading to a collection of ancillary data covering the range of emission of several molecular species (Table 1). Since the observed spectral windows of Sub-sample II overlap with the spectral windows of Sub-sample I, it was decided to carry out this project by merging the two samples to increase the statistical value of this study. Moreover, the project will include future studies on other molecular species detected in the available ancillary data. Thus, this paper is also a reference paper for the project, where we characterize some important physical properties of the sources needed for the analysis of the molecular emission of different species. 
\\ \indent
The merged sample has been already used to study the nitrogen fractionation in HCN and HNC by \citet{colzi2018b}. \\ \indent
The coordinates of the sources, the distance $d$, the velocity v$_{LSR}$, and the evolutionary classification are given in Table 3. The sources belonging to Sub-sample I are those that have an evolutionary classification, while for sources of Sub-sample II the last column in Table 3 is empty. All the $d$ are kinematic distances, with the exception of the distance estimate of G31.41+0.31 and G35.20-0.74 that are derived from parallax measurements \citep{immer2019,zhang2009}. When the distance ambiguity was not resolved (14\% of the full sample, 17\% of the sources with SED, and 14\% of the sources for which we detected CH$_3$CN) we adopted the near distance.  Since only in a low percentage of the sample the distance ambiguity is not resolved, this does not affect the interpretation of the data and the results, especially considering that in part of the discussion we will make use of distance-independent parameters. 

\section{Observations}

The observations of the CH$_3$CN(5$_K$--4$_K$) transition towards the Sub-sample I analyzed in this work are taken from the dataset described in \citet{fontani2015a}, but have not been previously analyzed. Due to time constraints the source G028-C3(MM11) was not observed in the run targeting the CH$_3$CN($5_{{K}}-4_{{K}}$) lines. However, this source is presented in this paper since the derivation of the physical parameters from the SED will be needed to analyze observations of other species, presented in future papers.\newline\indent
More details about the observations towards Sub-sample I can be found in \citet{fontani2015a}.\newline \indent
The sources in Sub-sample II were observed using the IRAM 30m telescope from 11 to 16 August 2014 (project 040-14, PI: S\'anchez-Monge). We simultaneously observed two bands at 3 and 2~mm covering some important rotational transitions of common species such as HCO$^+$, HCN, HNC, N$_2$H$^+$ and SiO, as well as transitions of complex organic molecules such as CH$_3$CN and CH$_3$OH. The observed frequencies are 85.7--93.5~GHz and 141.2--149.0~GHz (see Table~1). The atmospheric conditions were stable during the observing period, with precipitable water vapor usually between 4 and 8~mm. Pointing was checked every 1.5 
or 2 hours on nearby quasars. The spectra were obtained with the fast Fourier transform spectrometers (FTS) providing a broad frequency coverage of 16~GHz in total at a resolution of 200~kHz. This spectral resolution corresponds to 0.7 and 0.4~km~s$^{-1}$ for the 3~mm and 2~mm bands, respectively.\newline \indent
The data of both sub-samples were calibrated with the chopper wheel technique, with a calibration uncertainty of $\sim20-30\%$. The spectra were obtained in antenna temperature units, $T^{\ast}_{\rm A}$, and then converted to main beam brightness temperature, $T_{\rm MB}$, via the relation $T^{\ast}_{\rm A} = T_{\rm MB}\eta_{\rm MB}$, where $\eta_{\rm MB}= \mathrm{B_{eff}/F_{eff}}$
is 0.84\footnote{see https://publicwiki.iram.es/IRAM30mEfficiencies} for CH$_3$CN($5_{{K}}-4_{{K}}$) lines.\\ \indent
The spectroscopic data were taken from the Cologne Database for Molecular Spectroscopy\footnote{https://cdms.astro.uni-koeln.de} (CDMS, \citealt{cdms2001,cdms2005}). The entry of methyl cyanide in the CDMS is based on the spectroscopic work of \citet{muller2015}\footnote{For a complete documentation see https://cdms.astro.uni-koeln.de/cgi-bin/cdmsinfo?file=e041505.cat .}. The main spectroscopic parameters of the $K$-transitions are given in Table 2, 
where we can see that the $K$=3 transition has two components at the same frequency. This is due to the fact that for CH$_3$CN $K=3n$ the transitions belong to both the A1 and A2 species (for $n=0$ only to A1), while all the other transitions belong to E species.

\begin{table}
  \caption{Parameters of the $K$-ladder of CH$_{3}$CN(5$-$4): frequency, quantum numbers $K_{\rm{u}}$ and $K_{\rm{l}}$ ($K$ number of the upper and lower state, respectively, where the sign differentiate the A1 and A2 species for the $K=3$ component),  upper state energy $E_{\rm{u}}$, degeneracy of the upper level $g_{\rm{u}}$, and integrated intensity of the line at 300\,K, $I$.}
    \centering
    \begin{tabular}{cccccc}
    \hline\noalign{\smallskip}
      $\nu$ &  K$_{\rm{u}}$ &  K$_{\rm{l}}$ & $E_{\rm{u}}$ & $g_{\rm{u}}$ & $\mathrm{log(}I\mathrm{/[nm^{2}\,MHz])}$\\
      $\mathrm{[MHz]}$ &   & & [K] & & \\
    \hline\noalign{\smallskip}

   91958.7263  &   4    &     4 & 123.2& 22 &-3.8674  \\      
   91971.1307  &   3    &    -3 & 73.2 & 22 &-3.5450  \\      
   91971.1307  &  -3    &     3 & 73.2 & 22 &-3.5450  \\      
   91979.9946  &   2    &     2 & 37.4 & 22 &-3.3752  \\      
   91985.3144  &   1    &     1 & 16.0 & 22 &-3.2861  \\      
   91987.0879  &   0    &     0 &  8.8 & 22 &-3.2580  \\      
    \hline
    \end{tabular}
  
    \label{tab:ch3cnfrequency}
\end{table}

%old spectroscopy
%0 & 91.9871 & 13.2 & -4.36\\
%1 & 91.9853 & 20.4 & -4.38\\
%2 & 91.9800 & 41.8 & -4.44\\
%3 & 91.9711 & 77.5 & -4.55\\
%4 & 91.9587 & 127.5 & -4.80\\
%new spectroscopy CDMS
%  91958.7263  0.0001   3   85.5791 22  415051303 5 4 0       4 4 0       
%   91971.1307  0.0001  3   50.8290 22  415051303 5 3 0       4-3 0       
%   91971.1307  0.0001  3   50.8290 22  415051303 5-3 0       4 3 0       
%   91979.9946  0.0001  3   26.0019 22  415051303 5 2 0       4 2 0       
%   91985.3144  0.0001  3   11.1034 22  415051303 5 1 0       4 1 0       
%   91987.0879  0.0001  3    6.1368 22  415051303 5 0 0       4 0 0       

\setlength{\tabcolsep}{5pt}

%\longtab[1]{

\renewcommand{\arraystretch}{1.2}
%\begin{longtable}{clccccccl}
\begin{table*}[]
\centering
\caption{Sources coordinates, heliocentric distances, line-of-sight velocities, and evolutionary phases (e.p.) for the first 10 sources. The full table is available at the CDS.\\}
\label{tablesourcescoord}
\begin{tabular}{clccccccl}

\hline
 &Source& $\alpha({\rm J2000})$ & $\delta({\rm J2000})$ & l & b & d & v$_{\rm LSR}$ & e.p. \\
 & & [h m s] & [$^{\circ}$ $\arcmin$ $\arcsec$] & [$^{\circ}$] & [$^{\circ}$] & [kpc] & [$\mathrm{km\,s^{-1}}$] & \\ 
\hline
%\endfirsthead
%\caption{Continued.} \\
%\hline
%&Source& $\alpha({\rm J2000})$ & $\delta({\rm J2000})$ & l & b & d & v$_{\rm LSR}$ & e.p.  \\
%&& [h m s] & [$^{\circ}$ $\arcmin$ $\arcsec$] & [$^{\circ}$] & [$^{\circ}$] & [kpc] & [$\mathrm{km\,s^{-1}}$] & \\ 
%\hline
%\endhead
%\hline
%\endfoot
%\endlastfoot

1 &I00117--MM1 & 00 14 26.1& +64 28 44.0& 118.96 & +1.89 & 1.8$^{a}$ &$-$36.3 &HMPO \\
%1 & \textcolor{red} 00117+6412 M1 & 00 14 26.1 & +64 28 44.0& 118.96 & +1.89 & 1.8$^{a}$ \\
2 &I00117--MM2 & 00 14 26.3& +64 28 28.0& 118.96 & +1.89 & 1.8$^{a}$&$-$36.3 &HMSC \\
%3 &I04579--VLA1 & 05 01 39.9& +47 07 21.0& 160.14 & +3.16 & 2.5$^{b}$& \\
%3 & \textcolor{red} 04579+4703 & 05 01 39.9& +47 07 21.0& 160.14 & +3.16 & 2.5$^{a}$ \\
3 &AFGL5142--MM & 05 30 48.0& +33 47 54.0& 174.20 & $-$0.07 & 1.8$^{b}$&$-$3.9 &HMPO \\
4 &AFGL5142--EC & 05 30 48.7& +33 47 53.0& 174.20 & $-$0.07 & 1.8$^{b}$&$-$3.9 &HMSC$^{w}$ \\
5 &05358--mm3& 05 39 12.5& +35 45 55.0& 173.48 & +2.45 & 1.8$^{b}$&$-$17.6 &HMSC$^{w}$ \\
6 &05358--mm1& 05 39 13.1& +35 45 51.0& 173.48 & +2.45 & 1.8$^{b}$&$-$17.6 &HMPO\\
7 &G5.89--0.39 & 18 00 30.5& $-$24 04 01.0& 5.89 & $-$0.39 & 1.3$^{b}$& +9.0&UC HII \\
8 & G008.14+0.22 & 18 03 01.3& $-$21 48 05.0& 8.14 & +0.22 & 3.4$^{c}$& +20.1&\\
9&18089$-$1732M1 & 18 11 51.4& $-$17 31 28.0& 12.89 & +0.49 & 3.6$^{b,\ast}$ &+32.7 &HMPO \\
%10& \textcolor{red} 18089$-$1732 M1& 18 11 51.5& $-$17 31 29.0& 12.89 & +0.49 & 3.6$^{c}$ \\
10&18089$-$1732 M4& 18 11 54.0& $-$17 29 59.0& 12.92 & +0.49 & 3.6$^{d,\ast}$ &+33.8 & \\

\hline
%\end{longtable}
\end{tabular}
\tablefoot{For the sources present in the Hi-GAL catalog the distance values were taken from Elia et al. (2017) (distance match $<10\arcsec$). The sources of Sub-sample I present an evolutionary classification in column 9, while for sources of Sub-sample II column 9 is empty. For the other sources: a) \citet{molinari1996}; b) \citet{fontani2011}; c) \citet{faundez2004}; d) \citet{sridharan2002}. %e) \citet{hill2005}; f) \citet{immer2019};  g) \citet{zhang2009}; h) \citet{pandian2009};  i) M\'ege et al in prep.; j) \citet{sanchezmonge2013b}; k) \citet{sanchezmonge2008}. 
$\ast$) ambiguity in the distance estimate not resolved, the adopted distance is the $d_{\rm{near}}$; $\ast\ast$) during observations v$_{\rm LSR}$ differed from the correct value reported in this Table. $w$: HMSC defined as \textit{warm} \citep{fontani2011}. %j) \citet{ando2011}
%\textcolor{red}{$^{ab}$Sanchez-Monge et al. 2008 (Mueller et al. 2002); $^{c}$Sridharan et al. 2002 or Beuther et al. 2002; $^{d}$Pandian et al. ??; $^{e}$Fontani et al. 2011; $^{f}$Mege et al. (in prep).; $^{g}$Faundez et al ?; $^{h}$Hill et al. ?; $^{i}$Immer er al. (2019); $^{j}$Beltran et al. 2014; $^{k}$Sanchez-Monge et al. 2013; $^{l}$Sanchez-Monge et al. 2011;\\ }
}
\end{table*}
%}%end longtab[]

\section{Analysis}
\subsection{Spectral Energy Distributions (SEDs)}
\indent To determine the column density of H$_2$ and the dust temperature, $T_{\mathrm{dust}}$, in the sources of the sample we analysed the continuum SED of these objects. The  $N_{\rm{H_{2}}}$ also allows us to obtain an estimate of the mass of the targets, $M$ (see Sect. \ref{sect:massandL}), and to determine the abundances of methyl cyanide, $X_{\rm{CH_3CN}} = N_{\rm{CH_3CN}}/ N_{\rm{H_2}}$ (see Sect. \ref{sect:ch3cn}).
\subsubsection{Continuum flux densities}
\indent The SEDs were built using the maps from the Hi-GAL survey (\textit{Herschel} Infrared GALactic plane
survey, \citealt{molinari2010,molinari2016,elia2017}) in the four bands at $160\,\mu\rm{m}$, $250\,\mu\rm{m}$, $350\,\mu\rm{m}$ and $500\,\rm{\mu m}$ of the PACS and SPIRE instruments of \textit{Herschel}. To remove the background emission, all the maps have been smoothed to the same resolution of $5^{\prime}$ (starting from a resolution of $\sim14\arcsec$, 24\arcsec, 31\arcsec, and 44\arcsec \,for maps from $160\,\mu\rm{m}$ to $500\,\rm{\mu m}$, respectively - \citealt{traficante2011}) and we have subtracted the smoothed maps from the original ones. % used the smoothed maps as background. 
Further details on this method can be found in Appendix C of \citet{zahorecz2016}. We performed 2D Gaussian fits of the continuum emission of the sources at 250$\,\mu$m,  to determine their mean angular dimension $\theta = \sqrt{\theta_{\rm{a}}\theta_{\rm{b}}}$, where  $\theta_{\rm{a}}$ and  $\theta_{\rm{b}}$ are the FWHM of the 2D Gaussian along the major and minor axis. At this wavelength the optical depth is smaller with respect to that at 160$\,\mu$m, and the angular resolution is higher than the angular resolution of the maps at larger wavelengths. The source FWHM was resolved for $\sim65$\% of the sample (45 sources). For the remaining 21 sources ( $\sim30$\% of the sample)  we adopted $\theta=23.9\arcsec$, the HPBW at  250$\,\mu$m. For three sources - G028--C1(MM9), G034--F2(MM7), and G034--G2(MM2) - the fit does not properly converge. We thus estimated the source size by measuring the dimension of the contour at which the intensity is a factor of two lower than the peak intensity.
These values are more uncertain, and are indicated with the letter "V" in the flag of $\theta$, given in Table 4.\\\indent The flux densities at different wavelengths have been extracted from a region of 45\arcsec of angular diameter around the peak position of the sources (see Table \ref{tablesourcescoord}), with the exception of G5.89--0.39, G008.14+0.22, 19413+2332M1, 20343+4129M1, and NGC7538--IRS1, for which we have derived values of $\theta>30\arcsec$. For these sources, we have extracted the flux from a region of 90\arcsec of angular diameter, to avoid a significant flux loss. The maps at $250\,\rm{\mu m}$ of the sources 18089-1732, 19095+0930, G014.33-0.65, G024.78+0.08, G031.41+0.31, G035.20-0.74, and G5.89-0.39 were saturated. However, the small number of pixels affected by this problem in each source has allowed us to reconstruct the total fluxes using CuTEx \citep{molinari2011} assuming a 2D gaussian shape of the sources brightness distribution.\\ \indent We have extracted the flux densities also from the \textit{Herschel} $70\,\rm{\mu m}$ band, and used them to calculate the luminosity of the sources (see Sect. 4.1.3). These values have not been used to constrain the fit of the SEDs, since a consistent part of the flux at this wavelength is likely contaminated by the emission of very small grains, having temperatures different from those of larger grains \citep{compiegne2010}. Column 3 of Table 4 lists the 69 sources for which the \textit{Herschel} maps were available. For these sources we completed the SEDs using maps from the APEX Telescope Large Area Survey of the Galaxy (ATLASGAL, \citealt{schuller2009,csengeri2014}) at $870\,\rm{\mu m}$  or, if not available, from the SCUBA Legacy Catalogues\footnote{http://www.cadc-ccda.hia-iha.nrc-cnrc.gc.ca/community/scubalegacy/} at $850\,\rm{\mu m}$ \citep{difrancesco2008}. For only five of these 69 targets, both the 870 and 850$\,\mathrm{\mu m}$ maps are not available. 
Column 4 of Table 4 indicates for each source if the flux used to complete the SED is at 850 or $870\,\rm{\mu m}$. The calibration errors on the fluxes of \textit{Herschel} bands have been assumed to be 5\% \citep{balog2014,bendo2013}, while for ATLASGAL and SCUBA Legacy Catalogue fluxes we assumed calibration errors of 15\% and 20\%, respectively (see \citealt{csengeri2014,difrancesco2008}).

\subsubsection{Spectral Energy Distribution fitting}
\begin{figure}
\begin{picture}(100,430)
\centering
\put(10,200){\includegraphics[trim=0 10 0 0, clip,width=0.9\columnwidth]{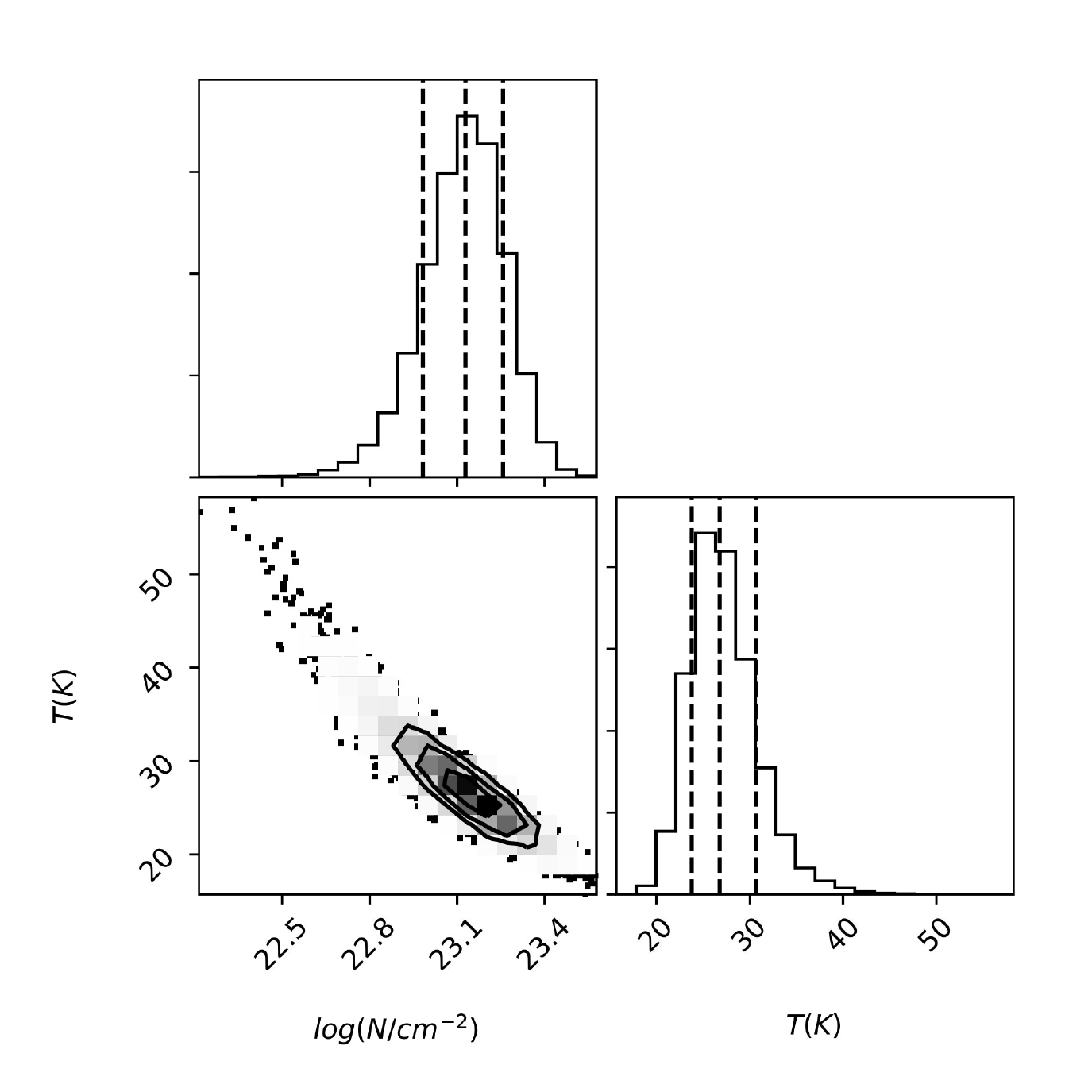}}
\put(10,40){\includegraphics[trim=0 0 0 58 , clip,width=0.95\columnwidth]{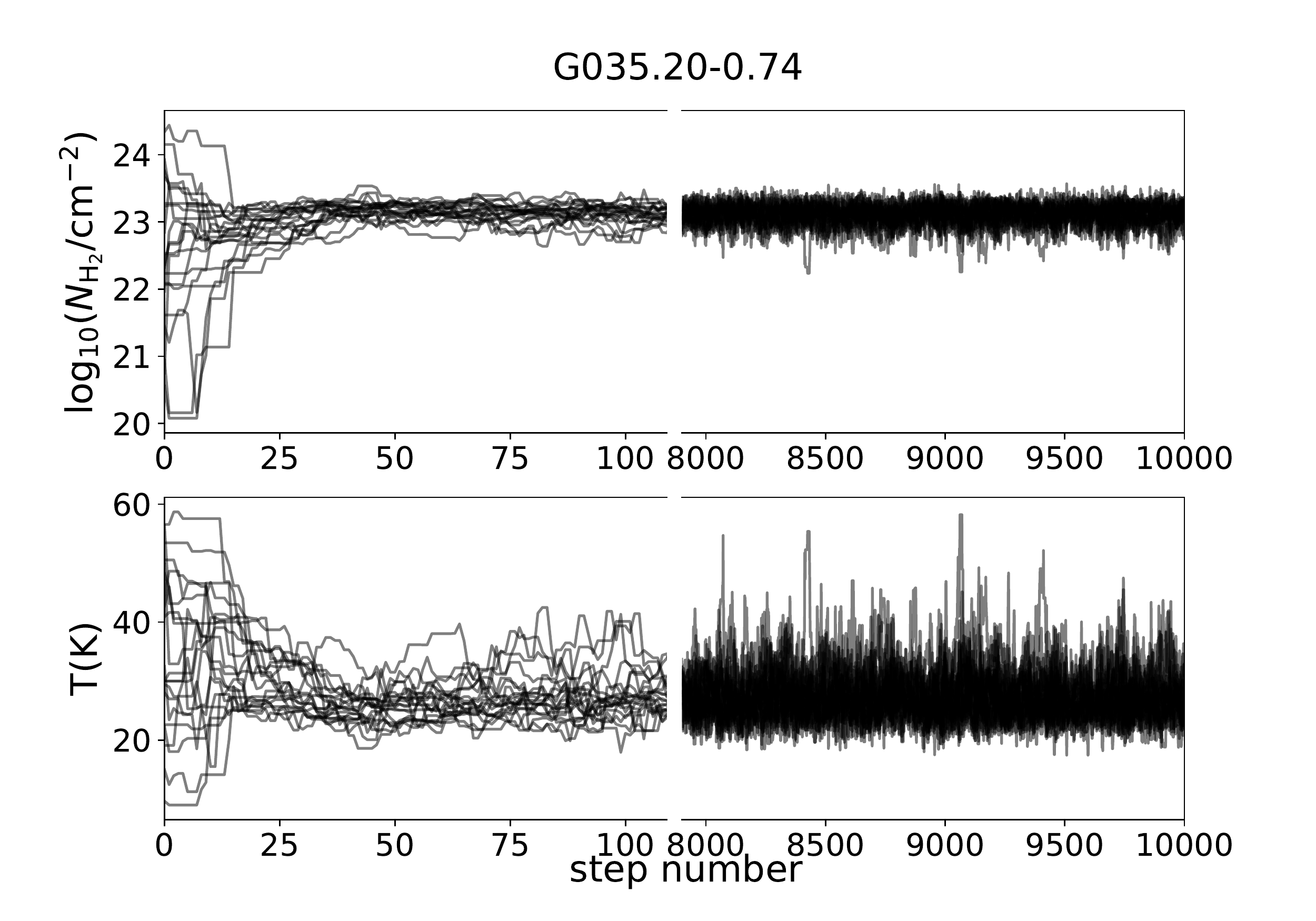}}
\put(112,423){G035.20-0.74}
\end{picture}
\vspace{-1.4cm}
\caption{\textit{Three upper panels}: corner plot for the source G035.20-0.74 of the two parameters $N(\rm{H_2})$ and $T_{\rm{dust}}$ free to vary in the MCMC used to fit the SED. The %gray scale in the bottom-left panel and the 
units of the y-axes of the other two panels are counts. \textit{Two lower panels}: walkers of the two parameters $N_{\rm{H_2}}$ and $T_{\rm{dust}}$, for the source G035.20-0.74.}
\label{fig:exampleMCMC}
\end{figure}
\begin{figure}
    \centering
    \includegraphics[width=0.8\columnwidth]{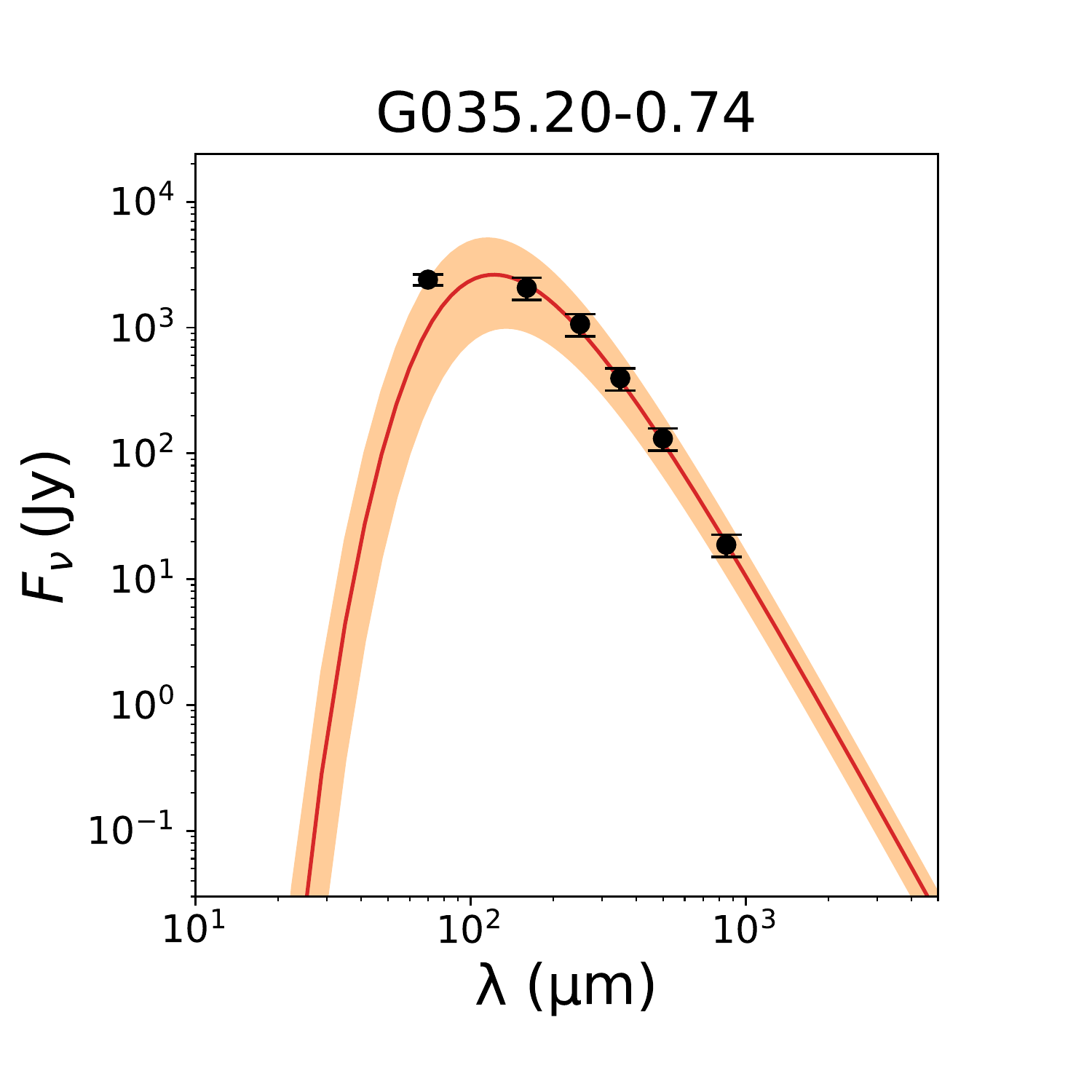}
    \caption{SED of the source G035.20-0.74. The red line is the best fit to the SED, obtained using the parameters $N_{\rm{H_2}}$ and $T_{\rm{dust}}$ given in Table 4. %\ref{tableSED}.
    The orange area represents the variation from the best fit if the two parameters are varied inside the uncertainties. The point at 70$\,\mu \rm{m}$ has not been used to constrain the fit of the SED, but it has been used to derive the luminosity of the sources.}
    \label{fig:exampleSED}
\end{figure}
\setlength{\tabcolsep}{5pt}

%\longtab[2]{
\label{tableSED}
\renewcommand{\arraystretch}{1.2}
%\begin{longtable}{clcccccc}
\begin{table*}

\caption{Availability of Herschel photometric data, SED fit results for the parameters $N(\mathrm{H_2})$ and $T_{\mathrm{dust}}$, angular dimension $\theta$ of the source from the 2D fitting at 250$\,\mu$m and flag on the estimate of $\theta$ for the first 10 sources of the sample. The SED fitting has been performed assuming $\beta=2$. The full table is available via the CDS.\\}
    \centering
    \begin{tabular}{clcccccc}

\hline
 &Source& 70  $\rm{\mu\,m}$ - 500  $\rm{\mu\,m}$ & 850 $\rm{\mu\,m}^{a}$/870  $\rm{\mu\,m}^{b}$ & $N(\mathrm{H_2})$ & $T_{\mathrm{dust}}$ & $\theta$ &  \\
 &  & & & [$10^{22}$ cm$^{-2}$] & [K] &[''] &  \\ 
\hline
1     &  I00117--MM1& Y & - & $(1.6\pm0.7)\times10^{22}$ & $21\pm4$ & 27.8 & R\\
%1    & \textcolor{red} 00117+6412 M1\\
2     &  I00117--MM 2& Y & - & $(2.3\pm1.0)\times10^{22}$ & $19\pm3$ & 23.9 & NR\\
%3     &  I04579--VLA1& - & - & - & - \\
%3    & \textcolor{red} 04579+4703\\
3     &  AFGL5142--MM   & Y & 850 & $(1.2\pm0.5)\times10^{23}$ & $26\pm5$ & 22.2 & R\\
4     &  AFGL5142--EC& Y & 850 & $(8.0\pm4.0)\times10^{22}$ & $27\pm6$ & 23.9 & NR\\
5     &  05358--mm3    & - & - & - & - \\
6     &  05358--mm1    & - & - & - & - \\
7     &  G5.89--0.39& Y & 870 & $(1.8\pm0.8)\times10^{23}$ & $27\pm5$ &  40.8 & R\\
8     & G008.14+0.22& Y & 850 & $(1.7\pm0.6)\times10^{23}$ & $27\pm4$ &  30.0 & R\\
9    &  18089$-$1732M1& Y & 870 & $(2.8\pm1.0)\times10^{23}$ & $27\pm4$ &  17.5 & R\\
%10   & \textcolor{red} 18089$-$1732 M1\\
10    &18089$-$1732 M4& Y & 850 & $(4.8\pm1.5)\times10^{22}$ & $22\pm3$ &  17.5 & R\\
\hline
\end{tabular}
\tablefoot{a) SCUBA Legacy Catalogues \citep{difrancesco2008}; b) ATLASGAL \citep{schuller2009,csengeri2014}; R = resolved sources; NR = unresolved sources for which we adopted $\theta = 23.9\arcsec$.}
\end{table*}
%\end{longtable}
%\tablefoot{}
%}%end longtab[]

\label{sect:sedfit}
\indent The SEDs have been modeled as a modified black body with
\begin{equation}
F_{\nu}= B_{\nu}(T_{\rm{dust}})\,(1-\rm{e}^{-\tau_{\nu}})\,\Omega\,,
\label{eq:greybody}
\end{equation}
where $F_{\rm{\nu}}$ is the observed
flux density at the frequency $\nu$, $B_{\rm{\nu}}(T_{\rm{dust}})$ is the Planck function at the
dust temperature $T_{\rm{dust}}$, $\tau_{\rm{\nu}}$ is the optical depth at the frequency $\nu$ and $\Omega$ is the source solid angle, given by $\Omega = \pi/(4 \ln2)\,\theta^2$. The optical depth can be written as:
\begin{equation}
\label{eq:tau}
\tau_{\rm{\nu}} = \mu_{\rm{H_2}}\,m_{\rm{H}}\,N_{\rm{H_2}}\,\kappa_{\nu_0}\,\Bigl(\frac{\nu}{\nu_0}\Bigr)^{\beta}\,,
\end{equation}
where $\mu_{\rm{H_2}}$ is the mean molecular weight, $m_{\rm{H}}$ is the mass of the hydrogen atom, $N_{\rm{H_2}}$ is the molecular hydrogen column density, $\kappa_{\nu_0}$ is the dust opacity at the reference frequency $\nu_{0}$ and $\beta$ is the spectral index. We have adopted a value of 2.8 for $\mu_{\rm{H_2}}$ \citep{kauffmann2008}, a dust opacity of $0.8\,\rm{cm^2\,g^{-1}}$ at $\nu_{0}=230\,\mathrm{GHz}$ \citep{ossenkopf1994}, assuming a gas-to-dust ratio of 100, and a value of $\beta=2$.\\

 \indent The SEDs have been fitted using a Monte Carlo Markov Chain (MCMC) algorithm that minimizes the $\chi^{2}$ modelling the observed data with Eq. (\ref{eq:greybody}), with  20 chains and a number of iterations equal to 10000. The two free parameters, $T_{\rm{dust}}$ and $N_{\rm{H_2}}$, have been constrained inside a range of $6-60\,\rm{K}$ and $10^{20}-10^{25}\,\rm{cm^{-2}}$ respectively. These ranges are consistent with mean values from \citet{elia2017}.
A first run with a lower number of iterations has been done with larger boundaries, and all the preliminary fit values have been found within the ranges given above, validating the choice of the ranges. The results of the best fit are given in Table 4: $T_{\rm{dust}}$ ranges from 9 to $36\,\rm{K}$, while the values of $N_{\rm{H_2}}$ cover about two orders of magnitude from $3\times10^{21}$ to $7\times10^{23}\,\rm{cm^{-2}}$. %\ref{tableSED}.
Figure \ref{fig:exampleMCMC} shows the path of the random walkers for the two free parameters and the corner plot with the results of the fit, for the source G035.20-0.74 as an example. The  SED and the best fit model are shown in Fig. \ref{fig:exampleSED}. The orange area shows the results of the models using values for $T_{\rm{dust}}$ and $N_{\rm{H_2}}$ within the uncertainties from the best-fit values. 

\subsubsection{Mass and luminosity estimates}
\label{sect:massandL}

\indent Two important physical parameters are the mass and the luminosity of the sources. The mass was derived using the molecular hydrogen column density derived in Sect. \ref{sect:sedfit}:
\begin{equation}
    \label{eq:mass_sed}
    M_{\rm{SED}} = \mu_{\rm{H_2}}\,m_{\rm{H}}\,N_{\rm{H_2}}\,d^{2}\,\Omega\,,
\end{equation}
where $d$ is the distance of the source. 

 The total flux $F$ in the far-IR band, for the 69 objects for which we have been able to construct the SEDs, has been calculated as the discrete integral of the best fit model of the SED between $10\,\rm{\mu m}$ and $3\,\rm{mm}$:
\begin{equation}
    F = \int F_{\rm{\lambda}}\,\rm{d}\lambda\,.
    \label{eq:Ftot}
\end{equation}
\indent For sources for which the ratio between the observed flux at 70$\,\rm{\mu m}$ and the flux at 70$\,\rm{\mu m}$ of the best-fit model of the SED is larger than 10\% ($\sim80\%$ of the sources), we added the flux excess to the total flux from Eq. (\ref{eq:Ftot}). The flux excess  has been calculated assuming that for $\lambda < \lambda_{\rm{p}}$, where $\lambda_{\rm{p}}$ is the wavelength at which $F_{\lambda}$ peaks (in our sample always larger than 70$\,\rm{\mu m}$), the flux of the source at any $\lambda$ is given by the value derived from the linear interpolation between the flux at $\lambda_{\rm{p}}$ and the flux at 70$\,\rm{\mu m}$, or between the flux at 70$\,\rm{\mu m}$ and the flux at $10\,\rm{\mu m}$, for $10\,{\rm{\mu m}}<\lambda<70\,{\rm{\mu m}}$.  The luminosity has been derived from the total flux using: 
\begin{equation}
    \label{eq:luminosity}
    L = 4\pi\,d^{2}\,F\,.
\end{equation}
\indent The modified blackbody emission of dust modeled is expected to underestimate the flux densities for wavelengths shorter than $70\,\rm{\mu m}$ for evolved sources such as HMPOs and UC HII regions (see SEDs in \citealt{konig2017}). We have calculated a part of the flux excess from the values of flux at 70$\,\rm{\mu m}$. However, what we included is not the totality of the flux excess. Thus the luminosities of HMPOs and UC HIIs are likely underestimated, because it is not possible to derive from the data collected here the part of flux excess at shorter wavelengths. For the cold and least evolved sources, HMSCs, the emission at shorter wavelengths is expected to be negligible and not to show excess from the observed SED. Table 5 lists  the luminosity $L$ and the ratio $L/M$. The ratio  $L/M$, which is a distance independent parameter, is an indicator of evolution since its value increases as a source evolves: in the first evolutionary phase more gas is converted into stars during the star formation process and the embedded source(s) becomes more luminous, thus the mass of the clump remains nearly constant while its luminosity increases, while in the second phase the young stellar object (YSO) is already on the zero-age main sequence (ZAMS) and starts to clean-up its surroundings, leading to a decrease of the mass of the clump  \citep{molinari2008,molinari2016,molinari2019, giannetti2017, elia2017, elia2021, konig2017}. Since the most evolved sources in our sample are UC HII regions, the expected $L/M$ increase in our sample would be related to the first accretion phase.
\begin{figure*}[h]
\centering
\includegraphics[trim = 22 0 10 25 , clip,height=5.7cm]{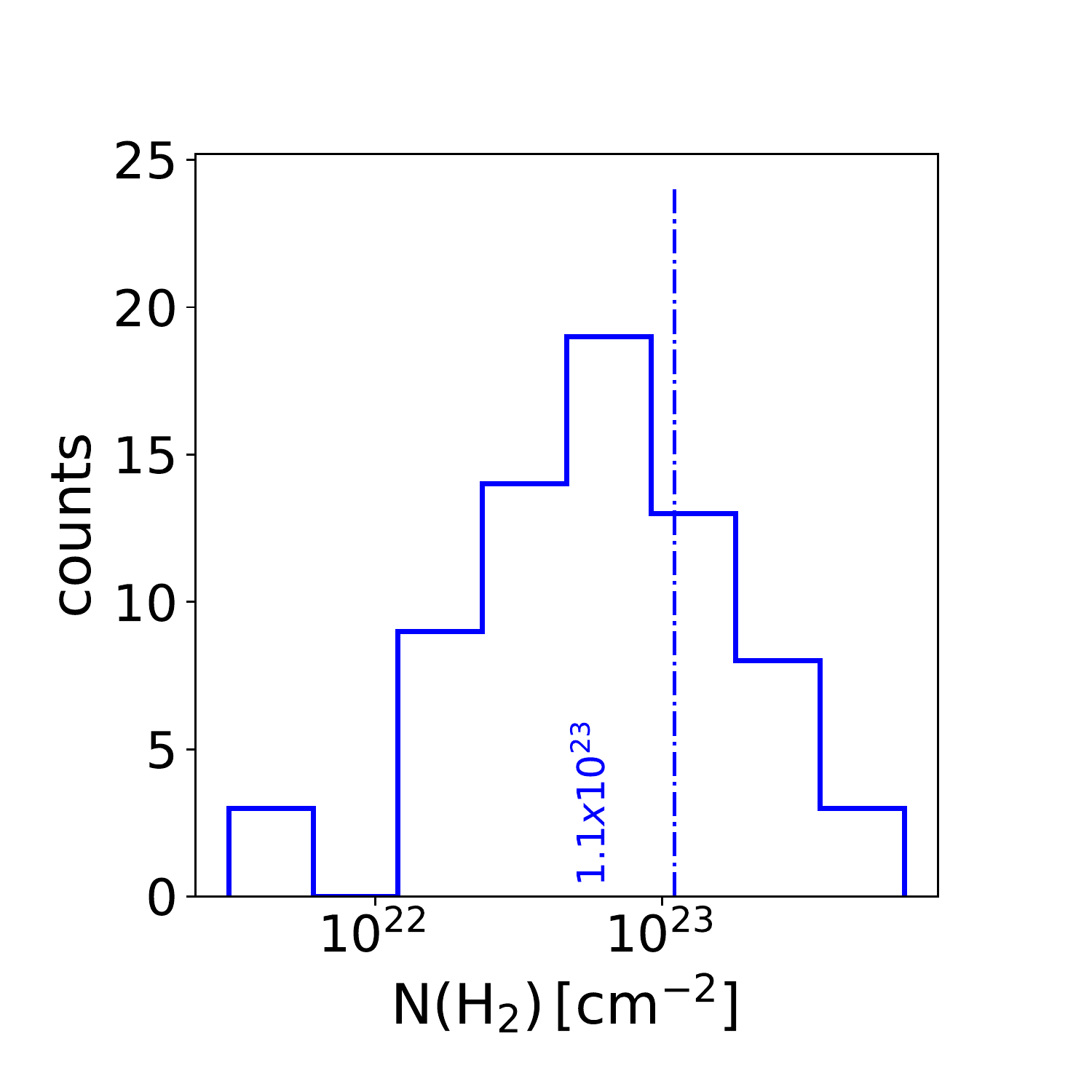}
\hspace{-0.2cm}\includegraphics[trim = 38 0 10 25 , clip, height=5.7cm]{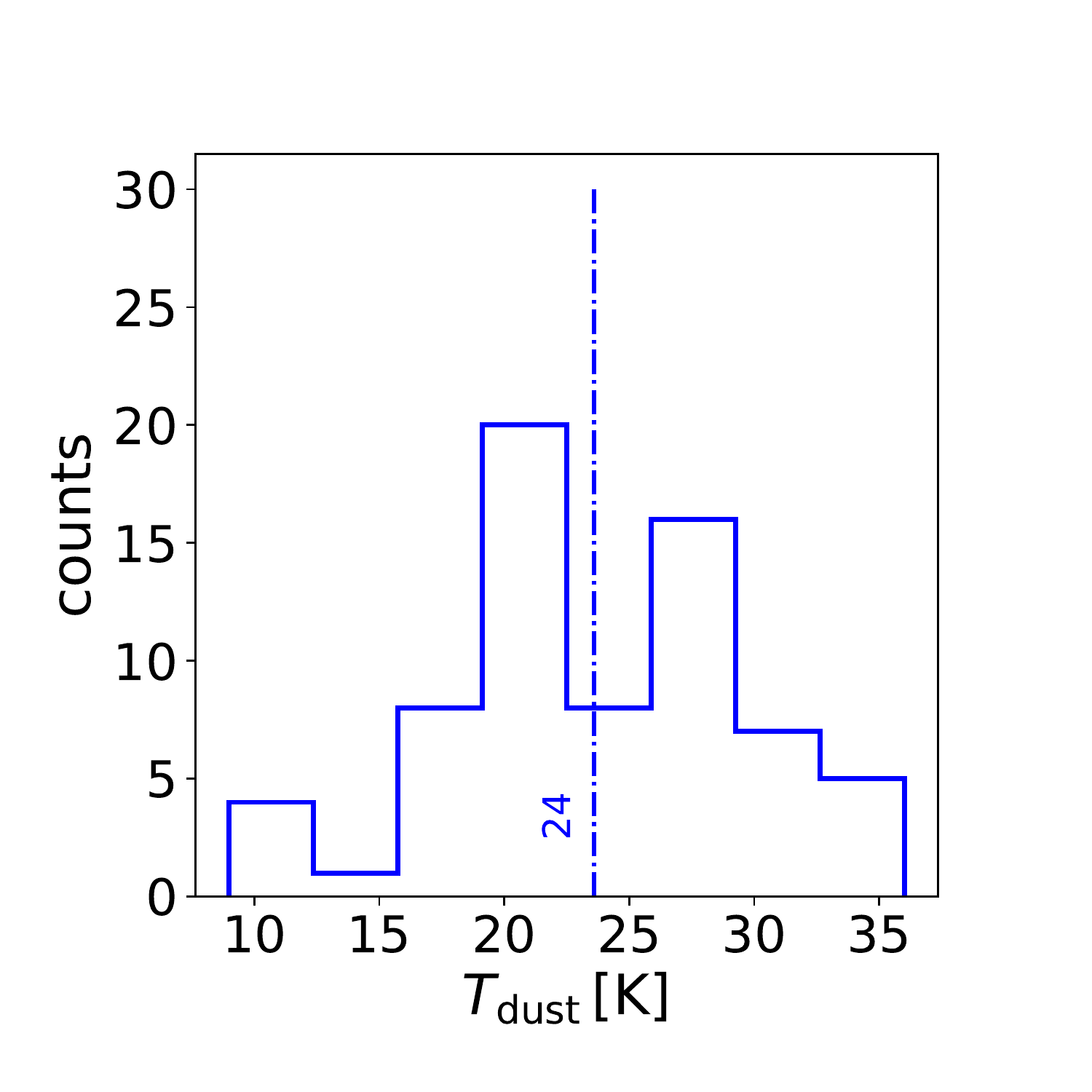}
\hspace{-0.2cm}\includegraphics[trim = 38 0 10 25 , clip, height=5.7cm]{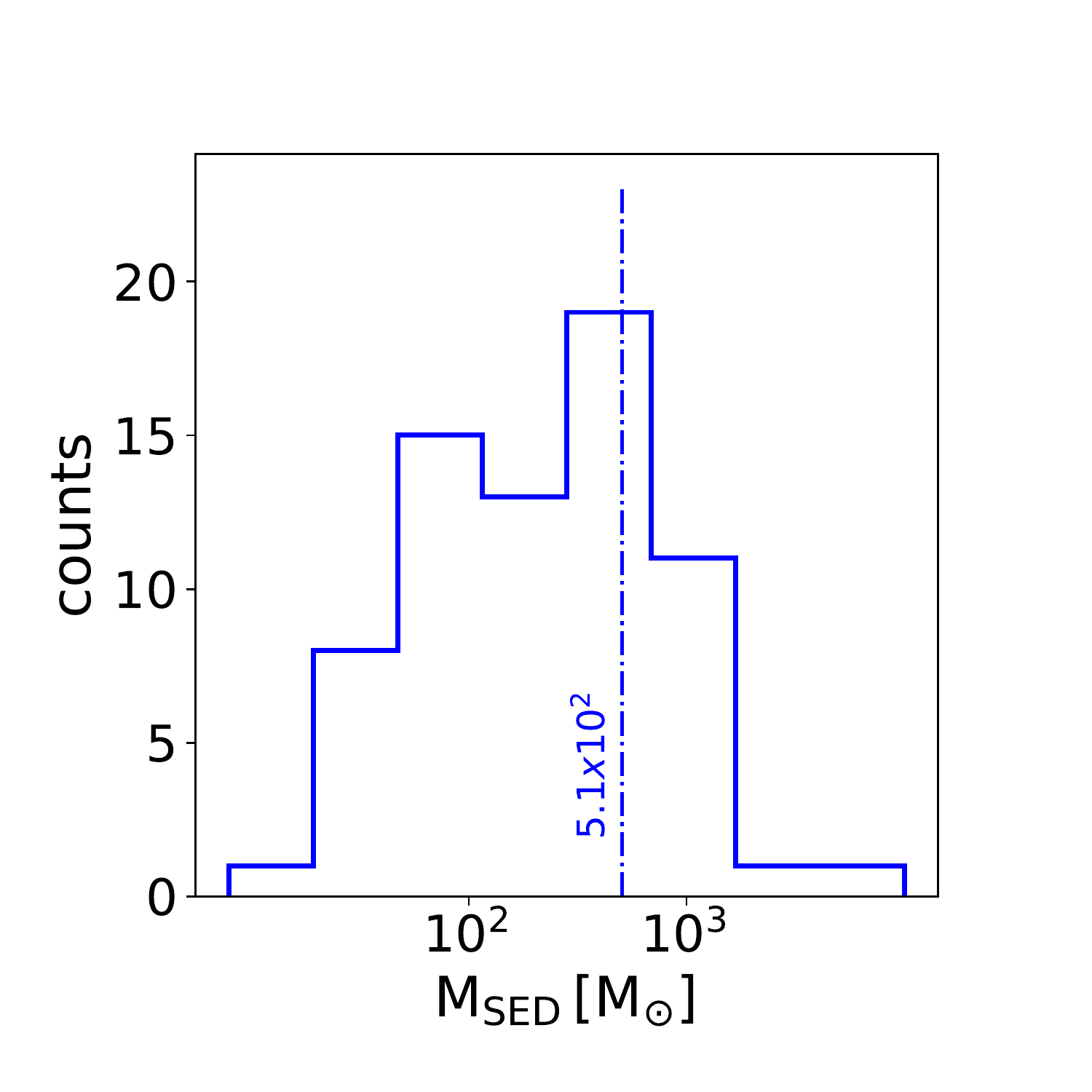}\\
\vspace{-0.2cm}\includegraphics[trim = 22 0 10 25, clip, height=5.7cm]{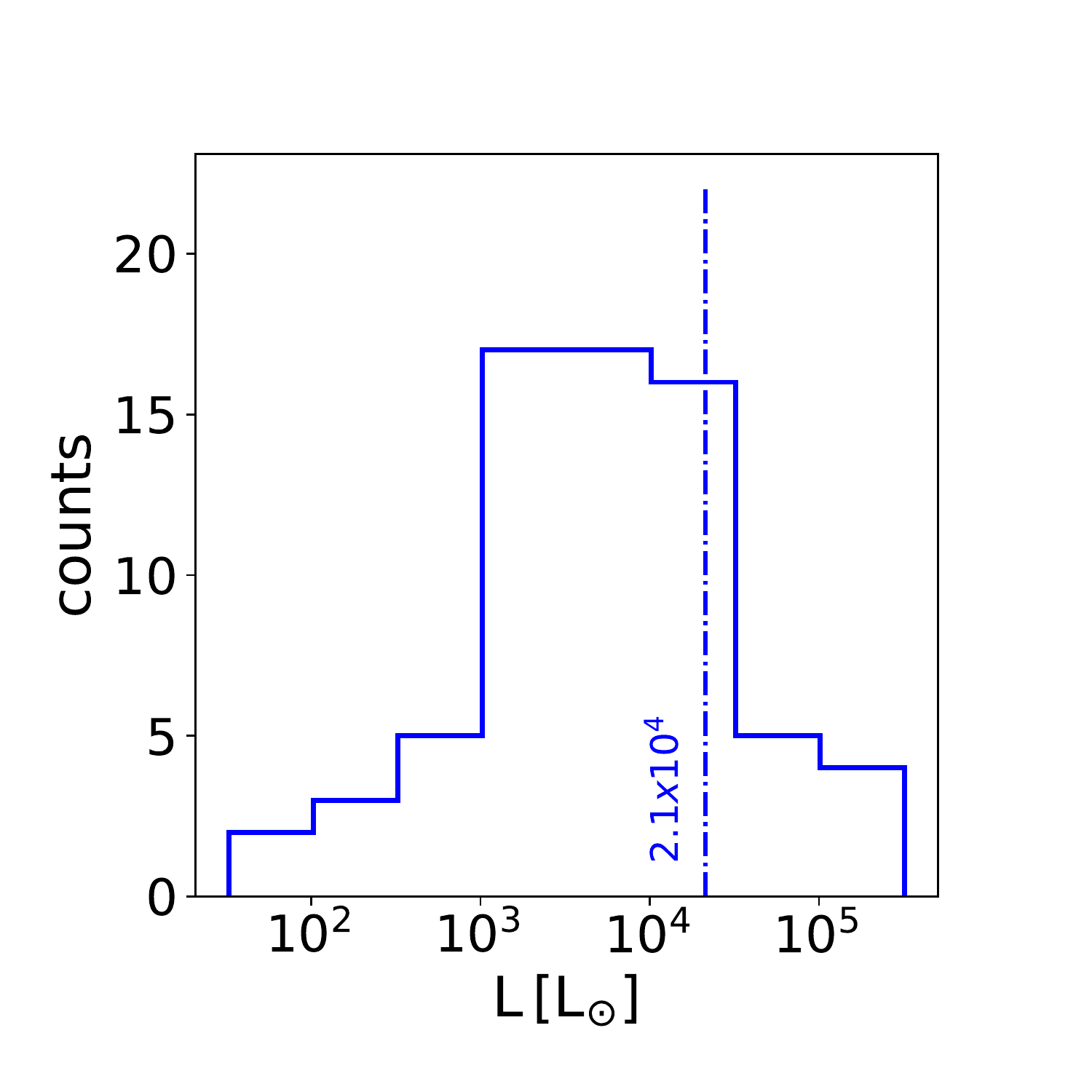}
\vspace{-0.2cm}\hspace{-0.2cm}\includegraphics[trim = 38 0 10 25 , clip, height=5.7cm]{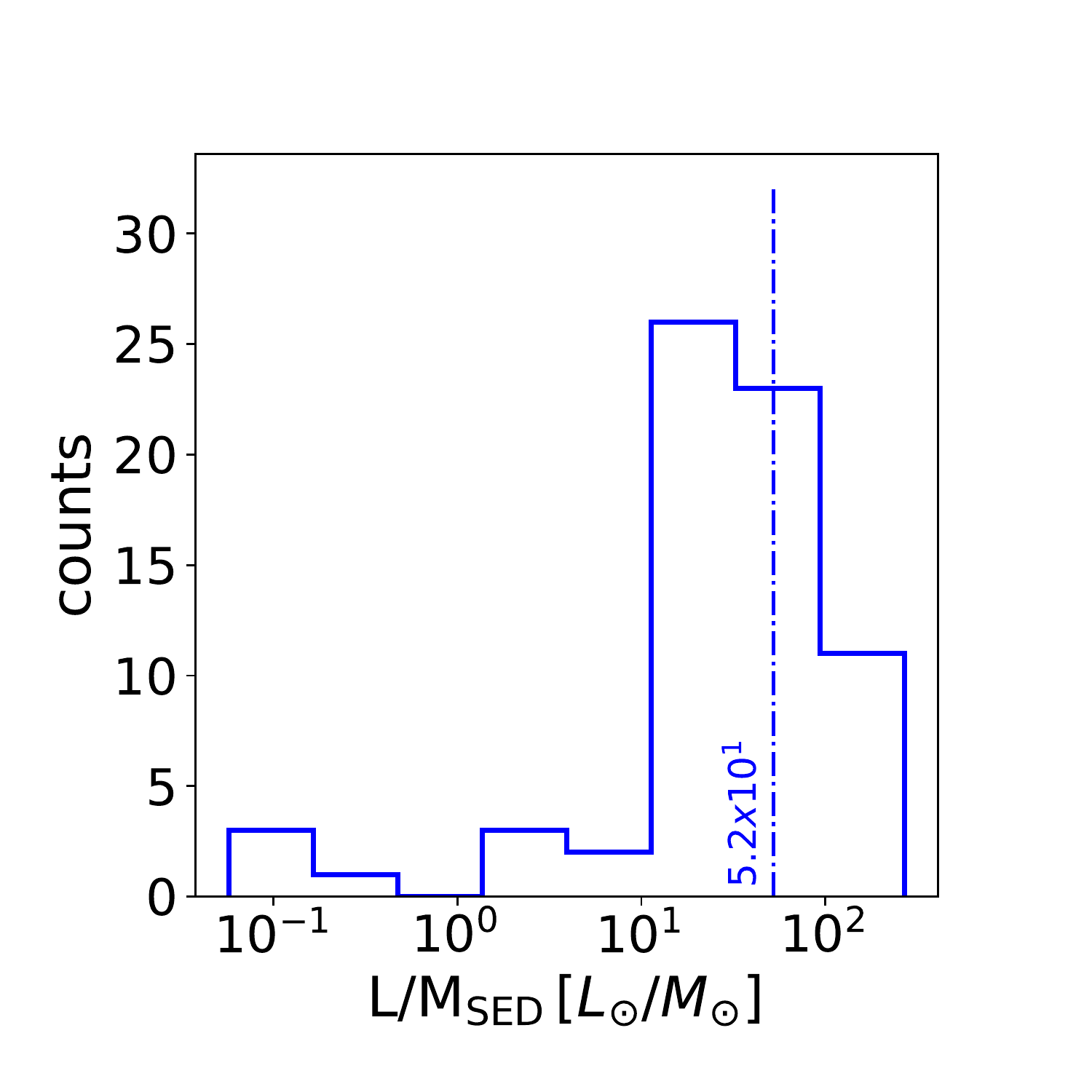}

\caption{Histograms of the distribution of H$_{2}$ column density, dust temperature, mass (\textit{upper panel from left to right}), luminosity, and luminosity-to-mass ratio (\textit{lower panel from left to right}), respectively, in the sample. The dotted vertical lines indicate the mean value of the distribution. Data are given in Tables 4 and 5. %\ref{tableSED} and \ref{tableMassandL}.
}
\label{fig:histo}
\end{figure*}
\setlength{\tabcolsep}{2pt}

%\longtab[3]{
\label{tableMassandL}
\renewcommand{\arraystretch}{1.2}
\begin{table}
\centering
\caption{Mass, luminosity and logarithm of the ratio $L/M$ in units of L$_{\odot}$/M$_{\odot}$ for the first 10 sources of the sample. Full table is available at CDS.\\}

\begin{tabular}{clccc}

\hline
 	&	Source	&	 $M_{\rm{SED}}$ &	 $L$ 	&	 log($L$/$M$) \\
 	&	 	&	 [$\rm{M_{\odot}}$] 	&	 [$\rm{L_{\odot}}$] 	&	 \\ 
\hline

1  	&	 I00117--MM1	& $24\pm11$ & $(6.1\pm1.3)\times10^{2}$ & $1.4\pm0.3$\\	
2  	&	 I00117--MM2	& $25\pm12$ & $(3.8\pm0.8)\times10^{2}$ & $1.2\pm0.3$\\
%3  	&	 I04579--VLA1 	&	-	&	 - 	&	- 	&	-	&	-\\
3  	&	 AFGL5142--MM 	&$110\pm50$ & $(5.4\pm1.1)\times10^{3}$ & $1.7\pm0.3$\\
4  	&	 AFGL5142--EC 	& $90\pm50$ & $(3.7\pm0.8)\times10^{3}$ & $1.6\pm0.3$\\
5  	&	 05358--mm3 	&	-	&	 - 	&	- 		\\
6  	&	 05358--mm1 	&	-	&	 - 	&	- 	\\
7  	&	 G5.89--0.39 	& $300\pm140$ & $(2.5\pm0.5)\times10^{4}$ & $1.9\pm0.3$\\
8  	&	 G008.14+0.22 	& $1100\pm400$ & $(6.5\pm1.4)\times10^{4}$ & $1.8\pm0.2$\\
9 	&	 18089-1732M1 	& $700\pm300$ & $(3.2\pm0.7)\times10^{4}$ & $1.7\pm0.2$\\
10 	&	 18089-1732M4 	& $110\pm40$ & $(3.1\pm0.7)\times10^{3}$ & $1.4\pm0.2$\\
\hline
\end{tabular}
\end{table}
%\end{longtable}
%\tablefoot{}
%}%end longtab[]

\subsection{CH$_3$CN analysis}
\label{sect:ch3cn}
The spectra at 3 mm of the CH$_3$CN(5$_{\rm{K}}-$4$_{\rm{K}}$) transitions towards the 85 sources 
presented in this work have been analyzed with the SLIM (Spectral Line Identification and Modeling) tool within the MADCUBA package\footnote{Madrid Data Cube Analysis (MADCUBA) is a software developed in the Center of Astrobiology (Madrid) to visualize and analyze data cubes and single spectra: https://cab.inta-csic.es/madcuba/ .} \citep{martin2019}.  At least one of the five $K$-components has been  detected towards 73 targets in the sample (85\% of the total sample), reported in Column 3 of Table 6, %\ref{tablech3cn}, 
while in Column 4 the $K$ components for which the signal-to-noise ratio S/N is larger than three are given. 
Before analysing the spectra, a first order baseline has been removed, determining the best fit of the baseline from free-channels around the lines.\\ \indent
To obtain the parameters of the molecular emission ($N_{\rm{CH_3CN}}$, $T_{\rm{ex}}$, FWHM and  V$_{\rm LSR}$), we used the AUTOFIT tool of MADCUBA-SLIM, which finds the best agreement between the observed spectra and the predicted LTE model, taking into account also the optical depth.

We performed the fit assuming that all the transitions are populated with the same $T_{\rm{ex}}$ %local thermodynamic equilibrium (LTE) 
and that the emission fills the beam, therefore the column densities of methyl cyanide have been computed inside a diameter of 26.6\arcsec, the IRAM 30m telescope beam ($\theta_{\rm b}$). We calculated the corrected column density, $N^{s}_{\rm{CH_3CN}}$, using the beam-dilution factor $\eta = \theta^2/(\theta^2+\theta_{\rm b}^2)$. The %LTE
single excitation temperature assumption is justified by the high densities of high-mass star-forming regions, comparable or larger than the values of the critical densities, $n_{\rm{crit}}$, for CH$_3$CN(5$_{{K}}-$4$_{{K}}$) transitions which are in the range $1-3\times10^{5}\,\rm{cm^{-3}}$ for $T$ in the range 20$-$140 K\footnote{The collisional coefficients have been taken from the CASSIS \citep{vastel2015} Collision Database: http://cassis.irap.omp.eu/download/collisions/files/CH3CN-H2-cdms.dat\,. The entry is based on \citet{Green1986}.}.\\

The results of the fit are given in Table 6, 
together with the CH$_3$CN abundance $X_{\rm{CH_3CN}}$.
The spectra are given in Appendix A in the online version.
For the source G31.41+0.31, the fit does not converge. This could be the combined result of high opacity at the center of this HMC, together with the presence of a temperature gradient \citep{beltran2005,beltran2018,cesa2011}, that makes a single $T_{\rm{ex}}$ fit impossible.\\  \indent For the sources in which only the component $K=0$ has been detected, the fit has been performed including also higher transitions, since their upper limits can give constraints during the fitting procedure. Moreover, even if not detected above $3\sigma$ level, in several cases the $K=1$ transition is close to the 2$\sigma$ level (or above) and its inclusion in the fit has allowed the determination of $T_{\rm{ex}}$. Only for I22134-B and 20332+4124 M1 we had to fix $T_{\rm{ex}}$ to 25\,K and 44\,K, respectively. These two values have been chosen from a visual inspection of the spectra, varying  $T_{\rm{ex}}$ and $N_{\rm{CH_3CN}}$ to search for a couple of parameter where the $K=0$ was visually well reproduced and the simulated higher $K$ components (not detected) with intensity below or comparable with the noise. \\ \indent For the sources for which at least the transitions $K=0,1$, and 2 were detected, we calculated a rotational diagram using the specific tool of MADCUBA-SLIM. The values of $T_{\rm{kin}}$  derived from the rotational diagram are consistent within the errors with the values of $T_{\rm{ex}}$ from AUTOFIT, except for two sources: 18454-0136 M1 and 19095+0930. For 18454-0136 M1 $T_{\rm{ex}}=35\pm6\,$K while $T_{\rm{kin}}=59\pm3\,$K. However, including in the rotational diagram the upper limit of the $K=3$ transition, the value of $T_{\rm{kin}}$ decreases to 22\,K. For 19095+0930, $T_{\rm{ex}}=57\pm4\,$K while $T_{\rm{kin}}=74\pm12\,$K. The inclusion of the upper limits of the $K=4$ transition does not change the results of the rotational diagram, thus this is the only source for which we have found a clear discrepancy between $T_{\rm{kin}}$ and $T_{\rm{ex}}$. The $T_{\rm{ex}}$ derived from the best agreement between the synthetic spectrum of all the (5$_{{K}}-$4$_{{K}}$) transitions and the observed spectrum is thus a reliable estimate of the $T_{\rm{kin}}$ of the sources. The goodness of the fit and the agreement between $T_{\rm{ex}}$ and $T_{\rm{kin}}$ indicates no need for a hotter component arising from a smaller region, when analyzing the CH$_3$CN(5$_{{K}}-$4$_{{K}}$) transitions. \citet{giannetti2017} also found that the (5$_{{K}}-$4$_{{K}}$) and (6$_{{K}}-$5$_{{K}}$) transitions are well reproduced by a single temperature fit, while to model the higher energy (19$_{{K}}-$18$_{{K}}$) transitions a second hotter component is needed.  \\ \indent

\subsection{Virial mass and virial parameter}
 
To discuss the gravitational stability of the targets we have also derived the virial masses, $M_{\rm{vir}}$, that for a spherical system are defined as \citep{maclaren1988}:
\begin{equation}
    M_{\rm{vir}} = \frac{k_1\sigma^{2} R}{G}\,,
    \label{eq:massvirial}
\end{equation}
where $\sigma$ is the three-dimensional velocity dispersion, R is the radius of the object, and $k_1=(5-2n)/(3-n)$ assuming a density profile $\rho\propto r^{-n}$. Assuming $n=2$ (i.e. $\rho\propto r^{-2}$) and a gaussian velocity distribution, Eq. (\ref{eq:massvirial}) can be written as:
\begin{equation}
     M_{\rm{vir}}=0.305\,d\,\theta\, \rm{FWHM^{2}}\,
 \end{equation}
  in units $\,M_{\odot}$, where $\theta$ is the angular dimension of the sources in arcsec, and FWHM is the linewidth in $\rm{km\,s^{-1}}$ (see \citealt{maclaren1988}). We then derived the virial parameter $\alpha = M_{\rm{vir}}/M_{\rm{SED}}$ \citep{kauffmann2013}, reported in Table 6. %\ref{tablech3cn}. 

\setlength{\tabcolsep}{2pt}

%\longtab[4]{

\renewcommand{\arraystretch}{1.2}
\begin{table*}
\centering
\label{tablech3cn}
\caption{K components of the CH$_3$CN($5_{\rm{K}}-4_{\rm{K}}$) transition detected with $S/N>3\sigma$, beam diluted column density $N_{\rm{CH_3CN}}$, column density corrected for the source size $N^{s}_{\rm{CH_3CN}}$, $T_{\rm{ex}}$,  FWHM, abundance X$_{\rm{CH_3CN}}$ w.r.t. H$_{2}$, and virial parameter $\alpha$ for the first 10 sources of the sample. The value of $N^{s}_{\rm{CH_3CN}}$ is not given for sources for which the maps in the range 70-500\,$\mu$m were not available. The full table is available via the CDS.%\ref{tableSED}. 
\\}
\begin{tabular}{clccccccccc}

\hline
 &Source	& det. & K & log$_{10}$($N_{\rm{CH_3CN}}$) & log$_{10}$($N^{s}_{\rm{CH_3CN}}$)& $T_{\rm{ex}}$ & FWHM & V$_{\rm{LSR}}$ & log$_{10}(X_{\rm{CH_3CN}})$& $\alpha$  \\
 & &  & & [cm$^{-2}$]&[cm$^{-2}$] & [K] & [$\rm{km\,s^{-1}}$] & [$\rm{km\,s^{-1}}$] & &\\
\hline

1  & I00117--MM1		&Y	&0-2 &	$12.10\pm0.12$	& $12.38\pm0.12$ &	$54\pm13$& $2.2\pm0.3$	&$-$36.2 & $-9.8\pm0.3$ &	$3.1\pm1.7$\\
2  & I00117--MM2		&Y	&0-2 &	$11.78\pm0.17$	& $12.13\pm0.18$ &	$36\pm12$& $1.4\pm0.3$	&$-$36.3 & $-10.2\pm0.3$ & $1.0\pm0.7$ \\
%3  & I04579--VLA1		&-$^{a}$&	&	-   	&	-				& - 	&	- \\ &
3  & AFGL5142--MM		&Y	&0-3 &	$13.20\pm0.03$	& $13.59\pm0.04$ &	$53\pm3$& $3.92\pm0.10$	&$-$3.8 &  $-9.49\pm0.13$ & $1.6\pm0.7$ \\
4  & AFGL5142--EC		&Y	&0-3 &	$13.16\pm0.03$	& $13.51\pm0.04$ &	$55\pm3$& $4.05\pm0.09$	&$-$3.7 &  $-9.39\pm0.18$ & $2.4\pm1.3$ \\
5  & 05358--mm3			&Y	&0-3 &	$12.85\pm0.03$	& - &	$37\pm3$& $3.43\pm0.10$	&$-$17.1 &	- &	- \\
6  & 05358--mm1			&Y	&0-3 &	$12.81\pm0.03$	& -	&   $36\pm3$& $3.27\pm0.10$	&$-$17.2 &	- &	- \\
7  & G5.89--0.39		&Y	&0-4 &	$14.05\pm0.03$	& $14.21\pm0.04$	&   $58\pm4$& $4.68\pm0.11$	&+9.9	& $-9.1\pm0.2$ & $1.2\pm0.6$ \\
8  & G008.14+0.22		&Y	&0-3 &  $13.01\pm0.05$  & $13.26\pm0.06$&   $53\pm5$& $5.0\pm0.3$   &+18.8 & $-9.97\pm0.17$ & $0.7\pm0.3$\\
9 & 18089-1732M1		&Y	&0-4 &	$13.75\pm0.04$	& $14.27\pm0.05$ &	$92\pm6$ & $4.86\pm0.11$&+34.0 & $-9.18\pm0.08$ & $0.7\pm0.3$\\
10 & 18089-1732M4		&Y	&0-2 &	$12.56\pm0.04$	& $13.08\pm0.05$ &	$36\pm4$& $3.8\pm0.2$	& +33.1 &$-9.60\pm0.08$ & $2.5\pm0.9$\\
%\end{longtable
\hline
\end{tabular}
\end{table*}
%\tablefoot{a) source not observed. $\ast$) K=1 component $>3\sigma$ and K=0 component between $2.5\sigma$ and $3\sigma$; $\ast\ast$) parameter kept fixed during the fit. }
%}%end longtab[]

%______________________________________________________________
\section{Discussion}
\subsection{Physical properties derived from the SEDs}
\begin{figure*}
    \centering
    \includegraphics[trim = 0cm 0.8cm 0.3cm 1.6cm , clip,height=5.6cm]{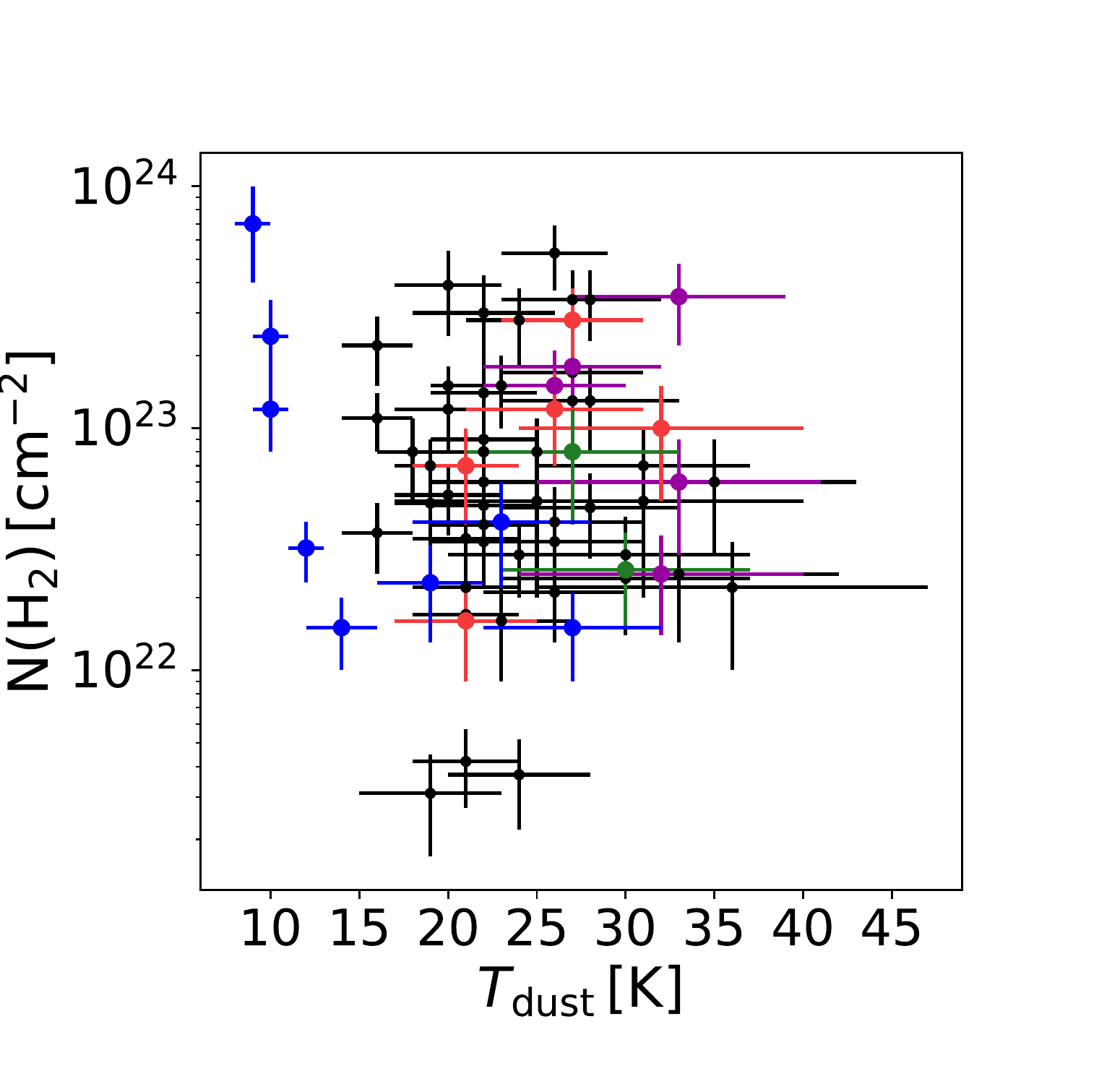} 
    \hspace{-8mm}
    \includegraphics[trim = 0cm 0.8cm 0.3cm 1.6cm , clip,height=5.6cm]{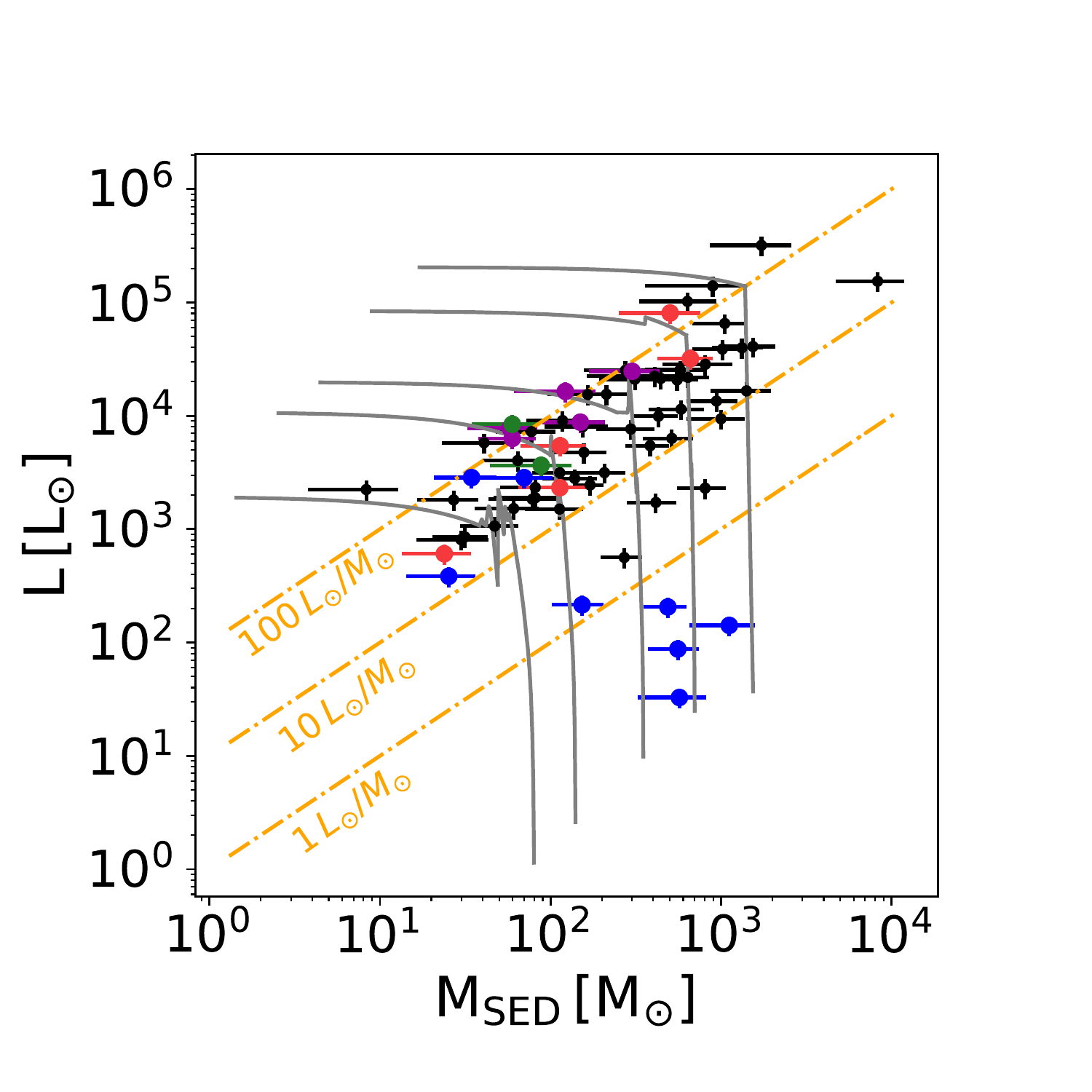}
    \hspace{-8mm}
    \includegraphics[trim = 0cm 0.8cm 0.3cm 1.6cm , clip,height=5.6cm]{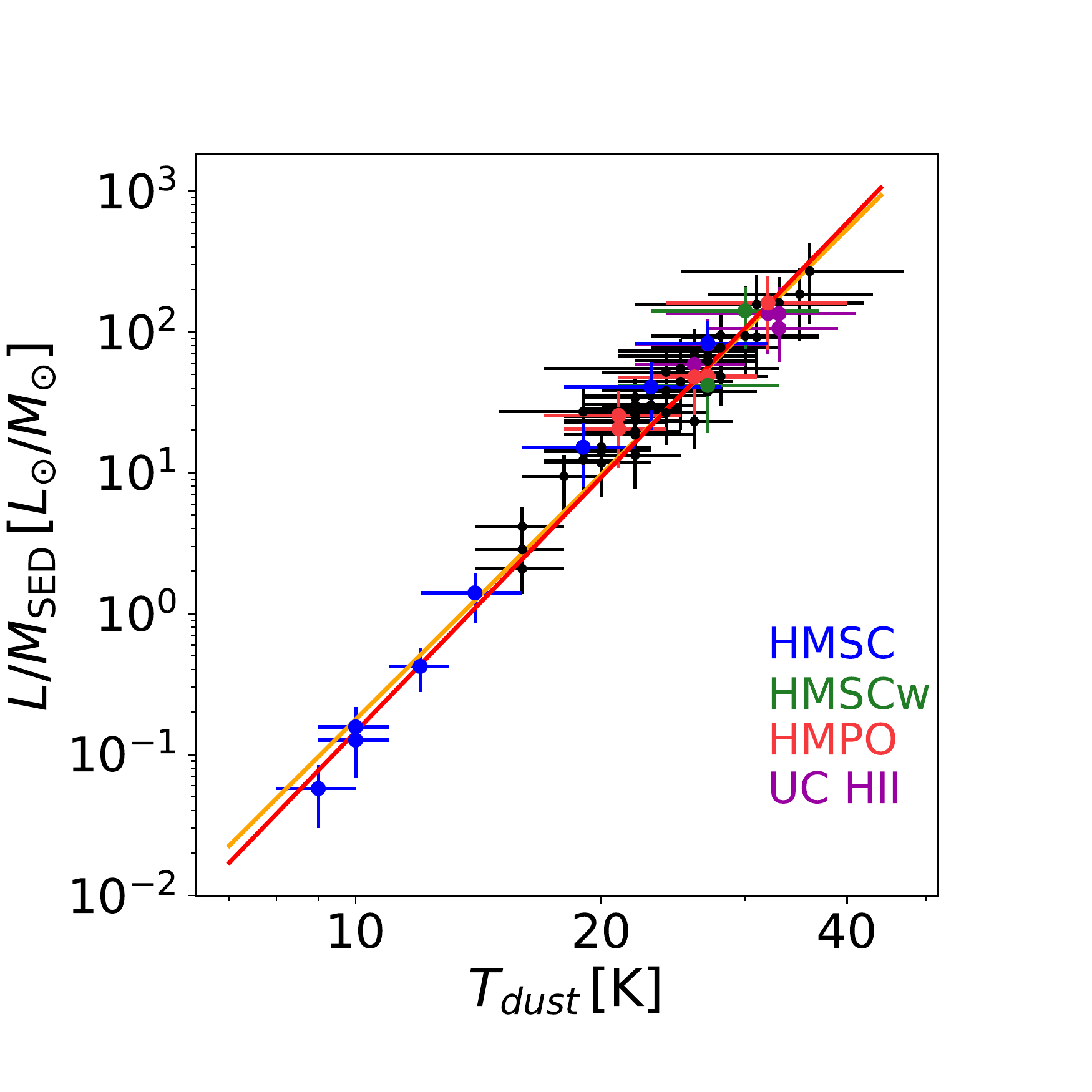}\\%TdustvsLsuM_class_new.eps}
    \caption{\textit{Left}: distribution of the sources of the sample in the plane $N(\rm{H_2})-T_{\rm dust}$ ; \textit{middle}: distribution of the sources of the sample in the plane $L-M_{\rm SED})$. Grey tracks are the evolutionary tracks from \citet{molinari2008}, while orange dashed lines indicate $L/M=1, 10$, and 100\,$L_{\odot}/M_{\odot}$; \textit{right}: plot of the evolutionary tracers $L/M$ as a function of $T_{\rm dust}$ and power-law relation best fit to the data given in orange. The legend for the classification color-code is given in the right panel. Sources that have not been classified yet are plotted in black.} 
    \label{fig:distributions}
\end{figure*}
\begin{figure*}

\centering
\includegraphics[trim = 13 0 10 25 , clip,height=5.7cm]{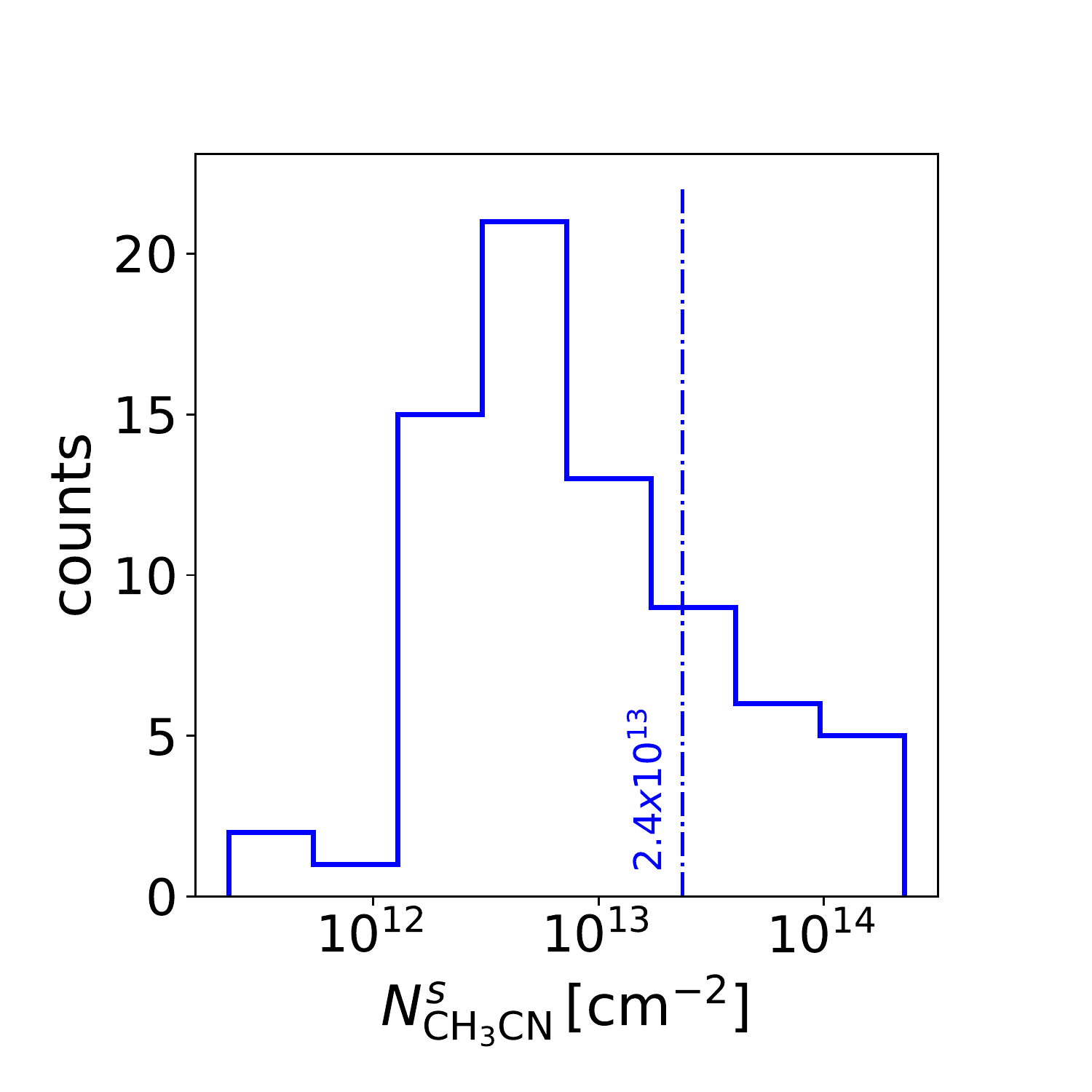}
\includegraphics[trim = 13 0 10 25 , clip,height=5.7cm]{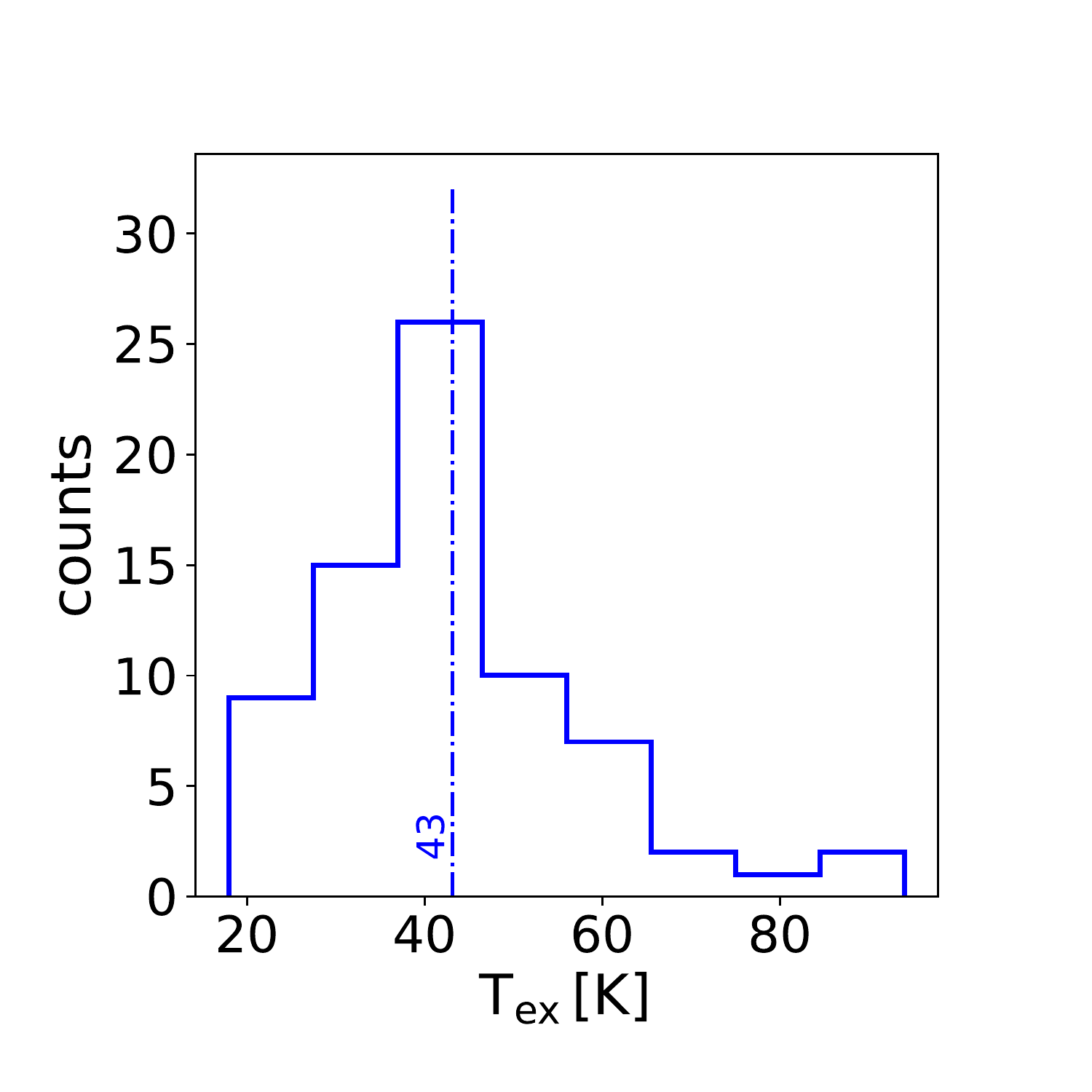}
\includegraphics[trim = 13 0 10 25 , clip,height=5.7cm]{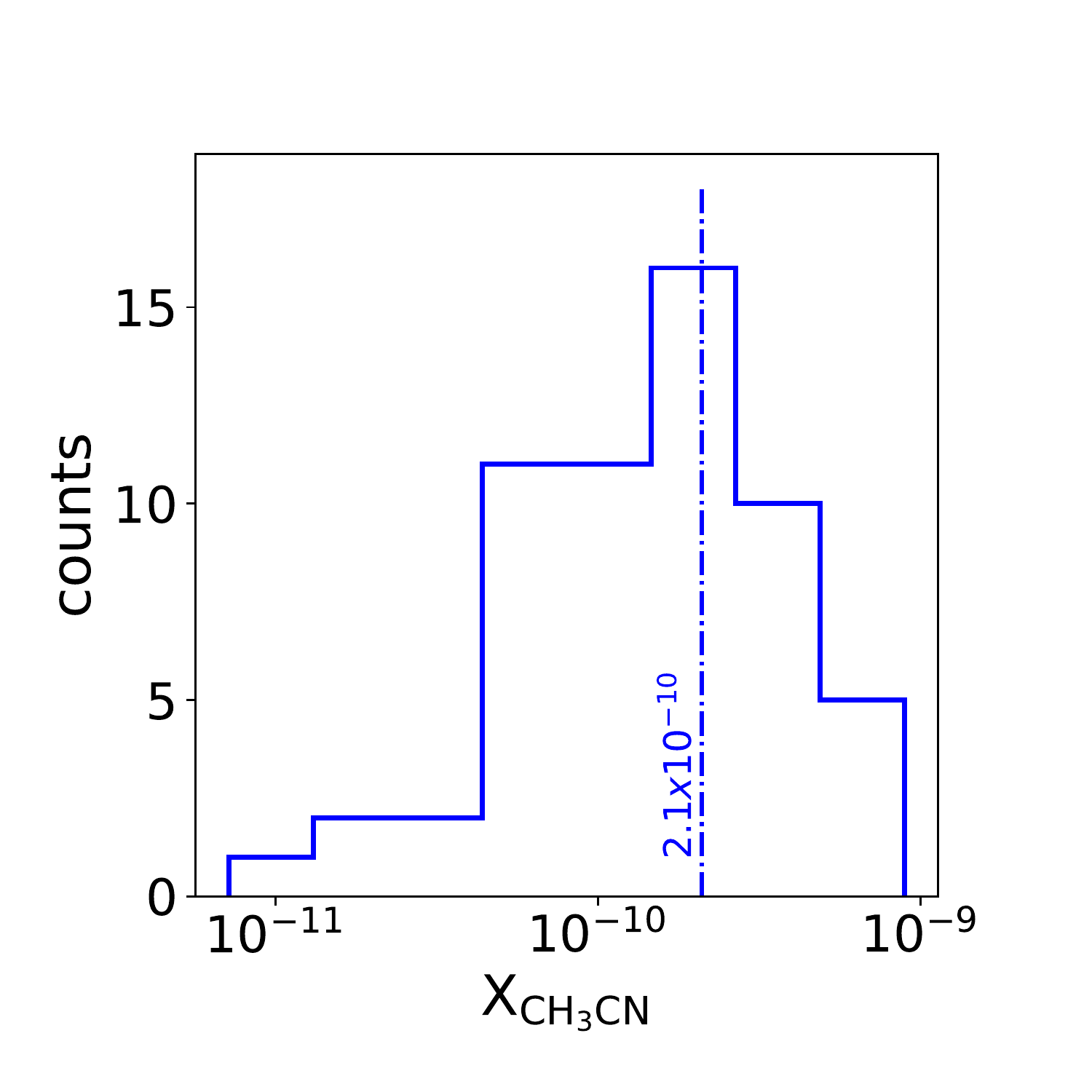}
\label{fig:histo_ch3cn}
\caption{Histograms of the distribution of column density of CH$_3$CN (\textit{left panel}), excitation temperature $T_{\rm ex}$ (\textit{middle panel}), and abundance $X_{\rm CH_3CN}$ (\textit{right panel}) within the sample. The dotted vertical lines indicate the mean values of the distributions. Data are given in Table 5. }
\end{figure*} 
\indent Figure \ref{fig:histo} shows the histograms of the distribution of molecular hydrogen column density, dust temperature, mass, luminosity, and luminosity-to-mass ratio for the sample. The values of $N_{\rm{H_2}}$ cover two orders of magnitude from $3\times10^{21}$ to $7\times10^{23}\,\rm{cm^{-2}}$, while $T_{\rm{dust}}$ ranges from 9 to $36\,\rm{K}$. These ranges are consistent with those found in works with similar or larger samples \citep{elia2017,konig2017}. 
The luminosity of the sources spans over four orders of magnitude, from $\sim30$ to $300\,000\,L_{\odot}$, while the mass ranges from $\sim30$ and $8\,000\,M_{\odot}$. Lastly, the luminosity-to-mass ratio $L/M$ covers more than three orders of magnitude from $6\times10^{-2}$ to $3\times10^{2}\,L_{\odot}/M_{\odot}$. The sources are located at different distances, from 0.7 to 13\,kpc, and the linear scale linear scales traced range from 
$\sim0.1$ and $\sim1.4\,\rm{pc}$, with a mean value of $\sim0.45\,$pc, and with only nine sources ($\sim13\%$) with diameters below 0.2\,pc. Thus, the considered structures are likely associated with clumps, and  only in a few cases with cores, i.e. smaller and denser structures embedded inside a clump. This  explains the broad ranges obtained  for the masses and the luminosities. \\ \indent Figure \ref{fig:distributions} shows the distribution of the sample in the space $N_{\rm{H_2}}-T_{\rm{dust}}$ and $L-M_{\rm{SED}}$, with different colors indicating the sources already classified (see last panel for the legenda). The plot of  $N_{\rm{H_2}}$ vs. $T_{\rm{dust}}$ does not highlight any trend between the two quantities and with evolutionary stage. The lack of any particular trend between surface density and dust temperature has been seen before on a sample of $\sim10^{5}$ sources in the work of \citet{elia2017} for $T_{\rm{dust}}>20\,\rm{K}$. On the last panel (\textit{right}) we show the relation between $L/M$ and $T_{\rm{dust}}$; the plot highlights a clear correlation between the two physical quantities. The best fit to the data, assuming a power-law: 
\begin{equation}
    L/M = a\,T_{\rm{dust}}^{b}\,,
\end{equation}
is given by $a = (2.9\pm1.5)\times10^{-7}\,L_{\odot}/M_{\odot}\,\rm{K^{-b}}$ and $b =(5.8\pm0.2)$, and is represented by the orange line in Fig. \ref{fig:distributions}. The red line in Fig.  \ref{fig:distributions} is the theoretical prediction for an optically thin modified blackbody \citep{elia2016}:
\begin{equation}
    L/M = \frac{8 \pi\, k_{\rm{B}}^{4+\beta}\, \zeta (4+\beta)\, \Gamma (4+\beta)\,\kappa_{\rm{ \nu_{0}}}}{h^{3+\beta} c^{2}\nu_{0}^{\,\beta}}\,T_{\rm dust}^{4+\beta}\,
\end{equation}
where $k_{\rm{B}}$ is the Boltzmann constant, $h$ is the Planck constant, $c$ is the speed of light, $\Gamma$ is the Euler gamma function, and $\zeta$ is the Riemann zeta function. 
With our assumption on $\beta$ and the dust opacity coefficient it can be written as:
\begin{equation}
L/M = 1.45\times10^{-7}\,T_{\rm{dust}}^{6}\,L_{\odot}/M_{\odot}\,,
\end{equation}
consistent with the best fit.\\ \indent
The positive correlation seen is consistent with $L/M$ tracing the evolution of the sources, as already discussed in previous studies \citep{molinari2008, urquhart2014,urquhart2018,giannetti2017, elia2017, elia2021}. These studies indicated that massive young stellar objects (YSO) - HMPO, UC HII or more evolved sources that have already dissipated their envelope - are characterized by a $L/M$ ratio between $\sim1\,L_{\rm \odot}/M_{\rm \odot}$ and $\sim100\,L_{\rm \odot}/M_{\rm \odot}$. %($L/M>22\,L_{\rm \odot}/M_{\rm \odot}$ for candidate UC HII regions, see \citealt{elia2021}). 
Lower values, $L/M < 1\,L_{\rm \odot}/M_{\rm \odot}$, are associated with earliest stages in which no star formation is yet present or is in a very early evolutionary stage. Alternatively, these cores could only contain embedded low-mass YSOs. However, \citet{elia2017} showed that pre-stellar objects have a
narrower distribution in $L/M$, compared to protostellar
sources that are widely distributed, and highlighted the presence of
a statistically significant overlap of the distributions in the two classes. It is thus possible to find HMPOs with values lower than  $\sim1\,L_{\rm \odot}/M_{\rm \odot}$, or HMSCs with higher values.  % in which no star-formation processes are present or is either in its very early stages or only low-mass YSOs are forming. 
\citet{molinari2019} confirmed that the %bolometric luminosity (and consequently the 
$L/M$ ratio is a good tracer of the evolutionary stage of star formation, running a grid of 20 million synthetic protocluster models. From Fig. 9 in \citet{molinari2019}, the  models in which at least one young stellar object (YSO) is a zero age main sequence (ZAMS) star, are in good agreement with $L/M>1\,L_{\rm \odot}/M_{\rm \odot}$, confirming that $L/M < 1\,L_{\rm \odot}/M_{\rm \odot}$ indicates the earliest stage of star formation. \\ \indent  
 The position of the HMSC sources in the plot $L$ vs. $M_{\rm SED}$ and $L/M$ vs. $T_{\rm dust}$ is consistent with values of $L/M \lesssim 1\,L_{\rm \odot}/M_{\rm \odot}$, with the exception of the sources I00117-MM2, I20293-WC and I22134-B. Among these, the high $L/M$ value in I00117-MM2 can be explained in terms of contamination of the fluxes used to build the SED by the nearby source I00117-MM1 (HMPO), which is only $\sim 16\arcsec$ apart. A similar contamination can be present in the following couples of sources due to their proximity: AFGL5142-MM and AFGL5142-EC ($\sim11\arcsec$), I19035-VLA1 and 19035+0641M1 ($\sim8\arcsec$), 19410+2336M1 and 19410+2336 ($\sim9\arcsec$), I22134-VLA1 and I22134-G ($\sim21\arcsec$), and 23033+5951 and 23033+5951M1 ($\sim11\arcsec$). However, these are compact and centrally peaked sources, therefore the molecular emission inside the telescope beam would be dominated by the main source, and not by the nearby companions. These sources also constitute the $\sim44\%$ of Sub-sample I, thus a large fraction of the classified sub-sample. For these reasons we decided not exclude them from the analysis. From the sources in Sub-sample I, UC HII regions are associated with larger values of the luminosity-to-mass ratio than HMPO (with only one exception), thus giving another confirmation of the clear growth of $L/M$ with evolution. The majority of the sources of the Sub-sample II (\textit{black dots}) have values of $L/M$ mostly larger than $10\,L_{\rm \odot}/M_{\rm \odot}$. Therefore it is likely that these sources are in more evolved evolutionary phases, being either HMPOs or UCHII regions. A small number of sources in Sub-sample II are well studied sources in evolved stages, like e.g. the HMCs G31.41+0.31 and G24.78+0.08 (e.g. \citealt{cesa2011,beltran2011bis}), for which we have derived values of $\sim20-25\,L_{\odot}/M_{\odot}$. However from this figure and from the analysis in Sect. 5.2, five unclassified sources of Sub-sample II show values of $L/M$ closer to the values of HMSCs and are likely to be themselves starless cores or very early HMPOs (see Fig. 9).\newline \indent 
 An accurate classification of the sources in the Sub-sample II, obtained from the fluxes of these sources in the mid-IR and at cm wavelengths, will be presented in a following paper.     
\begin{figure*}
\centering
\label{fig:AAAch3cn}
\includegraphics[trim = 0.3cm 20 10 2cm , clip,height=5.8cm]{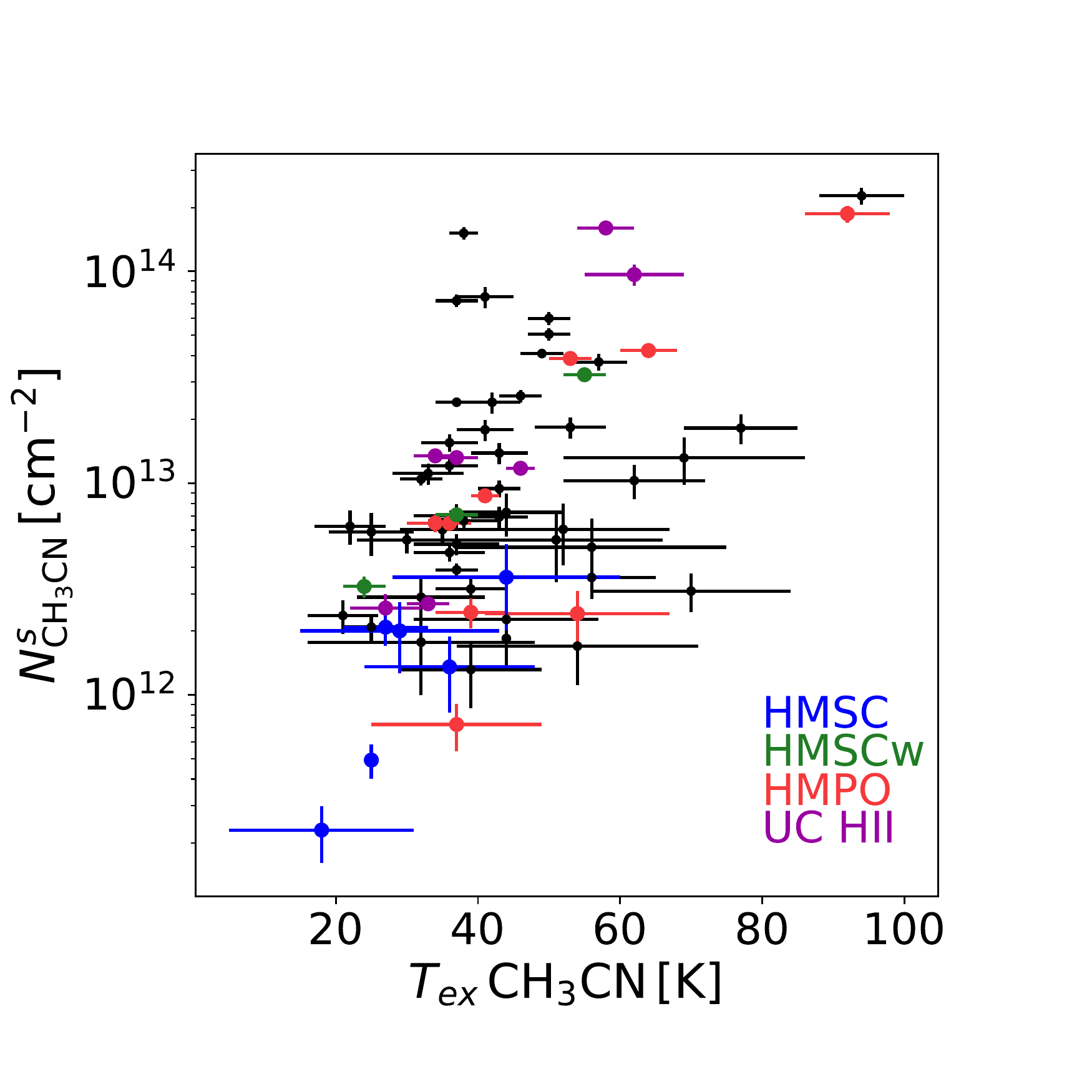}
\hspace{-9mm}
\includegraphics[trim = 0.8cm 20 10 2cm , clip,height=5.8cm]{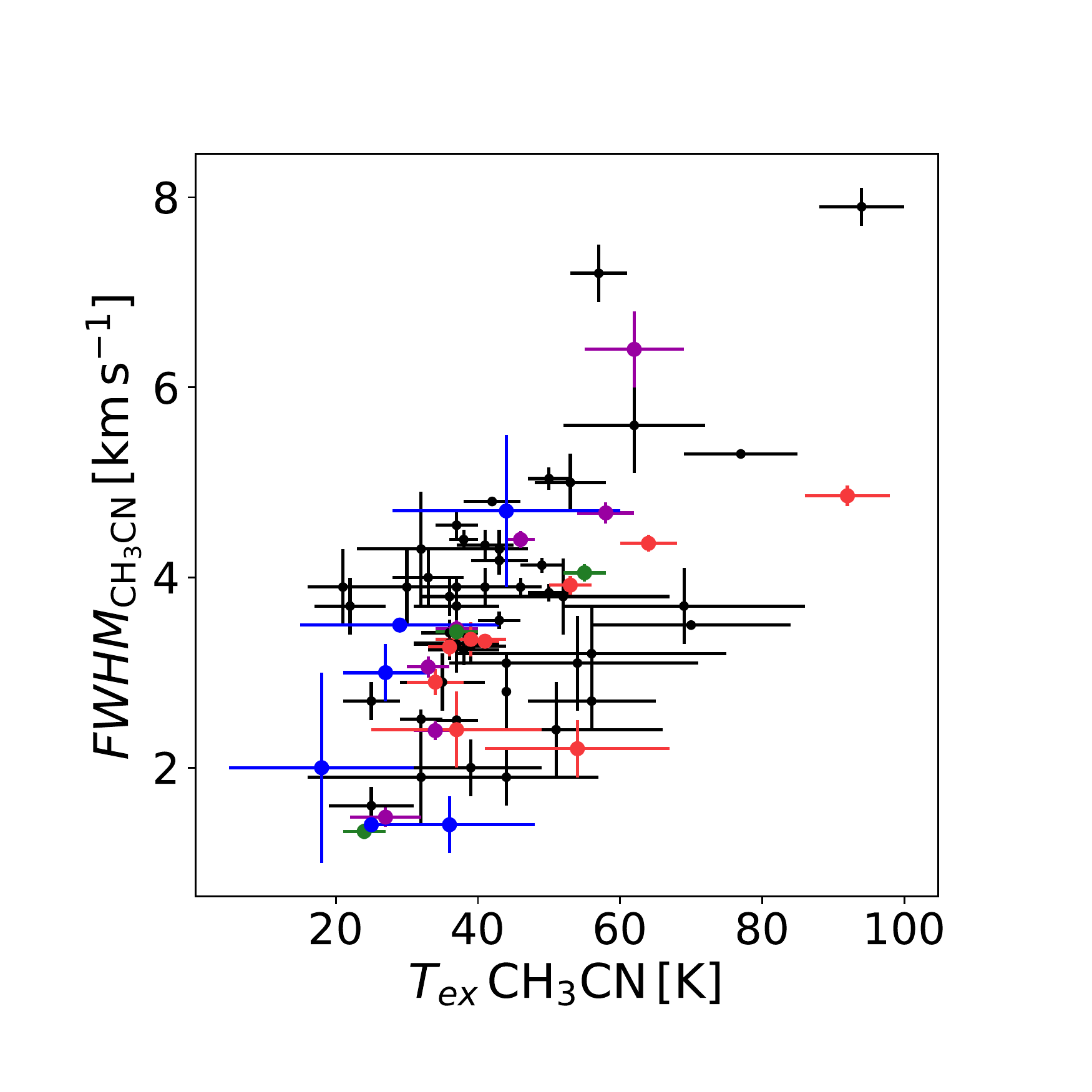}
\hspace{-9mm}
\includegraphics[trim = 0.6cm 20 10 2cm , clip,height=5.8cm]{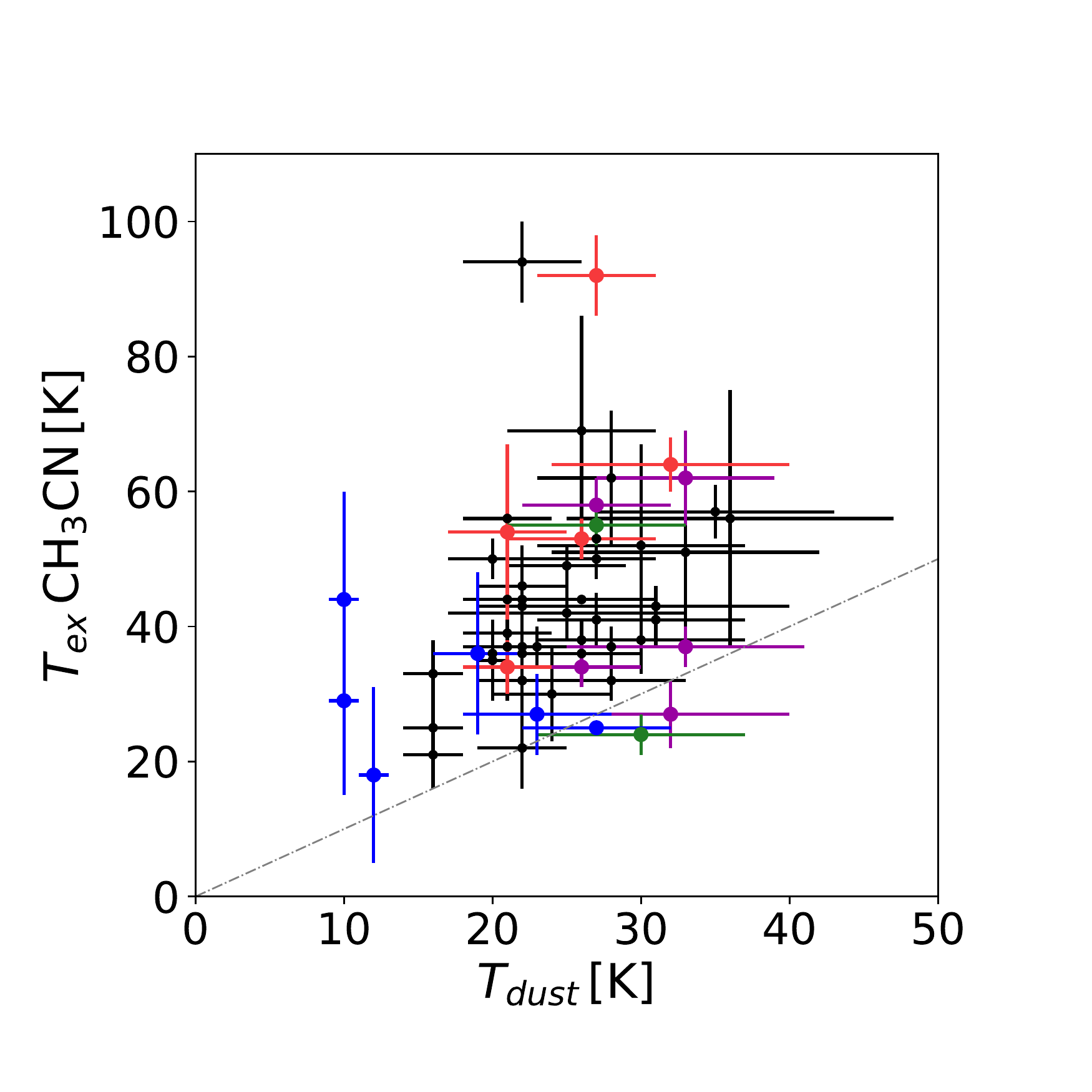}\\

\caption{\textit{Left}: plot of the column densities of CH$_3$CN as a function of the excitation temperature $T_{\rm ex}$; \textit{middle}: plot of the FWHM of CH$_3$CN as a function of the excitation temperature $T_{\rm ex}$; \textit{right}: plot of the excitation temperature $T_{\rm ex}$ of methyl cyanide vs. dust temperature $T_{\rm{dust}}$. The grey line indicates $T_{\rm{dust}}=T_{\rm ex}$. }
\end{figure*}

\begin{table*}
  \caption{Mean values of %beam diluted column density, 
 column density corrected for beam-dilution,  abundance, and excitation temperature of CH$_{3}$CN in the TOPG\"ot sample.}
    \centering
    \begin{tabular}{rc|cccc|c}
    \hline\noalign{\smallskip}
     & total sample & \multicolumn{4}{c}{ Sub-sample I} & Sub-sample II\\
     & &  HMSC & HMSC$^{w}$ & HMPO & UC HII & \\
     n$^{\circ}$ sources & (72/58)$^{\ast}$ &(6/6) &(3/2) &(9/5) & (7/5) & (47/40)\\ 
    \hline\noalign{\smallskip}
        %N$_{\rm{CH_3CN}}$ [cm$^{-2}$] & $9.9\times10^{12}$ & $6.7\times10^{11}$ & $7.7\times10^{12}$ & $1.3\times10^{13}$ & $2.1\times10^{13}$ & $9.2\times10^{12}$\\
        N$^{s}_{\rm{CH_3CN}}$ [cm$^{-2}$] & $2.4\times10^{13}$ & $1.6\times10^{12}$ & $1.4\times10^{13}$ & $3.2\times10^{13}$ & $4.3\times10^{13}$ & $2.2\times10^{13}$ \\
        X$_{\rm{CH_3CN}}$  &$2.1\times10^{-10}$ & $3.0\times10^{-11}$ & $2.7\times10^{-10}$ & $3.3\times10^{-10}$ & $3.2\times10^{-10}$ &  $2.1\times10^{-10}$\\
        T$_{\rm{ex}}$ [K] & 43 & 30 & 39 & 50 & 42 & 44\\
      
    \hline
    \end{tabular}
  \tablefoot{ The third row reports two numbers: the first number (largest) indicates the number of sources over which we derived the mean values of $N_{\mathrm{CH_3CN}}$ and $T_{\rm{ex}}$. The second number indicates the number of sources over which we derived the mean values of $X_{\mathrm{CH_3CN}}$. The latter value is smaller because we could not analyze the SED for all the sources for which we detected CH$_3$CN. $\ast$) the total number of sources for which CH$_3$CN has been detected is 73, but for G31.41+0.31 we could not derive the physical parameters.}
    \label{tab:meanvalues}
\end{table*}

\begin{table*}
  \caption{Mean values of column density and excitation temperature of CH$_{3}$CN for different evolutionary classes in literature works.}
    \centering
    \begin{tabular}{rcccccl}
    \hline\noalign{\smallskip}
    reference & $N^{s}_{\rm{CH_3CN}}$ & $T_{\rm{ex}}$ & $\theta^{a}$ & telescope  & type of sources\\
     & [cm$^{-2}$] & [K] & [\arcsec] & \\
    \hline\noalign{\smallskip}
    \citet{hung2019} & $5.2\times10^{15}$&102 & -$^{\ast}$ & SMT & EGO$^{\ast\ast}$\\
     \hline\noalign{\smallskip}
    \citet{giannetti2017}$^{b}$ & $1.2\times10^{13}$ & 33 & -$^{\ast}$ & IRAM30m, APEX, Mopra & 70$\rm{\mu m}$-weak   \\
        &$3.5\times10^{13}$ & 32 &-$^{\ast}$ & & IR-weak\\
        & $4.4\times10^{13}$& 43&-$^{\ast}$ & & IR-bright\\
        &$7.4\times10^{13}$ &53 &-$^{\ast}$ & & HII\\
    \hline\noalign{\smallskip}
    \citet{rosero2013}$^{c}$ & $8.0\times10^{13}$ & 145 & 10 & SMT & HMPO \\
     \hline\noalign{\smallskip}
    \citet{purcell2006}$^{d}$ & $5.0\times10^{12}$ & 30 &36 & Mopra &  no maser cores\\
    & $2.3\times10^{13}$ & 55 &36 &  &  maser cores\\
     &$3.0\times10^{13}$ & 57 & 36& & UC HII\\
     \hline\noalign{\smallskip}
    \citet{araya2005}$^{e}$ & $4.4\times10^{14}$& 53 & 10&SEST & UC HII \\
      \hline
    \end{tabular}
  \tablefoot{a) $\theta$: assumed source size or the dimension of the beam if the emission has been assumed to fill the beam; b) data of the cool component of methyl cyanide (for details see \citealt{giannetti2017}) taken from the VizieR On-line Data Catalog J/A+A/603/A33 \citep{dativizier}; c) G34.26+0.15 has been excluded from the mean value beeing the only UC HII region in the sample; d) data from the CH$_3$CN($5_{\rm{K}}-4_{\rm{K}}$) rotational diagram; e) data from CH$_3$CN($5_{\rm{K}}-4_{\rm{K}}$), and of the lowest available (J+1)$-$J transition if 5$-$4 is not available; ${\ast}$) corrected for beam dilution, evaluating the size of the emission of each source; ${\ast\ast}$) Extended Green Objects, emitting at 4.5$\,\mu\rm{m}$. This emission may come from H$_2$ ($\nu$ = 0$-$0, S(9, 10, 11)) or CO ($\nu$ = 1$-$0) band heads
 which can be excited by shocks from protostellar outflows \citep{cyganowski2008}.
  }
    \label{tab:meanlitt}
\end{table*}
\subsection{Physical properties derived from CH$_3$CN}
 
\begin{figure}
    \centering
    %\vspace{-1cm}
    \includegraphics[trim = 0 0.2cm 0 2cm , clip,width=8.1cm]{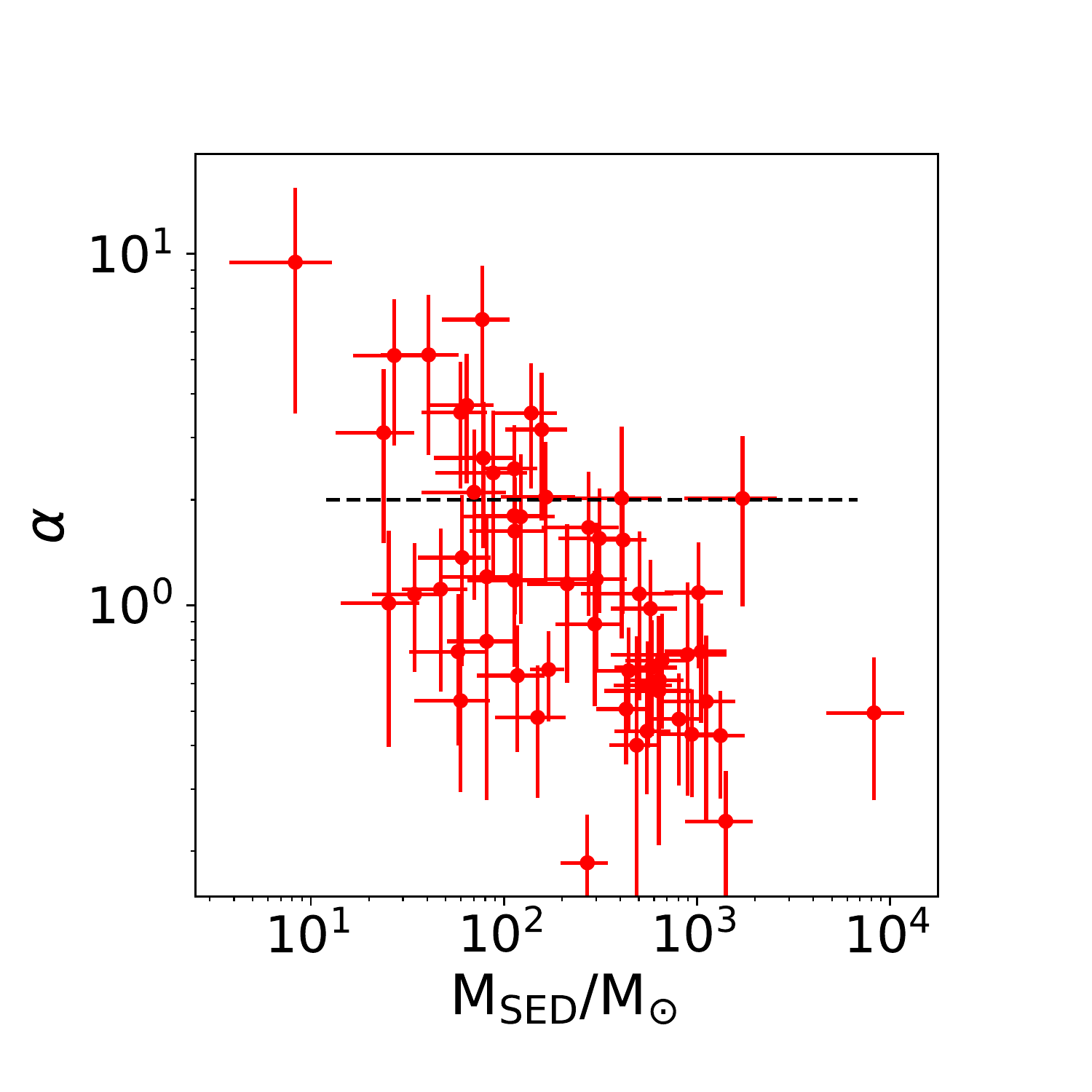}
    \caption{Plot of $\alpha=M_{\rm{vir}}/M_{\rm{SED}}$ as a function of $M_{\rm SED}$. The black dashed line represents $\alpha=2$. }
    \label{fig:alpha}
\end{figure}

We detected at least one of the CH$_3$CN(5$_{\rm{K}}-$4$_{\rm{K}}$) $K$-transitions in each of the $\sim 86\%$ of the sources observed in the sample. 
The histogram of the distributions of the column density $N^{s}_{\rm{CH_3CN}}$, excitation temperature $T_{\rm{ex}}$, and abundance $X_{\rm{CH_3CN}}$ are given in Fig. 6. %\ref{fig:histo_ch3cn}.
The column densities cover a range of three orders of magnitude, from $\sim2\times 10^{11}\,\rm{cm^{-2}}$ to $\sim 2\times10^{14}\,\rm{cm^{-2}}$, while the abundances are in the range $\sim7\times 10^{-12}-9\times10^{-10}$.\\ \indent 
Figure 7 shows that both column density and line-width (FWHM) of CH$_3$CN increase with excitation temperature, $T_{\rm ex}$. This implies that warmer sources have a higher degree of turbulence, as expected, and also higher values of the column density. Moreover, Fig. 7 shows the correlation between $T_{\rm{dust}}$ and $T_{\rm{ex}}$ of CH$_3$CN. This also highlights that $T_{\rm{ex}}$ is always larger (or in a few cases equal) than $T_{\rm{dust}}$.\\ \indent 
The virial parameter $\alpha=M_{\rm{vir}}/M_{\rm{SED}}$ vs. $M_{\rm{SED}}$ is plotted in Fig. \ref{fig:alpha}. From its definition, the virial parameter can be written as $\alpha = a(2\,E_{\rm{kin}})/E_{\rm{pot}}$ \citep{bertoldi1992}, where $E_{\rm kin}$ is the kinetic energy, $E_{\rm pot}$ is the gravitational potential energy, and $a$ is a geometrical factor that depends on the symmetry and gas density distribution of the considered "cloud". For non-magnetized gas, the critical value of $\alpha$, $\alpha_{\rm crit}$, which is the value that separates gravitationally stable ($\alpha>\alpha_{\rm crit}$) from unstable objects ($\alpha<\alpha_{\rm crit}$), is $2\pm1$ \citep{kauffmann2013}. This implies that sources with $\alpha>$2 may expand due to their kinetic motion, while sources with $\alpha<$2 are gravitationally bound. In our sample, 72\% of the sources\footnote{of the sample of 58 sources for which we detected CH$_3$CN and were able to construct the SED and derive $M_{\rm vir}$.}  show values $\alpha<2$, while only  28\%  show values above 2. However, the latter sources may be confined by the pressure of the surrounding gas. On the other hand, the presence of  magnetic fields can give support to unstable objects, leading to a lower value of $\alpha_{\rm crit}$ which depends on the strength of the magnetic field itself (see e.g. \citealt{bertoldi1992} and \citealt{kauffmann2013}).
\newline \indent 
\begin{figure}
\centering
\includegraphics[trim = 0 20 10 2cm , clip,height=6.8cm]{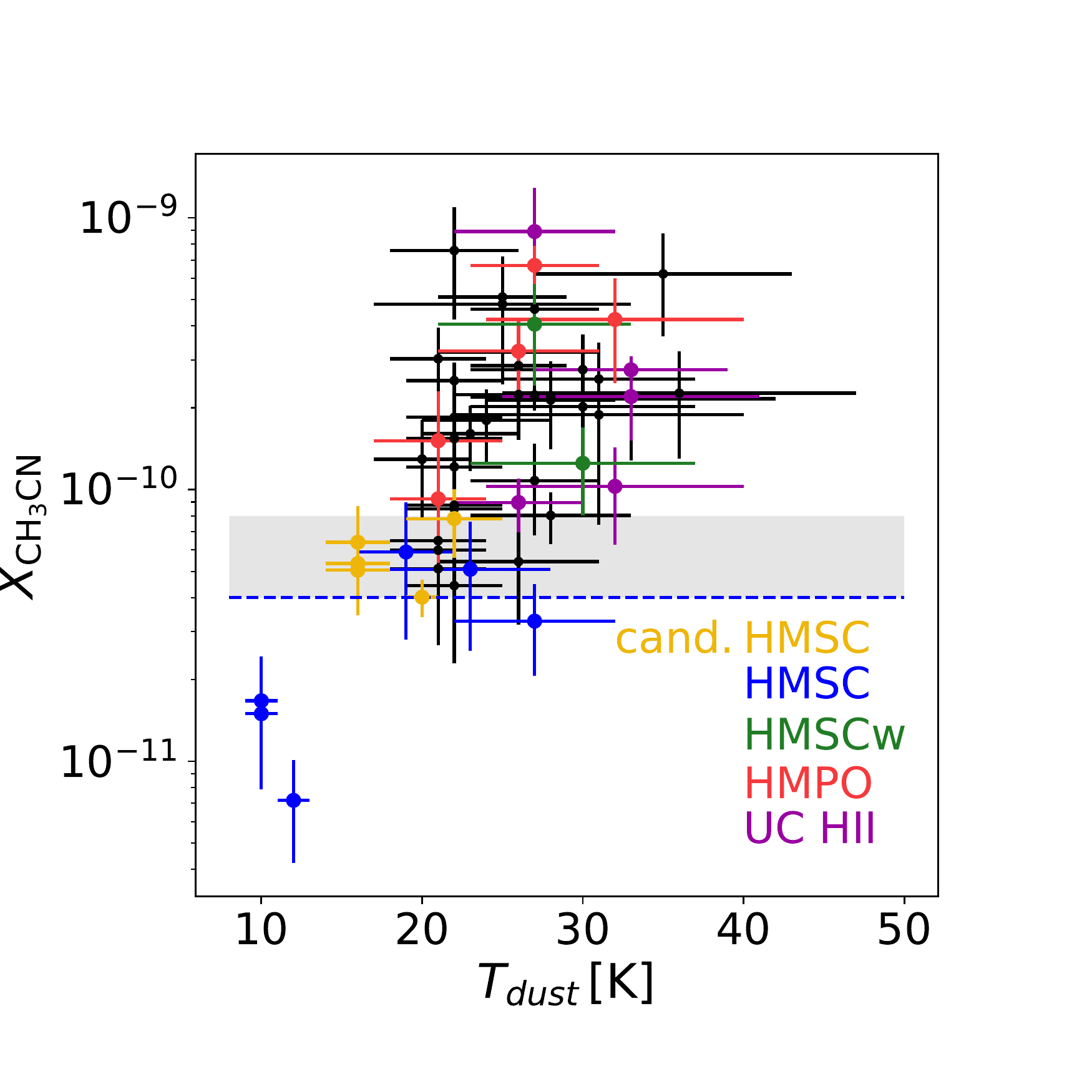}\\
\includegraphics[trim = 0 20 10 2cm , clip,height=6.8cm]{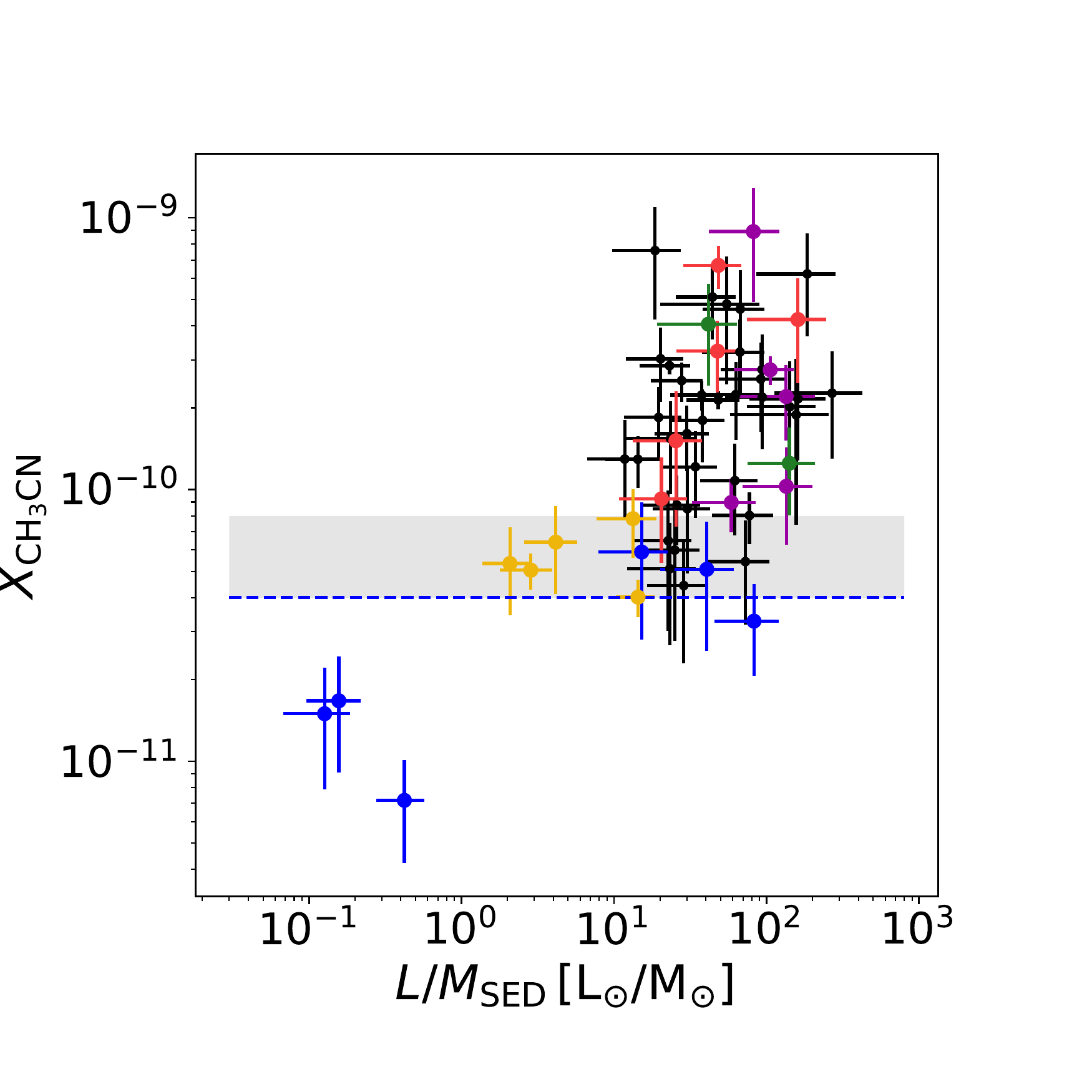}
\label{fig:ch3cn_Xevolution}
\caption{Distribution of the abundances of CH$_3$CN as a function of the two indicators of evolution $T_{\rm dust}$ (\textit{top panel}) and $L/M$ (\textit{bottom panel}). The yellow points represent the five sources of Sub-sample II identified as early evolutionary stage, candidates HMSCs or very early HMPOs. The blue dashed line indicates the conservative threshold for abundances of methyl cyanide of $4\times10^{-11}$ (only HMSCs), while the grey area indicates the abundance range in which the sources are most likely HMSCs or very early HMPOs.}
\end{figure}
The mean values of
$N^{s}_{\rm{CH_3CN}}$\footnote{For 14 sources detected in CH$_3$CN we do not have the emission maps to build the SED, and thus an estimate of $\theta$. Therefore, for these sources we have used the values of column density not corrected for the beam-dilution to calculate the mean value. For these sources we do not have an estimate of $N(\rm{H_2})$ and thus they were not included in the determination of the mean value of abundance.}, $T_{\rm{ex}}$, and  $X_{\rm{CH_3CN}}$ for the total sample are given in Table 7. %\ref{tab:meanvalues}.
From the mean values within the evolutionary classes of Sub-sample I, we can see that there is a clear positive trend of $N^{s}_{\rm{CH_3CN}}$ with evolutionary stage, with the highest difference of about one order of magnitude among HMSCs and more evolved sources. In the case of
 $X_{\rm{CH_3CN}}$, there is a clear increase from the earliest stage of star-formation to the more evolved stages, with at least one order of magnitude of difference between HMSC and HMSC$^{w}$ or more evolved sources, similarly to what was found for the column density. However, there is not a clear distinction in abundance between HMPOs and UC HII regions. An evolution of mean abundances with star-formation evolutionary phase has been also found by \citet{coletta2020} for other COMs, such as  methyl formate, dimethyl ether, and ethyl cyanide analyzing 39 sources belonging to our sample, with the clearest increasing trends found for methyl formate and dimethyl ether.\\ \indent
The mean values of $N^{s}_{\rm{CH_3CN}}$ for different evolutionary phases found by previous single-dish studies are given in Table 8. % \ref{tab:meanlitt}. 
The comparison is not straightforward considering the different assumptions,  
the different transitions observed, and the intrinsic differences between sources that can have different values of $N_{\rm{H_2}}$. \citet{hung2019} estimate the highest values of column densities, on average two orders of magnitude above the values found in our sample towards the most evolved sources. However, from Table 5 in their paper we can see that they corrected for filling factor $\eta\sim10^{-3}-10^{-4}$ in most cases, while in the sources presented in this paper  $\eta\sim7\times10^{-2}-1$. . 
For the comparison with the work of \citet{giannetti2017} we took the data from the VizieR On-line Data Catalog: J/A+A/603/A33 \citep{dativizier}. We considered only the cool component of methyl cyanide, that is able to reproduce the emission of the $\rm{CH_3CN(5_{K}-4_{K})}$ and  $\rm{(6_{K}-5_{K})}$ transitions, while a hotter component is needed for the higher energy transitions $\rm{(19_{K}-18_{K})}$ in their study. The column density is corrected for $\eta$, ranging from 1 to $\sim10^{-2}$. In Table 8 %\ref{tab:meanlitt}, 
we list the mean values over the different evolutionary phases.  These show an increase with evolutionary stage, confirming what we found with our study. The column densities of methyl cyanide towards the sample of \cite{giannetti2017} have been previously studied by \citet{sabatini2021} - considering both the cool and hot component - showing an upward trend with evolution of the mean column densities in the different evolutionary phases, and in agreement with the predictions of the chemical network presented in their work. However, the mean estimate of $N^{s}_{\rm{CH_3CN}}$  for HMSCs in this work is one order of magnitude below the mean values found by \citet{giannetti2017} for 70$\,\mu$m-weak sources. This may not be a real discrepancy if HMSCs in our work are on average less evolved than the selected 70$\,\mu$m-weak sources in \citet{giannetti2017}. 
The mean values found by \citet{rosero2013} for HMPOs are consistent with our estimates within a factor $\sim2.5$, while the values found by \citet{araya2005} in UC HII regions are on average a factor $\sim10$ larger than what we found in this paper. The mean values from \citet{purcell2006} are consistent with our estimates, with the biggest difference of a factor $\sim3$ found among the values for cores with no maser emission and radio-quiet, and our HMSCs.
 \citet{potapov2016} found a column density of methyl cyanide of $\sim6\times10^{11}\,\rm{cm^{-2}}$, in a cold dense core in TMC-1C. This is consistent with our estimates in HMSCs within a factor $\sim2$.\\ \indent
Methyl cyanide emission has been detected in all three evolutionary classes, from HMSCs to UC HII regions. This confirms what was found by \citet{olmi1996interf} and \citet{giannetti2017}, who showed
 that CH$_3$CN is not a tracer of evolved objects only: transitions associated with (relatively) low energies are detected also in very young objects, and thus are mostly tracers of dense gas.

In addition to the trend seen in Table 7 %\ref{tab:meanvalues}
for the mean values of $X_{\rm CH_3CN}$, in Fig. 8 we can see that the abundances of CH$_3$CN increase with both indicators of evolution, $T_{\rm{dust}}$, and the ratio $L/M$. The two plots show a clear correlation, although with some dispersion. From the Sub-sample I, we can see that the less evolved sources, HMSCs possibly not contaminated - i.e. HMSCs with $L/M<1\,L_{\rm \odot}/M_{\rm \odot}$ and $T_{\rm dust}<15\,\rm{K}$ - have abundances of $X_{\rm CH_3CN}$ that do not exceed the value of $2.0\times10^{-11}$, and that only HMSCs are found below $4.0\times10^{-11}$. For HMPOs and UC HII regions there is not a clear separation in the values of abundances. Moreover, we can see that five sources of the Sub-sample II show low values of $X_{\rm CH_3CN}$, close to that of HMSCs in the Sub-sample I. These sources also have low values of $L/M$ and $T_{\rm{dust}}$. These sources are G014.99$-$0.67, 18310$-$0825M3, 18445$-$0222M3, G015.02$-$0.62, and 18454$-$0136M1. The first three sources have $L/M$ between 2.0 and 4.0 $L_{\odot}/M_{\odot}$ and $T_{\rm dust}$ of 16\,K, while G015.02$-$0.62 and 18454$-$0136M1 have $L/M\sim12-15L_{\odot}/M_{\odot}$ and $T_{\rm dust}\sim20$\,K. However, G015.02$-$0.62 and G014.99$-$0.67 are not detected at $70\,\rm{\mu m}$, while the other three have flux densities consistent with those observed towards the HMSCs of Sub-sample I. 
These sources have masses in the range $\sim 1-8\times10^{2}\, M_{\odot}$ and are located within $\sim5\,\rm{kpc}$ from the Sun, but the distance ambiguity is not resolved for 18310$-$0825M3, 18445$-$0222M3, and  18454$-$0136M1, therefore these three sources may be located further away. However, both  $L/M$ and  $T_{\rm dust}$ are distance-independent and the assumption of the far distance would lead to larger values of mass. Note that there is no ambiguity in the distance of G015.02$-$0.62 and G014.99$-$0.67 ($\sim2\,\rm{kpc}$) and these sources have masses >100\,$M_{\odot}$, therefore the fact that they are not detected at $70\,\mu\rm{m}$ is not attributable to far and/or not massive enough sources. For this reason we conclude that these sources are likely in very early phases, possibly being HMSCs or early HMPOs, and that $X_{\rm CH_3CN}$ can be a very useful and practical tool to identify high-mass sources in the earliest stages of star formation, when a multiwavelength analysis to derive $T_{\rm{dust}}$ or $L/M$ is not possible. We give as conservative upper limit for HMSCs $X_{\rm CH_3CN}<4.0\times10^{-11}$, while the region between $4.0\times10^{-11}<X_{\rm CH_3CN}<7.0\times10^{-11}$ (grey area in Fig. 8) is likely populated by HMSCs or very early HMPOs.\\ \indent
\subsubsection{CH$_3$CN non-detections}
Of the 85 sources observed in CH$_3$CN($5_{{K}}-4_{{K}}$), we have 12 non-detections: 1 HMSC (G034$-$F2(MM7)) and 11 sources of the Sub-sample II, thus unclassified sources likely HMPOs or UC HII regions (see Sect. 5.1). 
\\ \indent To understand if there is a physical limit in the detection of CH$_3$CN, we have plotted the column densities of H$_2$ against the dust temperature in Fig. 9, highlighting in grey the nine sources that show no detection of methyl cyanide and for which we have been able to construct the SED. G034$-$F2(MM7) is the source that in the total sample shows the lowest temperature, being possibly the youngest object of the TOPG\"ot sample. It is thus possible that in this case not enough methyl cyanide has been formed in this source in order to be detected. Other three sources (18290$-$0924M2, G042.03+0.19 and G042.70$-$0.15) show very low values of $N_{\rm{H_2}}$ ($\sim 3.5\times10^{21}\,\rm{cm^{-2}}$): also in these cases the non-detections may be due to a limit in sensitivity. The remaining five cases are not of clear interpretation, in fact other sources with similar $N_{\rm{H_2}}$ show the emission of CH$_3$CN. Thus, in these five cases it is possible that the non-detections are related to a chemical differentiation of these sources with respect to the others.
\begin{figure}
    \centering
    \includegraphics[trim = 0 20 10 29 , clip,height=6.4cm]{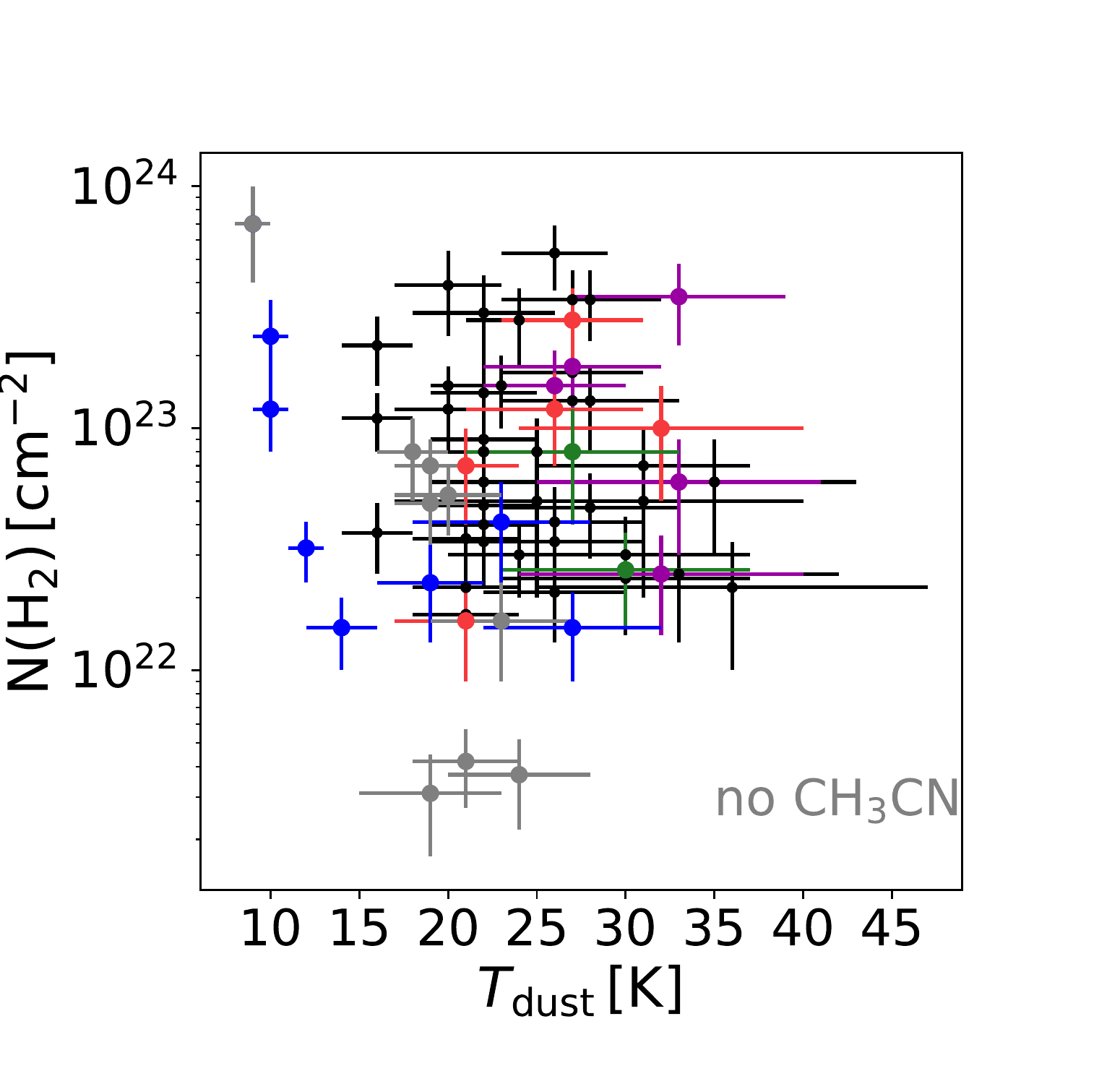}
    \caption{Distribution of sources in the plane  $N_{\rm{H_2}}-T_{\rm dust}$, with sources for which CH$_3$CN has not been detected highlighted in grey. }
    \label{fig:undetection}
\end{figure}
\section{Conclusions}
In this work we have presented the TOPG\"ot sample that arises from the combination of two separate sub-samples (Sub-sample I and Sub-sample II) of high-mass star-forming regions containing 86 sources. These sources have been observed with the IRAM 30m telescope in several spectral windows, allowing studies of different classes of molecules. In this first paper we have constructed the SEDs, derived the physical properties of the sample, and analyzed the emission of $\rm{CH_3CN(5_{K}-4_{K})}$ using MADCUBA.  We summarise below the main results of this study:
\begin{itemize}
    \item [-] We have built the SED for 69 of the 86 total sources in the sample (80\%). The derived $T_{\rm{dust}}$ and $N_{\rm{H_2}}$ are between $9-36\,\rm{K}$ and $\sim3\times 10^{21}-7\times10^{23}\,\rm{cm^{-2}}$, respectively. 
    \item [-] The luminosity spans over four orders of magnitude in the sample, from $\sim30$ to $3\times10^{5}\,L_{\odot}$, while masses vary between $\sim30$ to $8\times10^{3}\,M_{\odot}$. The luminosity-to-mass ratio $L/M$ covers three orders of magnitude from $6\times10^{-2}$ to $3\times10^{2}\,L_{\odot}/M_{\odot}$.
    \item [-]  The luminosity-to-mass ratio $L/M$, a robust evolutionary indicator as seen in previous studies \citep{molinari2008,molinari2016,molinari2019,urquhart2014,giannetti2017}, shows a tight positive correlation with $T_{\rm{dust}}$, well reproduced by a power-law ($L/M=a\,T_{\rm{dust}}^{b}$) as expected from \citet{elia2016}. The parameters of the best-fit are $a=(2.9\pm1.5)\times10^{-7}\, L_{\odot}/M_{\odot}\,\rm{K}^{-b}$ and $b=(5.8\pm0.2)$.
    \item [-] At least one of the $\rm{CH_3CN(5_{K}-4_{K})}$ K-components has been detected towards 73 sources (85\%), with 12 non-detections and one source not observed. The emission of methyl cyanide has been detected towards all the evolutionary stages, and the values of column density show a good agreement with previous studies, taking into account the difference in beam-filling factors and observations beams.
    \item [-] The mean values of  the column density of methyl cyanide show a clear positive trend with evolutionary stages. This behavior is also confirmed by the data of  \citet{giannetti2017} and their analysis by \citet{sabatini2021}. However, the mean values of observed abundances show an increase of one order of magnitude among HMSCs and more evolved sources, but no clear distinction between HMPOs and UC HII regions in the TOPG\"ot sample. An increase in abundances with evolutionary stages has been found by \citet{coletta2020} for other COMS such as methyl formate, dimethyl ether and ethyl cyanide, and it is a powerful tool to infer the evolutionary stage of a source without performing a multi-wavelength analysis.
    \item [-] From the comparison of values of $X_{\rm CH_3CN}$ in already classified sources of the Sub-sample I, we found five good candidates of HMSCs or very early HMPOs in Sub-sample II. The robustness of the $X_{\rm CH_3CN}$ value as a tracer of early evolutionary stages is confirmed by the low values of $L/M$, $T_{\rm{dust}}$, and flux densities at 70$\,\rm{\mu m}$ for these sources. This is an example of the importance of tools such as molecular evolutionary indicators. In particular, methyl cyanide is a widespread molecule which emits bright transitions, thus this result can be easily applied to identify sources in early stages.
    \item [-] We propose a conservative upper limit of $X_{\rm{CH_3CN}}$ of $4.0\times10^{-11}$ for clearly HMSCs and a range between $4.0\times10^{-11}$ and $7.0\times10^{-11}$ in which we could find HMSCs and possibly very early HMPOs.

\end{itemize}
\begin{acknowledgements}
The authors thank the anonymous referee for his/her comments, that improved this work. This work is based on observations carried out with the IRAM 30m telescope. IRAM is supported by INSU/CNRS (France), MPG (Germany) and IGN (Spain).
C.M. acknowledges support from the Italian Ministero dell'Istruzione, Universit\`a e Ricerca through the grant Progetti Premiali 2012 - iALMA (CUP C52I13000140001) and from the European Research Council (ERC) under the European Union’s  Horizon 2020 program, through the ECOGAL Synergy grant (grant ID 855130).
L.C. acknowledges financial support through Spanish grant PID2019-105552RB-C41 (MINECO/MCIU/AEI/FEDER) and from the Spanish State Research Agency (AEI) through the Unidad de Excelencia “Mar\'ia de Maeztu”-Centro de Astrobiología (CSIC-INTA) project No. MDM-2017-0737.
V.M.R. acknowledges support from the Comunidad de Madrid through the Atracci\'on de Talento Investigador Modalidad 1 (contratación de doctores con experiencia) Grant (COOL: Cosmic Origins Of Life; 2019-T1/TIC-15379).
This work was partly supported by the Italian Ministero dell Istruzione, Universit\`a e Ricerca through the grant Progetti Premiali 2012 – iALMA (CUP C$52$I$13000140001$), by the Deutsche Forschungs-gemeinschaft (DFG, German Research Foundation) - Ref no. FOR $2634$/$1$ TE $1024$/$1$-$1$, and by the DFG cluster of excellence Origins (www.origins-cluster.de). This project has received funding from the European Union's Horizon 2020 research and innovation programme under the Marie Sklodowska-Curie grant agreement No 823823 (DUSTBUSTERS) and from the European Research Council (ERC) via the ERC Synergy Grant {\em ECOGAL} (grant 855130).
We acknowledge support from the Gothenburg Centre of Advanced Studies in Science and Technology through the program “Origins of habitable planets” in 2016, when the TOPG\"ot project was initiated.

\end{acknowledgements}

%-------------------------------------------------------------------
\bibliographystyle{aa}
\bibliography{biblio.bib}

\begin{thebibliography}{96}
\expandafter\ifx\csname natexlab\endcsname\relax\def\natexlab#1{#1}\fi

\bibitem[{{Ag{\'u}ndez} {et~al.}(2015){Ag{\'u}ndez}, {Cernicharo},
  {Quintana-Lacaci}, {Velilla Prieto}, {Castro-Carrizo}, {Marcelino}, \&
  {Gu{\'e}lin}}]{agundez2015}
{Ag{\'u}ndez}, M., {Cernicharo}, J., {Quintana-Lacaci}, G., {et~al.} 2015,
  \apj, 814, 143

\bibitem[{{Araya} {et~al.}(2005){Araya}, {Hofner}, {Kurtz}, {Bronfman}, \&
  {DeDeo}}]{araya2005}
{Araya}, E., {Hofner}, P., {Kurtz}, S., {Bronfman}, L., \& {DeDeo}, S. 2005,
  \apjs, 157, 279

\bibitem[{{Arce} {et~al.}(2008){Arce}, {Santiago-Garc{\'\i}a}, {J{\o}rgensen},
  {Tafalla}, \& {Bachiller}}]{arce2008}
{Arce}, H.~G., {Santiago-Garc{\'\i}a}, J., {J{\o}rgensen}, J.~K., {Tafalla},
  M., \& {Bachiller}, R. 2008, \apjl, 681, L21

\bibitem[{{Balog} {et~al.}(2014){Balog}, {M{\"u}ller}, {Nielbock}, {Altieri},
  {Klaas}, {Blommaert}, {Linz}, {Lutz}, {Mo{\'o}r}, {Billot}, {Sauvage}, \&
  {Okumura}}]{balog2014}
{Balog}, Z., {M{\"u}ller}, T., {Nielbock}, M., {et~al.} 2014, Experimental
  Astronomy, 37, 129

\bibitem[{{Belloche} {et~al.}(2019){Belloche}, {Garrod}, {M{\"u}ller},
  {Menten}, {Medvedev}, {Thomas}, \& {Kisiel}}]{belloche2019}
{Belloche}, A., {Garrod}, R.~T., {M{\"u}ller}, H.~S.~P., {et~al.} 2019, \aap,
  628, A10

\bibitem[{{Beltr{\'a}n} {et~al.}(2011{\natexlab{a}}){Beltr{\'a}n}, {Cesaroni},
  {Neri}, \& {Codella}}]{beltran2011}
{Beltr{\'a}n}, M.~T., {Cesaroni}, R., {Neri}, R., \& {Codella}, C.
  2011{\natexlab{a}}, \aap, 525, A151

\bibitem[{{Beltr{\'a}n} {et~al.}(2004){Beltr{\'a}n}, {Cesaroni}, {Neri},
  {Codella}, {Furuya}, {Testi}, \& {Olmi}}]{beltran2004}
{Beltr{\'a}n}, M.~T., {Cesaroni}, R., {Neri}, R., {et~al.} 2004, \apjl, 601,
  L187

\bibitem[{{Beltr{\'a}n} {et~al.}(2018){Beltr{\'a}n}, {Cesaroni}, {Rivilla},
  {S{\'a}nchez-Monge}, {Moscadelli}, {Ahmadi}, {Allen}, {Beuther}, {Etoka},
  {Galli}, {Galv{\'a}n-Madrid}, {Goddi}, {Johnston}, {Klaassen},
  {K{\"o}lligan}, {Kuiper}, {Kumar}, {Maud}, {Mottram}, {Peters}, {Schilke},
  {Testi}, {van der Tak}, \& {Walmsley}}]{beltran2018}
{Beltr{\'a}n}, M.~T., {Cesaroni}, R., {Rivilla}, V.~M., {et~al.} 2018, \aap,
  615, A141

\bibitem[{{Beltr{\'a}n} {et~al.}(2011{\natexlab{b}}){Beltr{\'a}n}, {Cesaroni},
  {Zhang}, {Galv{\'a}n-Madrid}, {Beuther}, {Fallscheer}, {Neri}, \&
  {Codella}}]{beltran2011bis}
{Beltr{\'a}n}, M.~T., {Cesaroni}, R., {Zhang}, Q., {et~al.} 2011{\natexlab{b}},
  \aap, 532, A91

\bibitem[{{Beltr{\'a}n} {et~al.}(2005){Beltr{\'a}n}, {Cesaroni, R.}, {Neri,
  R.}, {Codella, C.}, {Furuya, R. S.}, {Testi, L.}, \& {Olmi,
  L.}}]{beltran2005}
{Beltr{\'a}n}, M.~T., {Cesaroni, R.}, {Neri, R.}, {et~al.} 2005, A\&A, 435, 901

\bibitem[{{Bendo} {et~al.}(2013){Bendo}, {Griffin}, {Bock}, {Conversi},
  {Dowell}, {Lim}, {Lu}, {North}, {Papageorgiou}, {Pearson}, {Pohlen},
  {Polehampton}, {Schulz}, {Shupe}, {Sibthorpe}, {Spencer}, {Swinyard},
  {Valtchanov}, \& {Xu}}]{bendo2013}
{Bendo}, G.~J., {Griffin}, M.~J., {Bock}, J.~J., {et~al.} 2013, \mnras, 433,
  3062

\bibitem[{{Bergman} \& {Hjalmarson}(1989)}]{bergman1989}
{Bergman}, P. \& {Hjalmarson}, A. 1989, {Methyl Cyanide (CH$_{3}$CN) in
  Molecular Cloud Cores}, ed. G.~{Winnewisser} \& J.~T. {Armstrong}, Vol. 331,
  124

\bibitem[{{Bergner} {et~al.}(2018){Bergner}, {Guzm{\'a}n}, {{\"O}berg},
  {Loomis}, \& {Pegues}}]{bergner2018}
{Bergner}, J.~B., {Guzm{\'a}n}, V.~G., {{\"O}berg}, K.~I., {Loomis}, R.~A., \&
  {Pegues}, J. 2018, \apj, 857, 69

\bibitem[{{Bertoldi} \& {McKee}(1992)}]{bertoldi1992}
{Bertoldi}, F. \& {McKee}, C.~F. 1992, \apj, 395, 140

\bibitem[{{Beuther} {et~al.}(2007){Beuther}, {Churchwell}, {McKee}, \&
  {Tan}}]{beuther2007}
{Beuther}, H., {Churchwell}, E.~B., {McKee}, C.~F., \& {Tan}, J.~C. 2007, in
  Protostars and Planets V, ed. B.~{Reipurth}, D.~{Jewitt}, \& K.~{Keil}, 165

\bibitem[{{Beuther} {et~al.}(2002){Beuther}, {Walsh}, {Schilke}, {Sridharan},
  {Menten}, \& {Wyrowski}}]{beuther2002}
{Beuther}, H., {Walsh}, A., {Schilke}, P., {et~al.} 2002, \aap, 390, 289

\bibitem[{{Bonnell} \& {Bate}(2006)}]{bonnell2006}
{Bonnell}, I.~A. \& {Bate}, M.~R. 2006, \mnras, 370, 488

\bibitem[{{Busquet} {et~al.}(2010){Busquet}, {Palau}, {Estalella}, {Girart},
  {S{\'a}nchez-Monge}, {Viti}, {Ho}, \& {Zhang}}]{busquet2010}
{Busquet}, G., {Palau}, A., {Estalella}, R., {et~al.} 2010, \aap, 517, L6

\bibitem[{{Cesaroni} {et~al.}(2011){Cesaroni}, {Beltr{\'a}n}, {Zhang},
  {Beuther}, \& {Fallscheer}}]{cesa2011}
{Cesaroni}, R., {Beltr{\'a}n}, M.~T., {Zhang}, Q., {Beuther}, H., \&
  {Fallscheer}, C. 2011, \aap, 533, A73

\bibitem[{{Cesaroni} {et~al.}(1999){Cesaroni}, {Felli}, {Jenness}, {Neri},
  {Olmi}, {Robberto}, {Testi}, \& {Walmsley}}]{cesaroni1999}
{Cesaroni}, R., {Felli}, M., {Jenness}, T., {et~al.} 1999, \aap, 345, 949

\bibitem[{{Churchwell} {et~al.}(1992){Churchwell}, {Walmsley}, \&
  {Wood}}]{churchwell1992}
{Churchwell}, E., {Walmsley}, C.~M., \& {Wood}, D.~O.~S. 1992, \aap, 253, 541

\bibitem[{{Codella} {et~al.}(2009){Codella}, {Benedettini}, {Beltr{\'a}n},
  {Gueth}, {Viti}, {Bachiller}, {Tafalla}, {Cabrit}, {Fuente}, \&
  {Lefloch}}]{codella2009}
{Codella}, C., {Benedettini}, M., {Beltr{\'a}n}, M.~T., {et~al.} 2009, \aap,
  507, L25

\bibitem[{{Coletta} {et~al.}(2020){Coletta}, {Fontani}, {Rivilla}, {Mininni},
  {Colzi}, {S{\'a}nchez-Monge}, \& {Beltr{\'a}n}}]{coletta2020}
{Coletta}, A., {Fontani}, F., {Rivilla}, V.~M., {et~al.} 2020, \aap, 641, A54

\bibitem[{{Colzi} {et~al.}(2018a){Colzi}, {Fontani}, {Caselli}, {Ceccarelli},
  {Hily-Blant}, \& {Bizzocchi}}]{colzi2018a}
{Colzi}, L., {Fontani}, F., {Caselli}, P., {et~al.} 2018a, \aap, 609, A129

\bibitem[{{Colzi} {et~al.}(2019){Colzi}, {Fontani}, {Caselli}, {Leurini},
  {Bizzocchi}, \& {Quaia}}]{colzi2019}
{Colzi}, L., {Fontani}, F., {Caselli}, P., {et~al.} 2019, \mnras, 485, 5543

\bibitem[{{Colzi} {et~al.}(2018b){Colzi}, {Fontani}, {Rivilla},
  {S{\'a}nchez-Monge}, {Testi}, {Beltr{\'a}n}, \& {Caselli}}]{colzi2018b}
{Colzi}, L., {Fontani}, F., {Rivilla}, V.~M., {et~al.} 2018b, \mnras, 478, 3693

\bibitem[{{Compi{\`e}gne} {et~al.}(2010){Compi{\`e}gne}, {Flagey},
  {Noriega-Crespo}, {Martin}, {Bernard}, {Paladini}, \&
  {Molinari}}]{compiegne2010}
{Compi{\`e}gne}, M., {Flagey}, N., {Noriega-Crespo}, A., {et~al.} 2010, \apjl,
  724, L44

\bibitem[{{Csengeri} {et~al.}(2014){Csengeri}, {Urquhart}, {Schuller}, {Motte},
  {Bontemps}, {Wyrowski}, {Menten}, {Bronfman}, {Beuther}, {Henning}, {Testi},
  {Zavagno}, \& {Walmsley}}]{csengeri2014}
{Csengeri}, T., {Urquhart}, J.~S., {Schuller}, F., {et~al.} 2014, \aap, 565,
  A75

\bibitem[{{Cyganowski} {et~al.}(2008){Cyganowski}, {Whitney}, {Holden},
  {Braden}, {Brogan}, {Churchwell}, {Indebetouw}, {Watson}, {Babler},
  {Benjamin}, {Gomez}, {Meade}, {Povich}, {Robitaille}, \&
  {Watson}}]{cyganowski2008}
{Cyganowski}, C.~J., {Whitney}, B.~A., {Holden}, E., {et~al.} 2008, \aj, 136,
  2391

\bibitem[{{Di Francesco} {et~al.}(2008){Di Francesco}, {Johnstone}, {Kirk},
  {MacKenzie}, \& {Ledwosinska}}]{difrancesco2008}
{Di Francesco}, J., {Johnstone}, D., {Kirk}, H., {MacKenzie}, T., \&
  {Ledwosinska}, E. 2008, \apjs, 175, 277

\bibitem[{{Elia} {et~al.}(2021){Elia}, {Merello}, {Molinari}, {Schisano},
  {Zavagno}, {Russeil}, {M{\`e}ge}, {Martin}, {Olmi}, {Pestalozzi}, {Plume},
  {Ragan}, {Benedettini}, {Eden}, {Moore}, {Noriega-Crespo}, {Paladini},
  {Palmeirim}, {Pezzuto}, {Pilbratt}, {Rygl}, {Schilke}, {Strafella}, {Tan},
  {Traficante}, {Baldeschi}, {Bally}, {di Giorgio}, {Fiorellino}, {Liu},
  {Piazzo}, \& {Polychroni}}]{elia2021}
{Elia}, D., {Merello}, M., {Molinari}, S., {et~al.} 2021, arXiv e-prints,
  arXiv:2104.04807

\bibitem[{{Elia} {et~al.}(2017){Elia}, {Molinari}, {Schisano}, {Pestalozzi},
  {Pezzuto}, {Merello}, {Noriega-Crespo}, {Moore}, {Russeil}, {Mottram},
  {Paladini}, {Strafella}, {Benedettini}, {Bernard}, {Di Giorgio}, {Eden},
  {Fukui}, {Plume}, {Bally}, {Martin}, {Ragan}, {Jaffa}, {Motte}, {Olmi},
  {Schneider}, {Testi}, {Wyrowski}, {Zavagno}, {Calzoletti}, {Faustini},
  {Natoli}, {Palmeirim}, {Piacentini}, {Piazzo}, {Pilbratt}, {Polychroni},
  {Baldeschi}, {Beltr{\'a}n}, {Billot}, {Cambr{\'e}sy}, {Cesaroni},
  {Garc{\'\i}a-Lario}, {Hoare}, {Huang}, {Joncas}, {Liu}, {Maiolo}, {Marsh},
  {Maruccia}, {M{\`e}ge}, {Peretto}, {Rygl}, {Schilke}, {Thompson},
  {Traficante}, {Umana}, {Veneziani}, {Ward-Thompson}, {Whitworth}, {Arab},
  {Band ieramonte}, {Becciani}, {Brescia}, {Buemi}, {Bufano}, {Butora},
  {Cavuoti}, {Costa}, {Fiorellino}, {Hajnal}, {Hayakawa}, {Kacsuk}, {Leto}, {Li
  Causi}, {Marchili}, {Martinavarro-Armengol}, {Mercurio}, {Molinaro},
  {Riccio}, {Sano}, {Sciacca}, {Tachihara}, {Torii}, {Trigilio}, {Vitello}, \&
  {Yamamoto}}]{elia2017}
{Elia}, D., {Molinari}, S., {Schisano}, E., {et~al.} 2017, \mnras, 471, 100

\bibitem[{{Elia} \& {Pezzuto}(2016)}]{elia2016}
{Elia}, D. \& {Pezzuto}, S. 2016, \mnras, 461, 1328

\bibitem[{{Fa{\'u}ndez} {et~al.}(2004){Fa{\'u}ndez}, {Bronfman}, {Garay},
  {Chini}, {Nyman}, \& {May}}]{faundez2004}
{Fa{\'u}ndez}, S., {Bronfman}, L., {Garay}, G., {et~al.} 2004, \aap, 426, 97

\bibitem[{{Fontani} {et~al.}(2015a){Fontani}, {Busquet}, {Palau}, {Caselli},
  {S{\'a}nchez-Monge}, {Tan}, \& {Audard}}]{fontani2015a}
{Fontani}, F., {Busquet}, G., {Palau}, A., {et~al.} 2015a, \aap, 575, A87

\bibitem[{{Fontani} {et~al.}(2015b){Fontani}, {Caselli}, {Palau}, {Bizzocchi},
  \& {Ceccarelli}}]{fontani2015b}
{Fontani}, F., {Caselli}, P., {Palau}, A., {Bizzocchi}, L., \& {Ceccarelli}, C.
  2015b, \apjl, 808, L46

\bibitem[{{Fontani} {et~al.}(2011){Fontani}, {Palau}, {Caselli},
  {S{\'a}nchez-Monge}, {Butler}, {Tan}, {Jim{\'e}nez-Serra}, {Busquet},
  {Leurini}, \& {Audard}}]{fontani2011}
{Fontani}, F., {Palau}, A., {Caselli}, P., {et~al.} 2011, \aap, 529, L7

\bibitem[{{Fontani} {et~al.}(2018){Fontani}, {Vagnoli}, {Padovani}, {Colzi},
  {Caselli}, \& {Rivilla}}]{fontani2018}
{Fontani}, F., {Vagnoli}, A., {Padovani}, M., {et~al.} 2018, \mnras, 481, 79

\bibitem[{{Furuya} {et~al.}(2008){Furuya}, {Cesaroni}, {Takahashi}, {Codella},
  {Momose}, \& {Beltr{\'a}n}}]{furuya2008}
{Furuya}, R.~S., {Cesaroni}, R., {Takahashi}, S., {et~al.} 2008, \apj, 673, 363

\bibitem[{{Giannetti} {et~al.}(2017{\natexlab{a}}){Giannetti}, {Leurini},
  {Wyrowski}, {Urquhart}, {Csengeri}, {Menten}, {Koenig}, \&
  {Guesten}}]{dativizier}
{Giannetti}, A., {Leurini}, S., {Wyrowski}, F., {et~al.} 2017{\natexlab{a}},
  {VizieR Online Data Catalog: Temperature evolution in massive clumps}

\bibitem[{{Giannetti} {et~al.}(2017{\natexlab{b}}){Giannetti}, {Leurini},
  {Wyrowski}, {Urquhart}, {Csengeri}, {Menten}, {K{\"o}nig}, \&
  {G{\"u}sten}}]{giannetti2017}
{Giannetti}, A., {Leurini}, S., {Wyrowski}, F., {et~al.} 2017{\natexlab{b}},
  \aap, 603, A33

\bibitem[{{Green}(1986)}]{Green1986}
{Green}, S. 1986, \apj, 309, 331

\bibitem[{{Hung} {et~al.}(2019){Hung}, {Liu}, {Su}, {He}, {Lee}, {Takahashi},
  \& {Chen}}]{hung2019}
{Hung}, T., {Liu}, S.-Y., {Su}, Y.-N., {et~al.} 2019, \apj, 872, 61

\bibitem[{{Immer} {et~al.}(2019){Immer}, {Li}, {Quiroga-Nu{\~n}ez}, {Reid},
  {Zhang}, {Moscadelli}, \& {Rygl}}]{immer2019}
{Immer}, K., {Li}, J., {Quiroga-Nu{\~n}ez}, L.~H., {et~al.} 2019, \aap, 632,
  A123

\bibitem[{{Johnston} {et~al.}(2015){Johnston}, {Robitaille}, {Beuther}, {Linz},
  {Boley}, {Kuiper}, {Keto}, {Hoare}, \& {van Boekel}}]{johnston2015}
{Johnston}, K.~G., {Robitaille}, T.~P., {Beuther}, H., {et~al.} 2015, \apjl,
  813, L19

\bibitem[{{Kalenskii} {et~al.}(2000){Kalenskii}, {Promislov}, {Alakoz},
  {Winnberg}, \& {Johansson}}]{kalenskii2000}
{Kalenskii}, S.~V., {Promislov}, V.~G., {Alakoz}, A., {Winnberg}, A.~V., \&
  {Johansson}, L.~E.~B. 2000, \aap, 354, 1036

\bibitem[{{Kauffmann} {et~al.}(2008){Kauffmann}, {Bertoldi}, {Bourke}, {Evans},
  \& {Lee}}]{kauffmann2008}
{Kauffmann}, J., {Bertoldi}, F., {Bourke}, T.~L., {Evans}, N.~J., I., \& {Lee},
  C.~W. 2008, \aap, 487, 993

\bibitem[{{Kauffmann} {et~al.}(2013){Kauffmann}, {Pillai}, \&
  {Goldsmith}}]{kauffmann2013}
{Kauffmann}, J., {Pillai}, T., \& {Goldsmith}, P.~F. 2013, \apj, 779, 185

\bibitem[{{Keto}(2007)}]{keto2007}
{Keto}, E. 2007, \apj, 666, 976

\bibitem[{{K{\"o}nig} {et~al.}(2017){K{\"o}nig}, {Urquhart}, {Csengeri},
  {Leurini}, {Wyrowski}, {Giannetti}, {Wienen}, {Pillai}, {Kauffmann},
  {Menten}, \& {Schuller}}]{konig2017}
{K{\"o}nig}, C., {Urquhart}, J.~S., {Csengeri}, T., {et~al.} 2017, \aap, 599,
  A139

\bibitem[{{Krumholz} {et~al.}(2009){Krumholz}, {Klein}, {McKee}, {Offner}, \&
  {Cunningham}}]{krumholz2009}
{Krumholz}, M.~R., {Klein}, R.~I., {McKee}, C.~F., {Offner}, S. S.~R., \&
  {Cunningham}, A.~J. 2009, Science, 323, 754

\bibitem[{{Loomis} {et~al.}(2018){Loomis}, {Cleeves}, {{\"O}berg}, {Aikawa},
  {Bergner}, {Furuya}, {Guzman}, \& {Walsh}}]{loomis2018}
{Loomis}, R.~A., {Cleeves}, L.~I., {{\"O}berg}, K.~I., {et~al.} 2018, \apj,
  859, 131

\bibitem[{{MacLaren} {et~al.}(1988){MacLaren}, {Richardson}, \&
  {Wolfendale}}]{maclaren1988}
{MacLaren}, I., {Richardson}, K.~M., \& {Wolfendale}, A.~W. 1988, \apj, 333,
  821

\bibitem[{{Mart{\'\i}n} {et~al.}(2019){Mart{\'\i}n}, {Mart{\'\i}n-Pintado},
  {Blanco-S{\'a}nchez}, {Rivilla}, {Rodr{\'\i}guez-Franco}, \&
  {Rico-Villas}}]{martin2019}
{Mart{\'\i}n}, S., {Mart{\'\i}n-Pintado}, J., {Blanco-S{\'a}nchez}, C.,
  {et~al.} 2019, \aap, 631, A159

\bibitem[{{McKee} \& {Tan}(2003)}]{mckee2003}
{McKee}, C.~F. \& {Tan}, J.~C. 2003, \apj, 585, 850

\bibitem[{{Minh} {et~al.}(2016){Minh}, {Liu}, \&
  {Galva{\'n}-Madrid}}]{minh2016}
{Minh}, Y.~C., {Liu}, H.~B., \& {Galva{\'n}-Madrid}, R. 2016, \apj, 824, 99

\bibitem[{{Mininni} {et~al.}(2018){Mininni}, {Fontani}, {Rivilla},
  {Beltr{\'a}n}, {Caselli}, \& {Vasyunin}}]{mininni2018}
{Mininni}, C., {Fontani}, F., {Rivilla}, V.~M., {et~al.} 2018, \mnras, 476, L39

\bibitem[{{Molinari} {et~al.}(2019){Molinari}, {Baldeschi}, {Robitaille},
  {Morales}, {Schisano}, {Traficante}, {Merello}, {Molinaro}, {Vitello},
  {Sciacca}, \& {Liu}}]{molinari2019}
{Molinari}, S., {Baldeschi}, A., {Robitaille}, T.~P., {et~al.} 2019, \mnras,
  486, 4508

\bibitem[{{Molinari} {et~al.}(1996){Molinari}, {Brand}, {Cesaroni}, \&
  {Palla}}]{molinari1996}
{Molinari}, S., {Brand}, J., {Cesaroni}, R., \& {Palla}, F. 1996, \aap, 308,
  573

\bibitem[{{Molinari} {et~al.}(2008){Molinari}, {Pezzuto}, {Cesaroni}, {Brand},
  {Faustini}, \& {Testi}}]{molinari2008}
{Molinari}, S., {Pezzuto}, S., {Cesaroni}, R., {et~al.} 2008, \aap, 481, 345

\bibitem[{{Molinari} {et~al.}(2016){Molinari}, {Schisano}, {Elia},
  {Pestalozzi}, {Traficante}, {Pezzuto}, {Swinyard}, {Noriega-Crespo}, {Bally},
  {Moore}, {Plume}, {Zavagno}, {di Giorgio}, {Liu}, {Pilbratt}, {Mottram},
  {Russeil}, {Piazzo}, {Veneziani}, {Benedettini}, {Calzoletti}, {Faustini},
  {Natoli}, {Piacentini}, {Merello}, {Palmese}, {Del Grand e}, {Polychroni},
  {Rygl}, {Polenta}, {Barlow}, {Bernard}, {Martin}, {Testi}, {Ali},
  {Andr{\'e}}, {Beltr{\'a}n}, {Billot}, {Carey}, {Cesaroni}, {Compi{\`e}gne},
  {Eden}, {Fukui}, {Garcia-Lario}, {Hoare}, {Huang}, {Joncas}, {Lim}, {Lord},
  {Martinavarro-Armengol}, {Motte}, {Paladini}, {Paradis}, {Peretto},
  {Robitaille}, {Schilke}, {Schneider}, {Schulz}, {Sibthorpe}, {Strafella},
  {Thompson}, {Umana}, {Ward-Thompson}, \& {Wyrowski}}]{molinari2016}
{Molinari}, S., {Schisano}, E., {Elia}, D., {et~al.} 2016, \aap, 591, A149

\bibitem[{{Molinari} {et~al.}(2011){Molinari}, {Schisano}, {Faustini},
  {Pestalozzi}, {di Giorgio}, \& {Liu}}]{molinari2011}
{Molinari}, S., {Schisano}, E., {Faustini}, F., {et~al.} 2011, \aap, 530, A133

\bibitem[{{Molinari} {et~al.}(2010){Molinari}, {Swinyard}, {Bally}, {Barlow},
  {Bernard}, {Martin}, {Moore}, {Noriega-Crespo}, {Plume}, {Testi}, {Zavagno},
  {Abergel}, {Ali}, {Anderson}, {Andr{\'e}}, {Baluteau}, {Battersby},
  {Beltr{\'a}n}, {Benedettini}, {Billot}, {Blommaert}, {Bontemps}, {Boulanger},
  {Brand}, {Brunt}, {Burton}, {Calzoletti}, {Carey}, {Caselli}, {Cesaroni},
  {Cernicharo}, {Chakrabarti}, {Chrysostomou}, {Cohen}, {Compiegne}, {de
  Bernardis}, {de Gasperis}, {di Giorgio}, {Elia}, {Faustini}, {Flagey},
  {Fukui}, {Fuller}, {Ganga}, {Garcia-Lario}, {Glenn}, {Goldsmith}, {Griffin},
  {Hoare}, {Huang}, {Ikhenaode}, {Joblin}, {Joncas}, {Juvela}, {Kirk},
  {Lagache}, {Li}, {Lim}, {Lord}, {Marengo}, {Marshall}, {Masi}, {Massi},
  {Matsuura}, {Minier}, {Miville-Desch{\^e}nes}, {Montier}, {Morgan}, {Motte},
  {Mottram}, {M{\"u}ller}, {Natoli}, {Neves}, {Olmi}, {Paladini}, {Paradis},
  {Parsons}, {Peretto}, {Pestalozzi}, {Pezzuto}, {Piacentini}, {Piazzo},
  {Polychroni}, {Pomar{\`e}s}, {Popescu}, {Reach}, {Ristorcelli}, {Robitaille},
  {Robitaille}, {Rod{\'o}n}, {Roy}, {Royer}, {Russeil}, {Saraceno}, {Sauvage},
  {Schilke}, {Schisano}, {Schneider}, {Schuller}, {Schulz}, {Sibthorpe},
  {Smith}, {Smith}, {Spinoglio}, {Stamatellos}, {Strafella}, {Stringfellow},
  {Sturm}, {Taylor}, {Thompson}, {Traficante}, {Tuffs}, {Umana}, {Valenziano},
  {Vavrek}, {Veneziani}, {Viti}, {Waelkens}, {Ward-Thompson}, {White},
  {Wilcock}, {Wyrowski}, {Yorke}, \& {Zhang}}]{molinari2010}
{Molinari}, S., {Swinyard}, B., {Bally}, J., {et~al.} 2010, \aap, 518, L100

\bibitem[{{Motte} {et~al.}(2018){Motte}, {Bontemps}, \& {Louvet}}]{motte2018}
{Motte}, F., {Bontemps}, S., \& {Louvet}, F. 2018, \araa, 56, 41

\bibitem[{{Mueller} {et~al.}(2002){Mueller}, {Shirley}, {Evans}, \&
  {Jacobson}}]{mueller2002}
{Mueller}, K.~E., {Shirley}, Y.~L., {Evans}, Neal~J., I., \& {Jacobson}, H.~R.
  2002, \apjs, 143, 469

\bibitem[{{M{\"u}ller} {et~al.}(2015){M{\"u}ller}, {Brown}, {Drouin},
  {Pearson}, {Kleiner}, {Sams}, {Sung}, {Ordu}, \& {Lewen}}]{muller2015}
{M{\"u}ller}, H. S.~P., {Brown}, L.~R., {Drouin}, B.~J., {et~al.} 2015, Journal
  of Molecular Spectroscopy, 312, 22

\bibitem[{{M{\"u}ller} {et~al.}(2005){M{\"u}ller}, {Schl{\"o}der}, {Stutzki},
  \& {Winnewisser}}]{cdms2005}
{M{\"u}ller}, H. S.~P., {Schl{\"o}der}, F., {Stutzki}, J., \& {Winnewisser}, G.
  2005, Journal of Molecular Structure, 742, 215

\bibitem[{{M{\"u}ller} {et~al.}(2001){M{\"u}ller}, {Thorwirth}, {Roth}, \&
  {Winnewisser}}]{cdms2001}
{M{\"u}ller}, H.~S.~P., {Thorwirth}, S., {Roth}, D.~A., \& {Winnewisser}, G.
  2001, \aap, 370, L49

\bibitem[{{{\"O}berg} {et~al.}(2015){{\"O}berg}, {Guzm{\'a}n}, {Furuya}, {Qi},
  {Aikawa}, {Andrews}, {Loomis}, \& {Wilner}}]{oberg2015}
{{\"O}berg}, K.~I., {Guzm{\'a}n}, V.~V., {Furuya}, K., {et~al.} 2015, \nat,
  520, 198

\bibitem[{{Olmi} {et~al.}(1996b){Olmi}, {Cesaroni}, {Neri}, \&
  {Walmsley}}]{olmi1996interf}
{Olmi}, L., {Cesaroni}, R., {Neri}, R., \& {Walmsley}, C.~M. 1996b, \aap, 315,
  565

\bibitem[{{Olmi} {et~al.}(1993){Olmi}, {Cesaroni}, \& {Walmsley}}]{olmi1993}
{Olmi}, L., {Cesaroni}, R., \& {Walmsley}, C.~M. 1993, \aap, 276, 489

\bibitem[{{Olmi} {et~al.}(1996a){Olmi}, {Cesaroni}, \&
  {Walmsley}}]{olmi1996singledish}
{Olmi}, L., {Cesaroni}, R., \& {Walmsley}, C.~M. 1996a, \aap, 307, 599

\bibitem[{{Ossenkopf} \& {Henning}(1994)}]{ossenkopf1994}
{Ossenkopf}, V. \& {Henning}, T. 1994, \aap, 291, 943

\bibitem[{{Pankonin} {et~al.}(2001){Pankonin}, {Churchwell}, {Watson}, \&
  {Bieging}}]{pankonin2001}
{Pankonin}, V., {Churchwell}, E., {Watson}, C., \& {Bieging}, J.~H. 2001, \apj,
  558, 194

\bibitem[{{Potapov} {et~al.}(2016){Potapov}, {S{\'a}nchez-Monge}, {Schilke},
  {Graf}, {M{\"o}ller}, \& {Schlemmer}}]{potapov2016}
{Potapov}, A., {S{\'a}nchez-Monge}, {\'A}., {Schilke}, P., {et~al.} 2016, \aap,
  594, A117

\bibitem[{{Purcell} {et~al.}(2006){Purcell}, {Balasubramanyam}, {Burton},
  {Walsh}, {Minier}, {Hunt-Cunningham}, {Kedziora-Chudczer}, {Longmore},
  {Hill}, {Bains}, {Barnes}, {Busfield}, {Calisse}, {Crighton}, {Curran},
  {Davis}, {Dempsey}, {Derragopian}, {Fulton}, {Hidas}, {Hoare}, {Lee}, {Ladd},
  {Lumsden}, {Moore}, {Murphy}, {Oudmaijer}, {Pracy}, {Rathborne}, {Robertson},
  {Schultz}, {Shobbrook}, {Sparks}, {Storey}, \& {Travouillion}}]{purcell2006}
{Purcell}, C.~R., {Balasubramanyam}, R., {Burton}, M.~G., {et~al.} 2006,
  \mnras, 367, 553

\bibitem[{{Remijan} {et~al.}(2004){Remijan}, {Sutton}, {Snyder}, {Friedel},
  {Liu}, \& {Pei}}]{remijan2004}
{Remijan}, A., {Sutton}, E.~C., {Snyder}, L.~E., {et~al.} 2004, \apj, 606, 917

\bibitem[{{Rivilla} {et~al.}(2020){Rivilla}, {Colzi}, {Fontani}, {Melosso},
  {Caselli}, {Bizzocchi}, {Tamassia}, \& {Dore}}]{rivilla2020}
{Rivilla}, V.~M., {Colzi}, L., {Fontani}, F., {et~al.} 2020, \mnras, 496, 1990

\bibitem[{{Rosero} {et~al.}(2013){Rosero}, {Hofner}, {Kurtz}, {Bieging}, \&
  {Araya}}]{rosero2013}
{Rosero}, V., {Hofner}, P., {Kurtz}, S., {Bieging}, J., \& {Araya}, E.~D. 2013,
  \apjs, 207, 12

\bibitem[{{Sabatini} {et~al.}(2021){Sabatini}, {Bovino}, {Giannetti}, {Grassi},
  {Brand}, {Schisano}, {Wyrowski}, {Leurini}, \& {Menten}}]{sabatini2021}
{Sabatini}, G., {Bovino}, S., {Giannetti}, A., {et~al.} 2021, arXiv e-prints,
  arXiv:2106.00692

\bibitem[{{S{\'a}nchez-Monge} {et~al.}(2008){S{\'a}nchez-Monge}, {Palau},
  {Estalella}, {Beltr{\'a}n}, \& {Girart}}]{sanchezmonge2008}
{S{\'a}nchez-Monge}, {\'A}., {Palau}, A., {Estalella}, R., {Beltr{\'a}n},
  M.~T., \& {Girart}, J.~M. 2008, \aap, 485, 497

\bibitem[{{S{\'a}nchez-Monge} {et~al.}(2011){S{\'a}nchez-Monge}, {Pandian}, \&
  {Kurtz}}]{sanchezmonge2011}
{S{\'a}nchez-Monge}, {\'A}., {Pandian}, J.~D., \& {Kurtz}, S. 2011, \apjl, 739,
  L9

\bibitem[{{Schuller} {et~al.}(2009){Schuller}, {Menten}, {Contreras},
  {Wyrowski}, {Schilke}, {Bronfman}, {Henning}, {Walmsley}, {Beuther},
  {Bontemps}, {Cesaroni}, {Deharveng}, {Garay}, {Herpin}, {Lefloch}, {Linz},
  {Mardones}, {Minier}, {Molinari}, {Motte}, {Nyman}, {Reveret}, {Risacher},
  {Russeil}, {Schneider}, {Testi}, {Troost}, {Vasyunina}, {Wienen}, {Zavagno},
  {Kovacs}, {Kreysa}, {Siringo}, \& {Wei{\ss}}}]{schuller2009}
{Schuller}, F., {Menten}, K.~M., {Contreras}, Y., {et~al.} 2009, \aap, 504, 415

\bibitem[{{Solomon} {et~al.}(1971){Solomon}, {Jefferts}, {Penzias}, \&
  {Wilson}}]{solomon1971}
{Solomon}, P.~M., {Jefferts}, K.~B., {Penzias}, A.~A., \& {Wilson}, R.~W. 1971,
  \apjl, 168, L107

\bibitem[{{Spezzano} {et~al.}(2017){Spezzano}, {Caselli}, {Bizzocchi},
  {Giuliano}, \& {Lattanzi}}]{spezzano2017}
{Spezzano}, S., {Caselli}, P., {Bizzocchi}, L., {Giuliano}, B.~M., \&
  {Lattanzi}, V. 2017, \aap, 606, A82

\bibitem[{{Sridharan} {et~al.}(2002){Sridharan}, {Beuther}, {Schilke},
  {Menten}, \& {Wyrowski}}]{sridharan2002}
{Sridharan}, T.~K., {Beuther}, H., {Schilke}, P., {Menten}, K.~M., \&
  {Wyrowski}, F. 2002, \apj, 566, 931

\bibitem[{{Tafalla} {et~al.}(2004){Tafalla}, {Myers}, {Caselli}, \&
  {Walmsley}}]{tafalla2004}
{Tafalla}, M., {Myers}, P.~C., {Caselli}, P., \& {Walmsley}, C.~M. 2004, \aap,
  416, 191

\bibitem[{{Tan} {et~al.}(2016){Tan}, {Kong}, {Zhang}, {Fontani}, {Caselli}, \&
  {Butler}}]{tan2016}
{Tan}, J.~C., {Kong}, S., {Zhang}, Y., {et~al.} 2016, \apjl, 821, L3

\bibitem[{{Traficante} {et~al.}(2011){Traficante}, {Calzoletti}, {Veneziani},
  {Ali}, {de Gasperis}, {di Giorgio}, {Faustini}, {Ikhenaode}, {Molinari},
  {Natoli}, {Pestalozzi}, {Pezzuto}, {Piacentini}, {Piazzo}, {Polenta}, \&
  {Schisano}}]{traficante2011}
{Traficante}, A., {Calzoletti}, L., {Veneziani}, M., {et~al.} 2011, \mnras,
  416, 2932

\bibitem[{{Urquhart} {et~al.}(2018){Urquhart}, {K{\"o}nig}, {Giannetti},
  {Leurini}, {Moore}, {Eden}, {Pillai}, {Thompson}, {Braiding}, {Burton},
  {Csengeri}, {Dempsey}, {Figura}, {Froebrich}, {Menten}, {Schuller}, {Smith},
  \& {Wyrowski}}]{urquhart2018}
{Urquhart}, J.~S., {K{\"o}nig}, C., {Giannetti}, A., {et~al.} 2018, \mnras,
  473, 1059

\bibitem[{{Urquhart} {et~al.}(2014){Urquhart}, {Moore}, {Csengeri}, {Wyrowski},
  {Schuller}, {Hoare}, {Lumsden}, {Mottram}, {Thompson}, {Menten}, {Walmsley},
  {Bronfman}, {Pfalzner}, {K{\"o}nig}, \& {Wienen}}]{urquhart2014}
{Urquhart}, J.~S., {Moore}, T.~J.~T., {Csengeri}, T., {et~al.} 2014, \mnras,
  443, 1555

\bibitem[{{Vastel} {et~al.}(2015){Vastel}, {Bottinelli}, {Caux}, {Glorian}, \&
  {Boiziot}}]{vastel2015}
{Vastel}, C., {Bottinelli}, S., {Caux}, E., {Glorian}, J.~M., \& {Boiziot}, M.
  2015, in SF2A-2015: Proceedings of the Annual meeting of the French Society
  of Astronomy and Astrophysics, 313--316

\bibitem[{{Zahorecz} {et~al.}(2016){Zahorecz}, {Jimenez-Serra}, {Wang},
  {Testi}, {T{\'o}th}, \& {Molinari}}]{zahorecz2016}
{Zahorecz}, S., {Jimenez-Serra}, I., {Wang}, K., {et~al.} 2016, \aap, 591, A105

\bibitem[{{Zhang} {et~al.}(2009){Zhang}, {Zheng}, {Reid}, {Menten}, {Xu},
  {Moscadelli}, \& {Brunthaler}}]{zhang2009}
{Zhang}, B., {Zheng}, X.~W., {Reid}, M.~J., {et~al.} 2009, \apj, 693, 419

\bibitem[{{Zhang} {et~al.}(1998){Zhang}, {Ho}, \& {Ohashi}}]{zhang1998}
{Zhang}, Q., {Ho}, P. T.~P., \& {Ohashi}, N. 1998, \apj, 494, 636

\bibitem[{{Zhang} {et~al.}(2002){Zhang}, {Hunter}, {Sridharan}, \&
  {Ho}}]{zhang2002}
{Zhang}, Q., {Hunter}, T.~R., {Sridharan}, T.~K., \& {Ho}, P. T.~P. 2002, \apj,
  566, 982

\end{thebibliography}
\clearpage
\begin{appendix}

\clearpage
\section{Methyl cyanide spectra}
\begin{minipage}{0.9\textwidth}
%\begin{figure}
\includegraphics[trim = 0.3cm 5cm 6.4cm 4cm , clip, width=16cm]{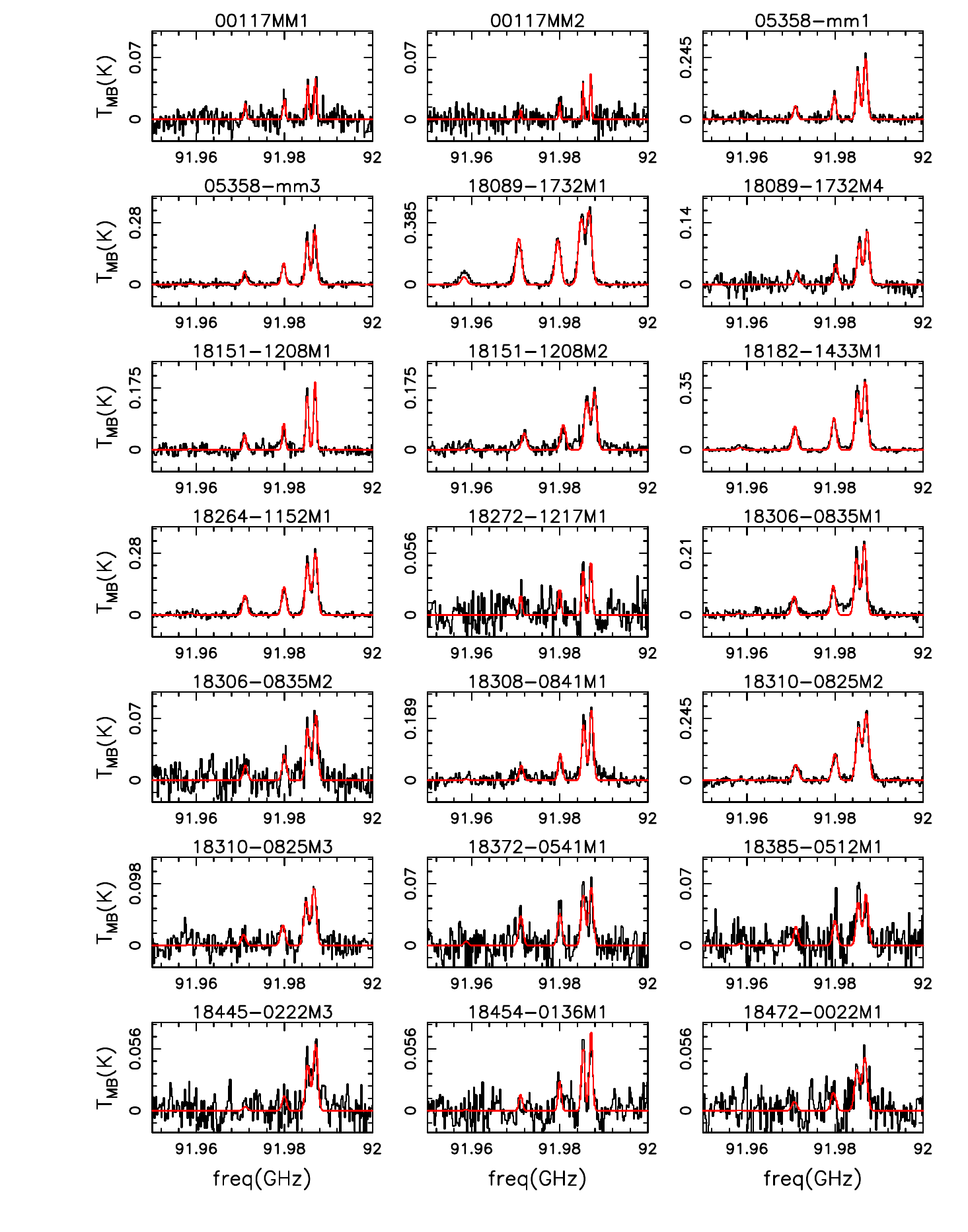}
\label{fig:spettri1}
\footnotesize{\flushleft \textbf{Fig. A.1} Spectra of CH$_3$CN(5$_{{K}}$-4$_{{K}}$) for the sources for which at least one transition has been detected. The red line is the synthetic spectrum obtained by the best fit within MADCUBA. For G31.41+0.31, the synthetic spectrum is not reported since the fit does not converge. V$_{\rm{LSR}}$ of sources G014.99-0.67 and 18445-0222M3 during observations mildly differ from the correct value. V$_{\rm{LSR}}$ has been corrected after observations, and given in Table 3.}

%\end{figure}
\end{minipage}
\clearpage
\begin{figure*}
\vspace{1cm}
\centering
\includegraphics[trim = 1cm 5cm 6cm 4cm , clip, width=17cm]{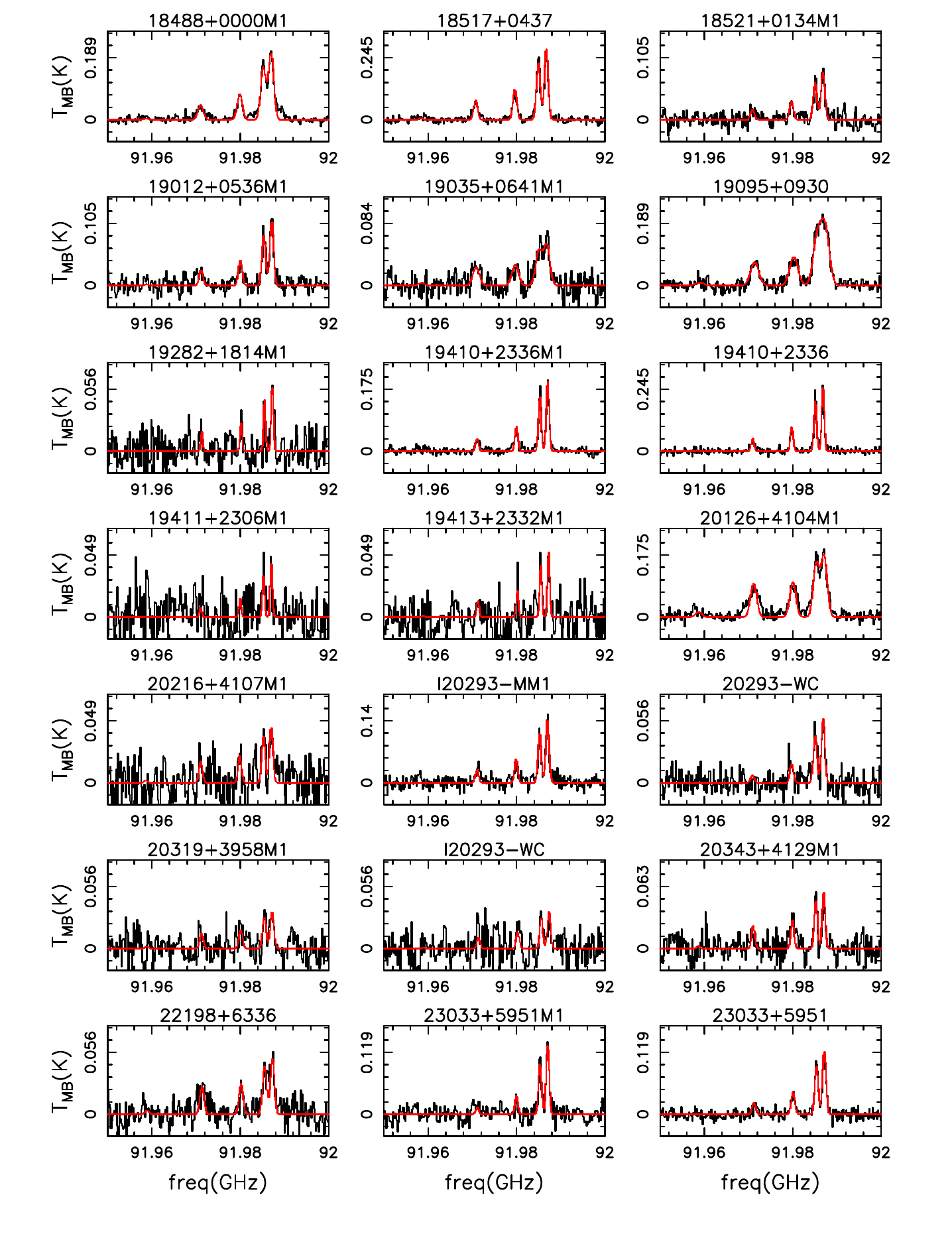}

\label{fig:spettri2}
\footnotesize{\flushleft \textbf{Fig. A.2} Continued}
\end{figure*}

\begin{figure*}
\vspace{1cm}
\centering
\includegraphics[trim = 1cm 5cm 6cm 4cm , clip, width=17cm]{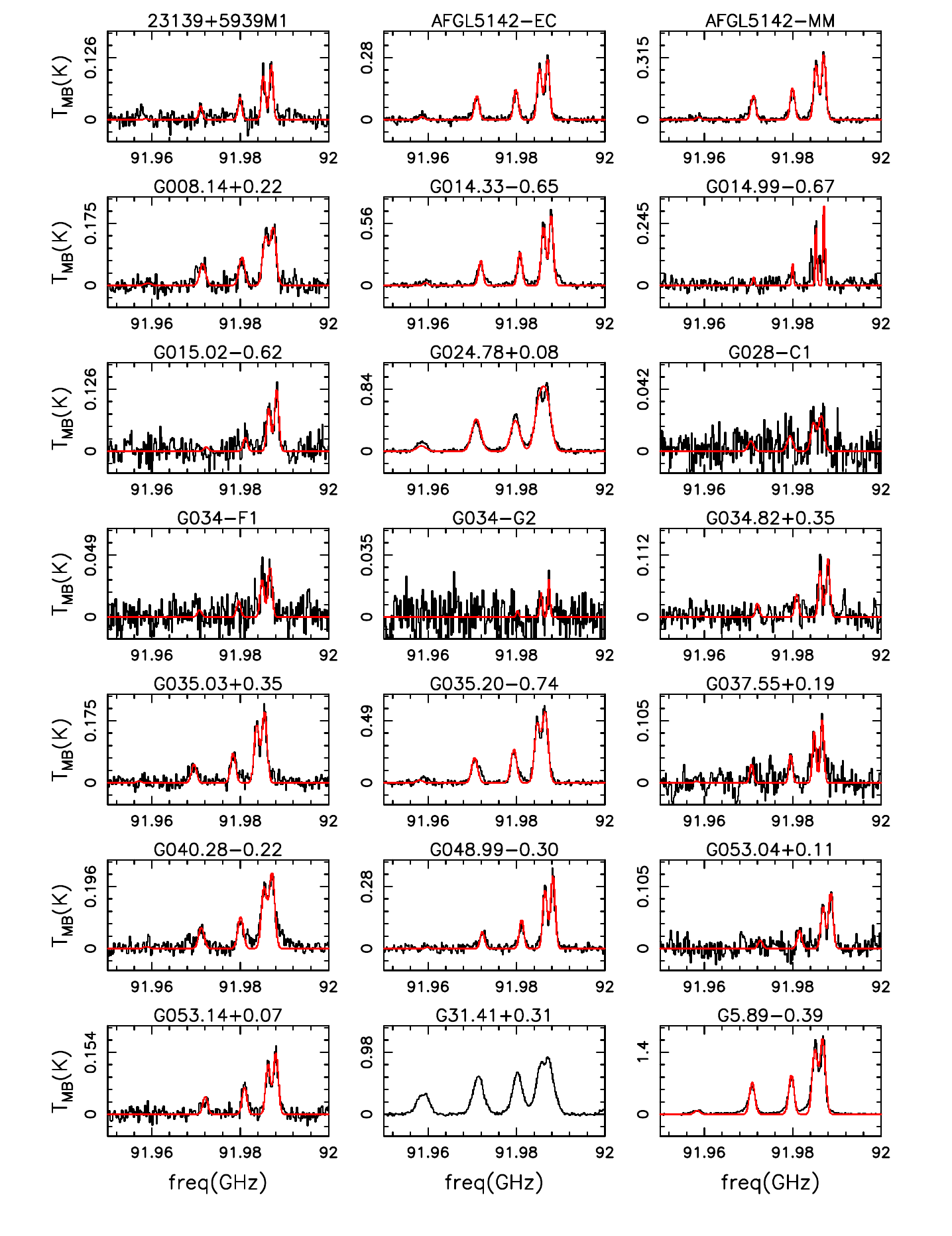}
\label{fig:spettri3}
\footnotesize{\flushleft \textbf{Fig. A.3} Continued}
\end{figure*}

\begin{figure*}
\vspace{1cm}
\centering
\includegraphics[trim = 1cm 13cm 6cm 4cm , clip, width=17cm]{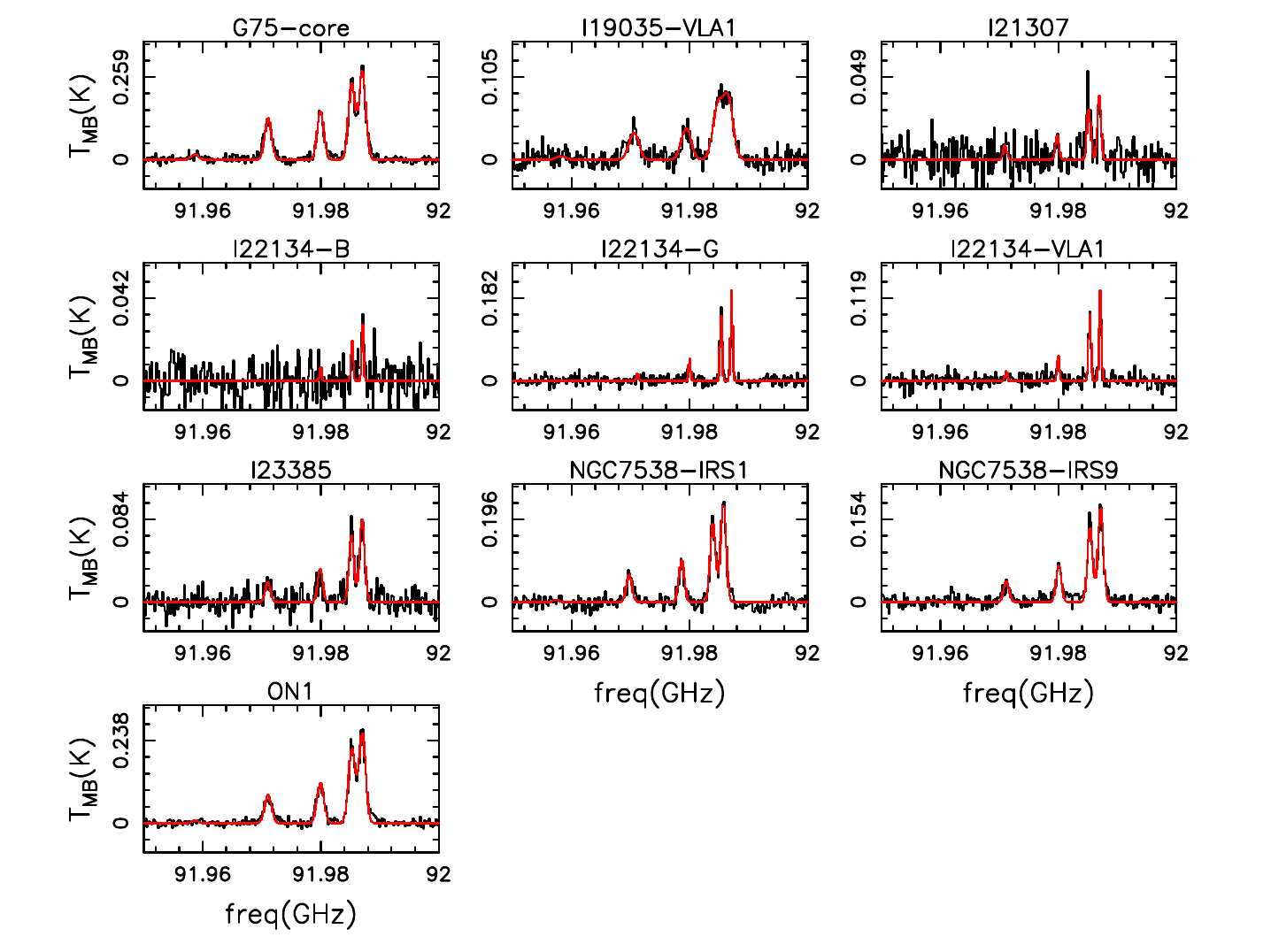}
\label{fig:spettri4}
\footnotesize{\flushleft \textbf{Fig. A.4} Continued}
\end{figure*}
\begin{figure*}
\vspace{1cm}
\centering
\includegraphics[trim = 1cm 10cm 6cm 4cm , clip, width=17cm]{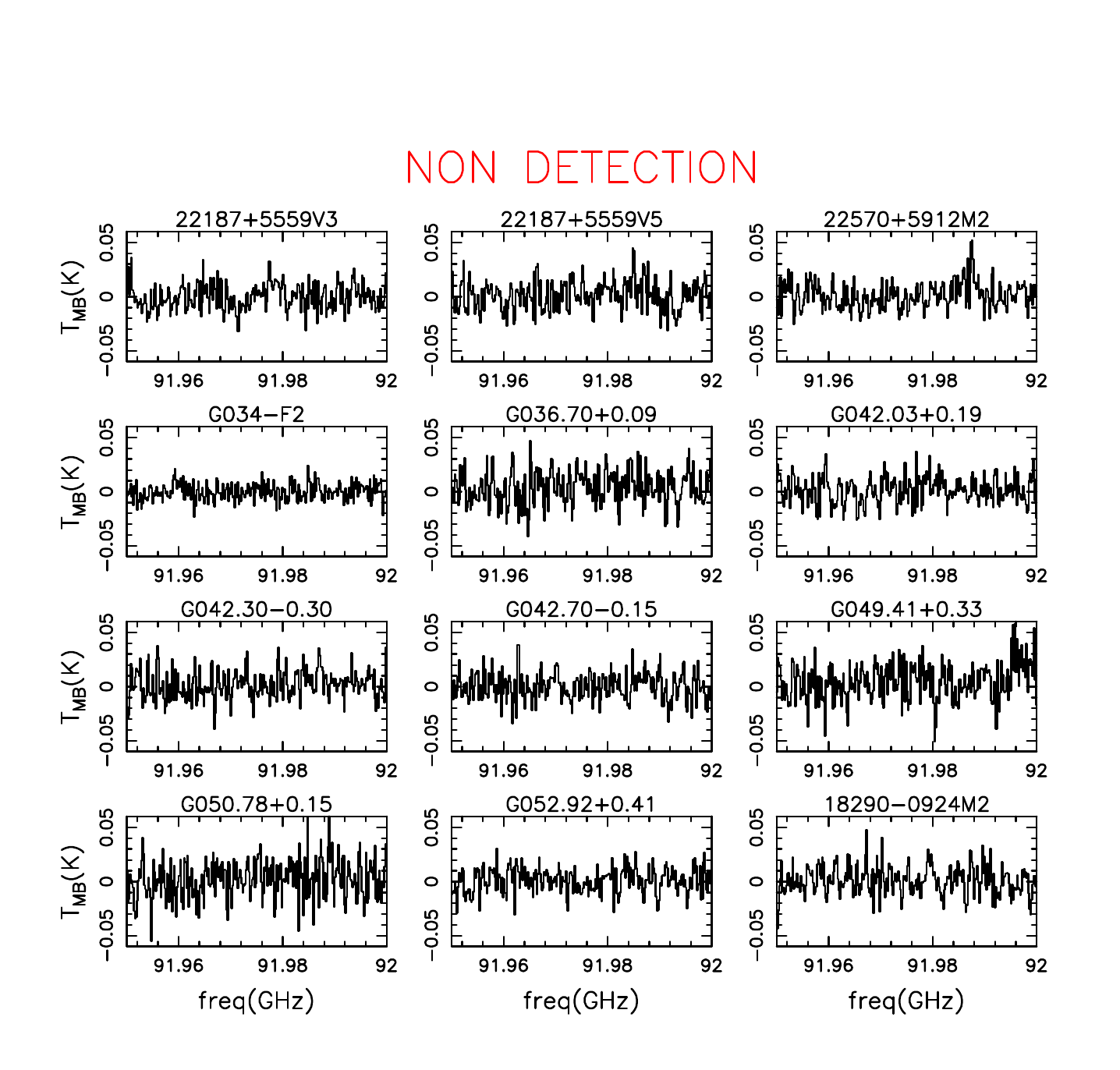}
\label{fig:spettri5}
\footnotesize{\flushleft \textbf{Fig. A.5} Spectra of CH$_3$CN(5$_{{K}}$-4$_{{K}}$) for the sources for which no transition has been detected. }
\end{figure*}
\end{appendix}
\end{document}